\documentclass[12pt]{article}
\usepackage[a4paper, margin=1in]{geometry} %

\usepackage{graphicx} %
\usepackage{titlesec}
\titleformat*{\subsection}{\large\bfseries}
\titleformat*{\subsubsection}{\small\bfseries}

\usepackage[T1]{fontenc}
\usepackage[utf8]{inputenc}
\usepackage{array}
\usepackage{longtable}
\usepackage{multirow}
\usepackage{setspace}
\usepackage{float}
\usepackage[british]{babel} %
\usepackage[autostyle=false,english=american]{csquotes} %
\usepackage{chngcntr}  %
\usepackage{xcolor}
\usepackage{pdflscape} %
\usepackage{fancyhdr}  %
\usepackage{booktabs}       %
\usepackage{threeparttable} 
\usepackage{caption}
\usepackage{subcaption}
\usepackage{adjustbox}
\usepackage{placeins}  %
\usepackage{needspace} %

\usepackage{amsmath}
\usepackage{amsfonts}
\usepackage{amsthm}
\usepackage{mathtools}
\newtheorem{proposition}{Proposition}
\theoremstyle{definition}
\newtheorem{remark}{Remark}

\makeatletter

\usepackage{microtype} %
\usepackage[hidelinks]{hyperref} %
\usepackage[capitalise]{cleveref}

\allowdisplaybreaks %

\usepackage[style=authoryear,maxcitenames=2,maxbibnames=99,backend=biber,natbib=true,hyperref=true,uniquelist=false,uniquename=false,url=false,doi=false,giveninits]{biblatex}
\renewbibmacro{in:}{}  %
\DeclareFieldFormat[article,periodical,incollection,inproceedings]{pages}{#1}  %
\renewbibmacro*{volume+number+eid}{%
  \printfield{volume}%
  \setunit*{}%
  \printfield[parens]{number}%
  \setunit{\addcomma\space}%
  \printfield{eid}}
\AtEveryBibitem{\clearfield{month}\clearfield{day}\clearlist{language}\clearfield{note}\clearfield{issn}}

\makeatother

\title{Evaluating Maternity Leave Reforms When RD Fails\thanks{von Bismarck-Osten: University College London (email: clara.bismarck-osten@ucl.ac.uk); Zeleneev: University College London and CeMMAP (email: a.zeleneev@ucl.ac.uk). We are particularly grateful to Sarah Cattan, who drew our attention to these reforms, and to Kirill Borusyak. We also thank Uta Sch\"onberg, \'Aureo de Paula, our discussant Yingying Dong, Takahiro Toriyabe, Riccardo D'Adamo, and participants at the 2026 AEA Annual Meeting. \vspace{0.3cm}
} }

\author{Clara von Bismarck-Osten \and Andrei Zeleneev} 

\date{October 2026}

\begin{document}

\maketitle
\thispagestyle{empty}

\vspace{0.4cm}

\begin{abstract}
\noindent This paper documents that expansions of UK maternity pay in the 2000s improved children's test scores while reducing maternal employment. Extending job protection alone generated similar employment effects but no detectable impact on children. None of these effects are recoverable with a conventional regression discontinuity (RD) design because eligibility is only imperfectly captured by the available data. We explain why existing corrections for measurement error in RD are not well suited and propose a simple imputation-based solution. The approach preserves RD's main advantages—nonparametric identification and transparency—and remains valid even when measurement error has economic content. 

\end{abstract}

\newpage
\section{Introduction}
\label{sec:introduction}

Although economic gender inequalities are largely driven by motherhood \parencite{kleven_children_2019,kleven_child_2019} and the early years are decisive for child development \parencite{heckman_skill_2006}, the consensus is that policy---even when targeted at early maternity---can do little to change either. This view rests on evaluations of family-policy reforms in Scandinavia and continental Europe, which struggle to document any effects \parencite{dahl_what_2016,schoenberg_expansions_2014,kleven_family_2024}, and on evidence pointing to strong cultural determinants of maternal behaviour \parencite{boelmann_wind_2021}.

Despite this consensus, there are two arguments for another reform evaluation. The first comes from the observation that most of the reforms considered in the literature are expansions of already generous benefits in strong welfare states, from which it may not be appropriate to draw general conclusions. The reforms in the United Kingdom considered in this paper are evaluated against a weaker benchmark, both in terms of the benefits themselves and the surrounding welfare environment. They also expanded job protection and pay sequentially, which allows for the effects of the two components to be distinguished. In three steps between 2000 and 2007, minimum job protection rose from 14 to 52 weeks and minimum paid leave from 18 weeks (six at 90\% of earnings, then 12 at a flat rate of \pounds 60.20 per week) to 39 weeks (six at 90\%, then 33 at up to \pounds 112.75 per week).

Beyond their substantive interest, the UK reforms serve as a warning of a methodological problem common to settings in which eligibility is only imperfectly captured by the available data. Although widely regarded as credible and transparent, the regression discontinuity (RD) designs used in many recent family-policy evaluations are not directly applicable to the reforms studied here and would not detect any effect if applied naively. Because maternity leave can begin before the child is born, each expansion covered mothers whose \emph{expected} date of birth fell on or after a cutoff. The datasets from which we take outcomes record only the actual date of birth, which leaves both the relevant running variable and treatment status unobserved.

It is tempting to remain within the RD framework, treating the actual date of birth as a mismeasured version of the true running variable and drawing on the methods developed for measurement error in RD. Measurement error smooths both the treatment probability and the conditional mean of the outcome as functions of the observed running variable, attenuating---and, in our setting, effectively eliminating---the discontinuities on which RD relies \parencite{davezies_regression_2017}. Most existing corrections rely on auxiliary information, such as instrumental variables or validation data, and require the measurement error to be non-differential ``noise'', conditionally independent of potential outcomes \parencite[see, e.g.,][]{hullegie2010effect,davezies_regression_2017,pei_devil_2017,bartalotti2021correction}. With age-based running variables, this assumption is hard to justify, because the ``measurement error'' has economic content of its own. In this case, the gap between the actual and the expected date of birth reflects premature or delayed delivery, which may itself affect children's outcomes. Another commonly employed, less demanding approach is the donut RD design \parencite[e.g.,][]{almond2011role,barreca2011saving}. It drops observations near the cutoff to restore correct treatment-status classification among the remaining observations, and then extrapolates the estimated regression functions to the cutoff to construct bias-aware confidence intervals for certain causal parameters \parencite{dong_when_2023,noack2023donut}. Because the measurement error has wide support in our setting, restoring correct classification would require dropping a large fraction of observations, and the resulting extrapolation makes inference uninformative and sensitive to the assumed smoothness of the regression functions.\footnote{See Section~\ref{ssec:challenges} for a more detailed discussion of the applicability of existing measurement-error methods for RD to our setting.}

To address the infeasibility of standard RD and the limitations of existing corrections, we propose a simple imputation-based approach. We use an untreated pre-reform cohort to impute average outcomes absent the reform under a parallel-trends assumption. We then recover the treated share at each actual birth date---the fraction of children whose expected birth date falls on or after the policy cutoff---from validation data on actual and expected birth dates. Finally, we recover the conditional average treatment effect on the treated (CATT) at each birth date with a positive treated share by dividing the difference between observed and imputed average outcomes by that share.

This imputation approach preserves important advantages of both RD and difference-in-differences (DiD) designs. As in RD, it nonparametrically identifies treatment effects conditional on the actual birth date. Moreover, unlike conventional RD, it identifies them at every birth date with a positive treated share rather than at the cutoff alone. As in DiD, it recovers counterfactual outcomes under a transparent parallel-trends assumption, whose observable implications we assess among births too early to be affected by the reform. The method is simple to implement and avoids the ill-posed inverse problems that arise in some nonparametric corrections for measurement error in RD \parencite[see, e.g.,][]{davezies_regression_2017}. Crucially for our application, it remains valid even when measurement error is differential.

We draw on three datasets. Children's outcomes come from the National Pupil Database, which records test scores at ages 5, 7 and 11 for the universe of children attending state schools in England. Mothers' outcomes come from the Quarterly Labour Force Survey, which records employment and hours worked. The validation data come from the Millennium Cohort Study, which records both the actual and the expected date of birth and from which we recover the treatment probability at each actual birth date.

Our main finding is that maternity-leave policy can improve children's outcomes where initial provision is low. A comparison of the three reforms, which extended different benefits, further suggests that the important component of maternity leave is pay rather than job protection. The clearest evidence of effects on children comes from the 2007 reform, primarily an extension of pay benefits, which raised age-5 test scores by 0.045 standard deviations, equivalent to roughly 25 additional days of age at testing. The effect is similar across maths and language and larger in rural and high-poverty areas. By age 11 it has fallen to the equivalent of about 15 days but remains statistically significant. The results are robust to the choice of pre-reform cohort, and placebo estimates for births too early to be affected by the reform are consistent with the parallel-trends assumption. Effects of this kind are rare in the literature and, when they are found at all, are largely confined to highly educated families \parencite{liu_duration_2010,danzer_paid_2018}. We attribute the difference to the lower level of provision from which the UK reforms started and to the precision afforded by our large administrative dataset. The 2003 reform, which extended both job protection and pay, raised test scores at age 5 by less than half as much as the 2007 reform, and its estimate at age 11 is too imprecise to be distinguished from zero. The 2000 extension of job protection alone had no measurable effect on children (we can rule out effects larger than about four days of age).

The results for mothers do not support the premise that maternity-leave expansions can improve children's outcomes and strengthen mothers' long-run labour-market attachment at the same time. All three reforms reduced maternal employment and hours worked, and although the point estimates become smaller, they remain negative up to six years after childbirth. The estimated labour-supply responses are broadly similar across reforms, despite their different effects on children—suggesting that additional time at home alone may not be sufficient to generate gains for children and pointing instead to a role for income support during leave.%

Our use of pre-reform cohorts builds on established practice. Evaluations of family-policy reforms introduced at birth-date cutoffs commonly use such cohorts to account for systematic differences in outcomes across birth dates, often within parametric RD-DiD regressions \parencite[e.g.,][]{dustmann_expansions_2012,schoenberg_expansions_2014,lalive2014parental,danzer_paid_2018}. Related nonparametric work on RD-DiD and difference-in-discontinuities designs uses restrictions on how confounding discontinuities evolve across cohorts to isolate the reform's effect at the cutoff \parencite{grembi_fiscal_2016,picchetti2024difference,leventer2024correcting}.\footnote{\citet{cattaneo2021extrapolating} provide another example of nonparametric RD identification away from the cutoff, using units facing different cutoffs and a parallel-trends-type assumption to recover counterfactual regression functions.} Our approach addresses a different problem in that treatment status itself is unobserved. We therefore supplement the comparison across cohorts with validation data that recover the treated share at each actual birth date. This connects our approach to fuzzy DiD, which adjusts comparisons of outcomes for differences in treatment rates \parencite{dechaisemartin_fuzzy_2018}.

Our approach may be useful in other RD settings where researchers cannot observe true treatment status because the relevant running variable is latent or measured imperfectly. It requires (i) a credible way of imputing counterfactual outcomes---here, a comparable untreated cohort---and (ii) auxiliary data that recover the treatment probability conditional on the observed running variable. Other settings that may meet these requirements include retirement or pension reforms that combine age cutoffs with imperfectly observed contribution histories \parencite{battistin2009retirement}, family-policy reforms for which eligibility is only imperfectly captured by the available data \parencite[see, e.g.,][]{carneiro_flying_2015,markussen_not_2023}, and fetal-origins designs, in which exposure to a shock is inferred from the date of birth although true exposure depends on gestational timing \parencite{almond_influenza_2006,almond_chernobyl_2009}. As in our setting, the measurement error in these applications need not be mere recording noise: it may reflect economic behaviour, such as the timing of retirement, and thus be systematically related to potential outcomes.

The rest of the paper is structured as follows. Section~\ref{sec:contextanddata} describes the institutional setting, situates the reforms in the existing literature, and introduces the data. Section~\ref{sec:methodology} presents the identification results and estimation procedures. Section~\ref{sec:findings} reports the effects of the reforms on child and maternal outcomes. Section~\ref{sec:conclusion} concludes.

\section{Context and data}
\label{sec:contextanddata}
\subsection{Context}
By European standards, the United Kingdom has had, and still has, relatively ungenerous maternity-pay provision. Leave and pay benefits were restricted to those most attached to the labour force; a reform in October 1994 entitled all pregnant employees (regardless of hours worked or length of service) to 14 weeks of maternity leave, of which the two weeks following childbirth were compulsory, while maternity pay remained conditional on tenure and earnings \parencite{gregg_employment_2007,lourie_family_1999,zabel_eligibility_2009}. The most important reforms to the legislation took place during the succession of Labour governments between 1997 and 2010, amid a wider policy agenda to (i) improve conditions for the working poor and (ii) reduce child poverty. These are the reforms studied in this paper.\footnote{The same agenda produced other family-oriented programmes, in particular the Working Families' Tax Credit (October 1999) and the Child and Working Tax Credits (April 2003). These programmes were not assigned by the expected date of birth and thus generate no discontinuities at the reform cutoffs; systematic differences between birth cohorts are absorbed by the cohort fixed effects introduced in Section~\ref{sec:methodology}.} Since then, successive governments have extended rights to fathers, but have kept overall maternity-leave provision largely unchanged; \ref{sec:reformsdetails} reviews the institutional history before and after the reforms in more detail. Relative to other European countries, the UK gives more weight to job protection than to wage replacement: job-protection guarantees are comparatively long, while pay rates are among the lowest in Europe \parencite{koslowski_17th_2021}.\footnote{In terms of the overall family friendliness of its policies, the UK was ranked 28th out of 31 OECD/EU countries by UNICEF \parencite{chzhen_are_2019}.}

\paragraph{The three reforms}
The first reform extended job-protection rights and relaxed their tenure requirements. Pregnant women with an expected date of birth on or after 30 April 2000 were the first to benefit from it.\footnote{Formally, each reform applied to women whose expected week of childbirth began on or after the threshold date specified in the relevant regulations: 30 April 2000 (SI 1999/3312), 6 April 2003 (SI 2002/2789), and 1 April 2007 (SI 2006/2014 and SI 2006/2379). Since each of these dates fell on a Sunday---the first day of an expected week of childbirth---the statutory rule coincides exactly with a rule based on the expected date of birth, which is the formulation we use throughout.} Leave available to all pregnant employees was extended from 14 to 18 weeks, and the tenure requirement for extended leave (the right to return within 29 weeks of childbirth) was reduced from two years to one. Statutory Maternity Pay was left unchanged \parencite{gregg_employment_2007,lourie_family_1999,zabel_eligibility_2009}.\footnote{The eligibility test for Maternity Allowance, the benefit for mothers who do not qualify for Statutory Maternity Pay, was reformed separately within the 2000 regime, for expected weeks of childbirth beginning on or after 20 August 2000; see \ref{sec:reformsdetails}.}

A second reform in 2003 extended job-protection rights and maternity pay. It benefited mothers with an expected date of birth on or after 6 April 2003. Leave available to all pregnant employees was extended from 18 to 26 weeks, and extended leave was lengthened from 29 weeks (counted from the week of childbirth) to 52 weeks in total, with the tenure requirement reduced from one year to 26 weeks of continuous employment (measured at the start of the 14th week before the expected week of childbirth). The reform also lengthened the flat-rate period of pay. After the six weeks paid at 90\% of previous earnings, mothers received eight additional weeks of flat-rate pay (from 12 to 20 weeks), under unchanged eligibility requirements.\footnote{The 2003 legislation also raised the flat rate from $\pounds 75$ to $\pounds 100$ per week. Because weekly rates apply by week of payment rather than by the eligibility cutoff, this increase also reached mothers who were already on leave and is therefore not part of the treatment assigned by the expected date of birth; see \ref{sec:reformsdetails}.} Similarly, Maternity Allowance (for mothers who earn too little or lack sufficient tenure with a single employer to qualify for Statutory Maternity Pay) was extended from 18 to 26 weeks. The same legislative package also introduced two weeks of paid paternity leave at the flat rate, available for children born on or after 6 April 2003 \emph{or} whose expected date of birth fell on or after that date.

A third reform in 2007 (benefiting mothers with an expected date of birth on or after 1 April 2007) extended maternity pay. It lengthened Statutory Maternity Pay from 26 to 39 weeks (the flat-rate period from 20 to 33 weeks), and Maternity Allowance likewise from 26 to 39 weeks. In addition, the reform removed the remaining 26-week tenure requirement for extended leave, making the full 52 weeks of job-protected leave available to all pregnant employees. Because this last change affected only women with fewer than 26 weeks of tenure (who, by the same tenure test, did not qualify for Statutory Maternity Pay either), we characterise the 2007 reform as primarily a pay reform. Table~\ref{tab:reforms_simple} summarises the three reforms.

\begin{table}[htb]
\begin{center}
\caption{The three UK maternity-leave reforms, 2000--2007}\label{tab:reforms_simple}
\begin{threeparttable}
\small
\setstretch{1}\renewcommand{\arraystretch}{1.15}
\begin{tabular}{l l >{\raggedright\arraybackslash}p{7.6cm}}
\toprule
Reform & Benefit affected & Main changes \\
\midrule
30 April 2000 & Job-protected leave & leave for all pregnant employees extended from 14 to 18 weeks \\
 & & tenure requirement for extended leave reduced from 2 years to 1 \\
\addlinespace
6 April 2003 & Job-protected leave & leave for all pregnant employees extended from 18 to 26 weeks \\
 & & extended leave lengthened to 52 weeks in total; tenure requirement reduced from 1 year to 26 weeks \\
 & Maternity pay & paid period extended from 18 to 26 weeks \\
 & Paternity leave & 2 weeks of paid paternity leave introduced \\
\addlinespace
1 April 2007 & Maternity pay & paid period extended from 26 to 39 weeks \\
 & Job-protected leave & full 52 weeks extended to all pregnant employees (tenure requirement removed) \\
\bottomrule
\end{tabular}
\begin{tablenotes}[flushleft]
\footnotesize
\item \textit{Notes:} Each reform applied to mothers with an expected date of birth on or after the indicated date. The flat weekly rate of maternity pay is not listed because it is uprated every April and applies by week of payment rather than by the expected date of birth; in April 2003 it was raised from \pounds 75 to \pounds 100 per week (paid at the lower of the flat rate and 90\% of earnings), far above routine upratings, and stood at \pounds 112.75 in April 2007. Table~\ref{tab:reforms_details} in \ref{sec:reformsdetails} reports the full set of eligibility requirements and entitlements under each regime, together with sources.
\end{tablenotes}
\end{threeparttable}

\end{center}
\end{table}

\begin{figure}[p]
  \centering
  \includegraphics[width=0.7\textwidth]{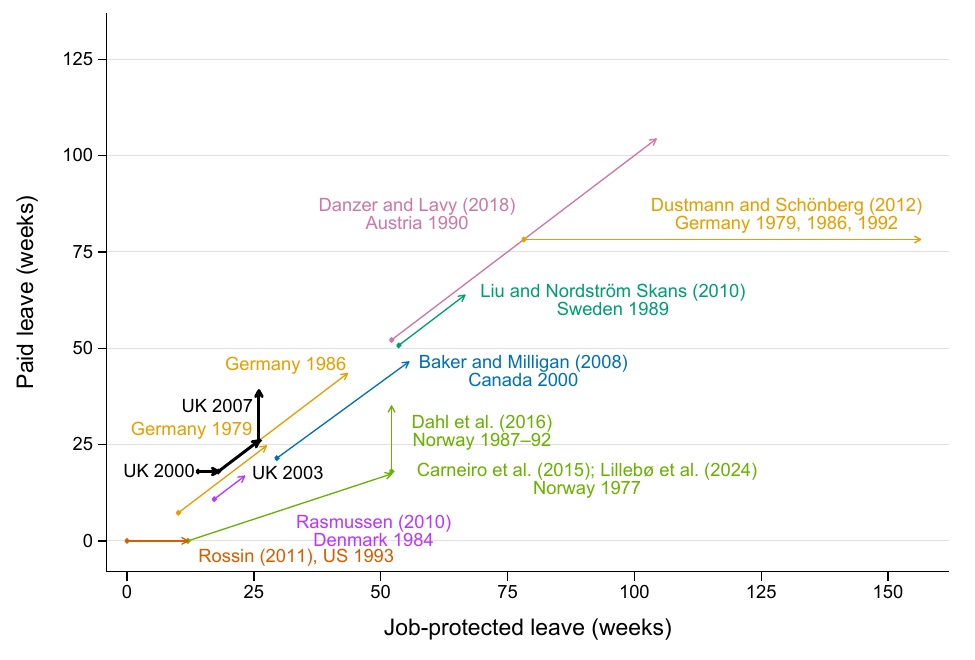} \\[0.05cm]
  {\small Panel A. Child outcomes} \\[0.35cm]
  \includegraphics[width=0.7\textwidth]{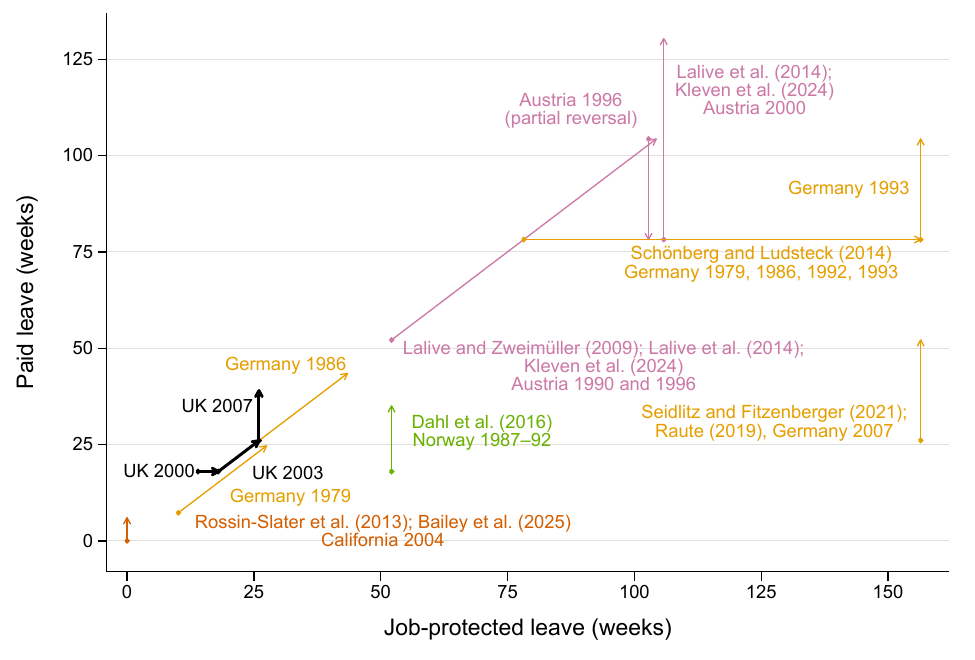} \\[0.05cm]
  {\small Panel B. Maternal outcomes} \\[0.35cm]
  \refstepcounter{figure}
  \label{fig:literature}
  {\small \textbf{Figure~\thefigure: The UK maternity-leave reforms relative to the literature}} \\[0.3cm]
  \begin{minipage}{0.9\textwidth}
      \begin{footnotesize}
          \begin{spacing}{1.15}
          \textit{Notes:} The figure locates the reforms studied in the literature on child outcomes (Panel A) and maternal outcomes (Panel B) in the space of job-protected leave (horizontal axis) and paid leave (vertical axis), both in weeks. Each arrow runs from the entitlement before a reform to the entitlement after it, labelled with the studies that evaluate it and with the country and year of the reform; where one study covers several reforms of a country, it is cited once. The black arrows are the three UK reforms studied in this paper; the downward arrow for Austria 1996 is the partial reversal of the 1990 extension. Months are converted at 4.35 weeks, and where two segments coincide, one is drawn slightly offset. Durations do not capture replacement rates or eligibility, and not all studies discussed in the text are shown. Figure~\ref{fig:top5} in \ref{sec:reformsdetails} restricts attention to studies published in the five leading economics journals.\nocite{seidlitz_effects_2021,rossin-slater_effects_2013,bailey_long-run_2025}
          \end{spacing}
      \end{footnotesize}
  \end{minipage}
\end{figure}

\paragraph{Relative to the literature}
Figure~\ref{fig:literature} locates the reforms studied in the literature in the space of pay and job-protection entitlements, and adds the three UK reforms.\footnote{We focus on natural experiments for the sake of comparability. A large complementary literature studies associations between early maternal return to work and child outcomes, finding negative associations mainly for children of highly educated mothers \parencite{waldfogel_effects_2002,berger_maternity_2005}. In addition, a randomised experiment providing monthly unconditional cash transfers to low-income families documents effects on infant brain activity associated with the development of subsequent cognitive skills \parencite{troller-renfree_impact_2022}.} First, most previously studied reforms changed both benefit types at once, moving diagonally in this space. The 2000 and 2007 UK reforms, by contrast, each move essentially along a single axis, while the 2003 reform bundles the two. Second, most of the existing evidence, in particular the studies published in the five leading economics journals (Appendix Figure~\ref{fig:top5}), comes from reforms far from the zero-benefit point, that is, from settings with an already generous benefit benchmark. The UK reforms start from a lower benchmark and are comparatively modest.

The literature on child outcomes (Panel A) suggests limited effects overall and provides little evidence separating the effects of job protection from those of paid leave. When positive effects are found, they tend to come from reforms introduced in low-welfare environments (United States: \cite{rossin_effects_2011}, who finds improvements in infant health concentrated among children of college-educated and married mothers). In higher-welfare settings, when effects are detected, they appear restricted to highly educated families (Sweden: \cite{liu_duration_2010}; Austria: \cite{danzer_paid_2018}). Whether the reforms extended job protection (Germany, 1992 expansion: \cite{dustmann_expansions_2012})\footnote{This study finds that a 1992 extension of leave coverage led to a 0.6 percentage point reduction in the share of children attending the academic track.}, paid leave (Denmark: \cite{rasmussen_increasing_2010}; Norway: \cite{dahl_what_2016}), or both (Canada: \cite{baker_maternal_2008}), the overall picture remains one of limited or no impact.\footnote{\textcite{carneiro_flying_2015} study a major Norwegian reform in 1977 that introduced four months of paid leave and one year of job protection, compared to a baseline of minimal rights. They report lasting effects on school dropout rates and earnings at age 30; a subsequent reanalysis finds that these estimates are not robust \parencite{markussen_not_2023}.}

Although the reforms studied in the literature on maternal outcomes (Panel B) move differently in the space of pay and job-protection entitlements, their results are relatively consistent. Maternity leave achieves its immediate goals: mothers spend longer at home after childbirth, and, particularly for reforms starting from little or no provision, a larger share of them subsequently returns to work rather than exiting employment. However, studies find no evidence of positive long-term effects on labour-market participation, earnings, or wage differentials. One outcome that does appear sensitive to maternity-leave reforms is future fertility \parencite{lalive_how_2009,raute_can_2019}.

\subsection{Data}\label{sec:Data}

\paragraph{On children's outcomes} To explore the impact of the reforms on children's outcomes, we rely on the National Pupil Database (NPD), an administrative database maintained by the Department for Education, covering the universe of pupils in state (or partially state-funded) schools in England. Our extract contains test scores at three stages of the school curriculum: age 5 (Early Years Foundation Stage Profile, EYFSP; not available in our extract for the cohorts used to evaluate the 2000 reform), age 7 (Key Stage 1, KS1), and age 11 (Key Stage 2, KS2). For each reform, we use the children born in the academic year of the reform (September to August) together with the preceding cohort, which serves as the control; each cohort comprises roughly 600,000 children, or around 1,600 per date of birth.{} The extract is enriched (at our request) with sociodemographic variables from the School Census and linked geographic information: free school meal eligibility, the Income Deprivation Affecting Children Index (IDACI) associated with the pupil's address, first language, and ethnicity, together with the rural--urban classification of the pupil's area. These variables are used in the heterogeneity analyses in Section~\ref{sec:findings}.

\paragraph{On maternal outcomes} To analyse the impact of the reforms on maternal outcomes, we rely on the Quarterly Labour Force Survey (QLFS), a large household survey recording the labour-market status of all household members together with their dates of birth, which allows us to link mothers to their children's dates of birth.{} Although the QLFS follows each household for up to five consecutive quarters, we use it as a series of cross-sections: for each reform, we retain mothers with a child born in the reform academic year or in the preceding (control) year, and measure their outcomes one to six years after childbirth, treating each horizon as a separate sample.\footnote{Sample sizes range from roughly 3,900 to 4,900 mothers per horizon; see the notes to Table~\ref{tab:mothers_hours}. {}} We consider two outcomes: an indicator for being employed, and the number of hours worked in the reference week, with mothers who are not employed recorded as zero hours (so that hours combine extensive- and intensive-margin responses).

\paragraph{On the distribution of actual and expected birth dates} Neither the child-outcome nor the maternal-outcome database records eligibility for the new benefits or its main determinant, the expected date of birth. To recover the treatment probability at each actual date of birth, we draw on the Millennium Cohort Study (MCS), which records both the actual and the expected date of birth for a sample of 18,818 children born in the United Kingdom between September 2000 and January 2002.\footnote{The MCS covers children born in England and Wales between September 2000 and August 2001, and in Scotland and Northern Ireland between November 2000 and January 2002; its sampling design oversamples disadvantaged areas and, in England, areas with high shares of ethnic minorities. {}} We use the MCS only to estimate the probability that a child born on a given date was \emph{expected} on or after the relevant cutoff (the treatment probability $p^t(x)$ of Section~\ref{sec:methodology}) and not to measure actual benefit receipt. The MCS children were born under the 2000 regime, one academic year after the 2000 reform-year cohort and before the 2003 and 2007 cohorts, so that the validation sample falls within the analysis period without coinciding with any of the analysis cohorts. Applying the estimated relationship between actual and expected birth dates to the reform cohorts therefore requires it to be stable over time, which we maintain as an assumption; \ref{sec:phat} states it precisely and presents an informal diagnostic (Figure~\ref{fig:stationarity}). Finally, while the NPD covers England only, the MCS covers the entire United Kingdom; we assume that the distribution of the gap between actual and expected dates of birth does not differ systematically across UK countries.

\section{Methodology}
\label{sec:methodology}
{}
\subsection{The setting and main identification challenge} %
\label{ssec:challenges}

To introduce the framework and define our causal estimands, let $i$ index a mother--child pair and $t$ the academic year in which the relevant policy reform takes effect. Let $Y_i^t(1)$ and $Y_i^t(0)$ denote the child's or mother's potential outcomes under the post-reform and pre-reform policy regimes, respectively. Let $D_i^t$ be a binary variable indicating whether mother--child pair $i$ is assigned to the post-reform policy regime. Throughout, we use ``treatment'' to refer to this policy-regime assignment rather than to actual leave-taking or receipt of maternity pay. Accordingly, the treatment effects considered below are causal effects of assignment to the post-reform policy regime.

Let $X_i^t$ and $E_i^t$ denote the child's actual and expected dates of birth, respectively, indexed by day within academic year $t$, which begins on 1 September. For example, if the child was born on 7 September but was expected to be born on 9 September, then $X_i^t=7$ and $E_i^t=9$. For each of the UK reforms studied here, assignment to the post-reform policy regime is determined by the expected date of birth:
\[
D_i^t=\mathbf{1}\{E_i^t\geq x_0\},
\]
where $x_0$ is the day index corresponding to the relevant reform cutoff. Thus, a mother--child pair is assigned to the post-reform regime if the child's expected date of birth falls on or after $x_0$, and to the pre-reform regime otherwise. We observe the realised outcome $Y_i^t=Y_i^t(D_i^t)$, the actual date of birth $X_i^t$, and other covariates. However, neither the expected date of birth $E_i^t$ nor the resulting policy-regime assignment $D_i^t$ is observed.

With $E_i^t$ interpreted as the unobserved running variable, the setup considered in this paper can be viewed as a sharp regression discontinuity (RD) design with a mismeasured running variable, where $X_i^t$ is the observed proxy for $E_i^t$ and $\varepsilon_i^t \coloneqq X_i^t-E_i^t$ plays the role of measurement error. It is well understood that measurement error in the running variable generally smooths the observed conditional expectation $\mathbb E[Y_i^t\mid X_i^t=x]$, attenuating---and, in our setting, effectively eliminating---the policy-induced discontinuity (e.g., \citealp{davezies_regression_2017}). Consequently, standard RD methods using $X_i^t$ as the running variable fail to recover the usual sharp-RD estimand ${\theta_{\rm RD}\coloneqq\mathbb E[Y_i^t(1)-Y_i^t(0)\mid E_i^t=x_0]}$. This problem has prompted the development of several approaches to RD with measurement error, some of which aim to recover $\theta_{\rm RD}$ (or its fuzzy-RD analogue) using auxiliary information or additional restrictions, while others focus on alternative estimands, such as $\theta \coloneqq\mathbb E[Y_i^t(1)-Y_i^t(0)\mid X_i^t=x_0]$.

While our setting fits within the RD with measurement error framework, it has two important features that limit the applicability of many existing approaches. First, the non-differential measurement error assumption $\bigl(Y_i^t(0),Y_i^t(1)\bigr)\perp X_i^t\mid E_i^t$, commonly imposed in this literature, is unlikely to hold in our application. Here, the ``measurement error'' $\varepsilon_i^t=X_i^t-E_i^t$ is not simply recording noise. It measures how far delivery occurs before or after the expected date. For example, premature delivery may itself affect children's potential outcomes, such as their standardised test scores. Approaches relying on restrictions of this kind (e.g., \citealp{hullegie2010effect,davezies_regression_2017,pei_devil_2017,bartalotti2021correction}) are therefore not well suited to our setting.

\begin{figure}[t]
  \centering
  \begin{minipage}[t]{0.48\textwidth}
    \centering
    \includegraphics[width=\textwidth]{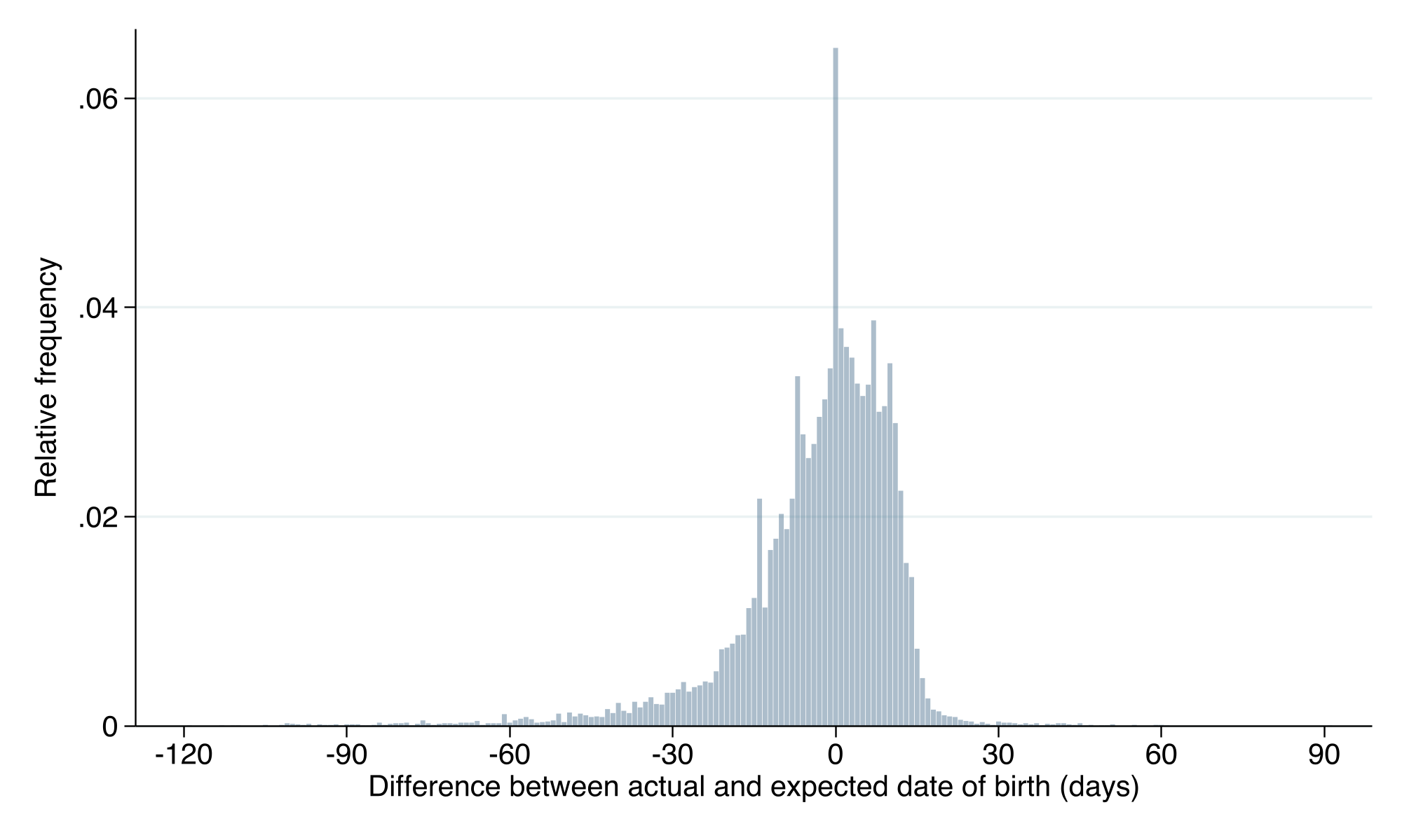} \\[0.05cm]
    {\small Panel A. Measurement error}
  \end{minipage}\hfill
  \begin{minipage}[t]{0.48\textwidth}
    \centering
    \includegraphics[width=\textwidth]{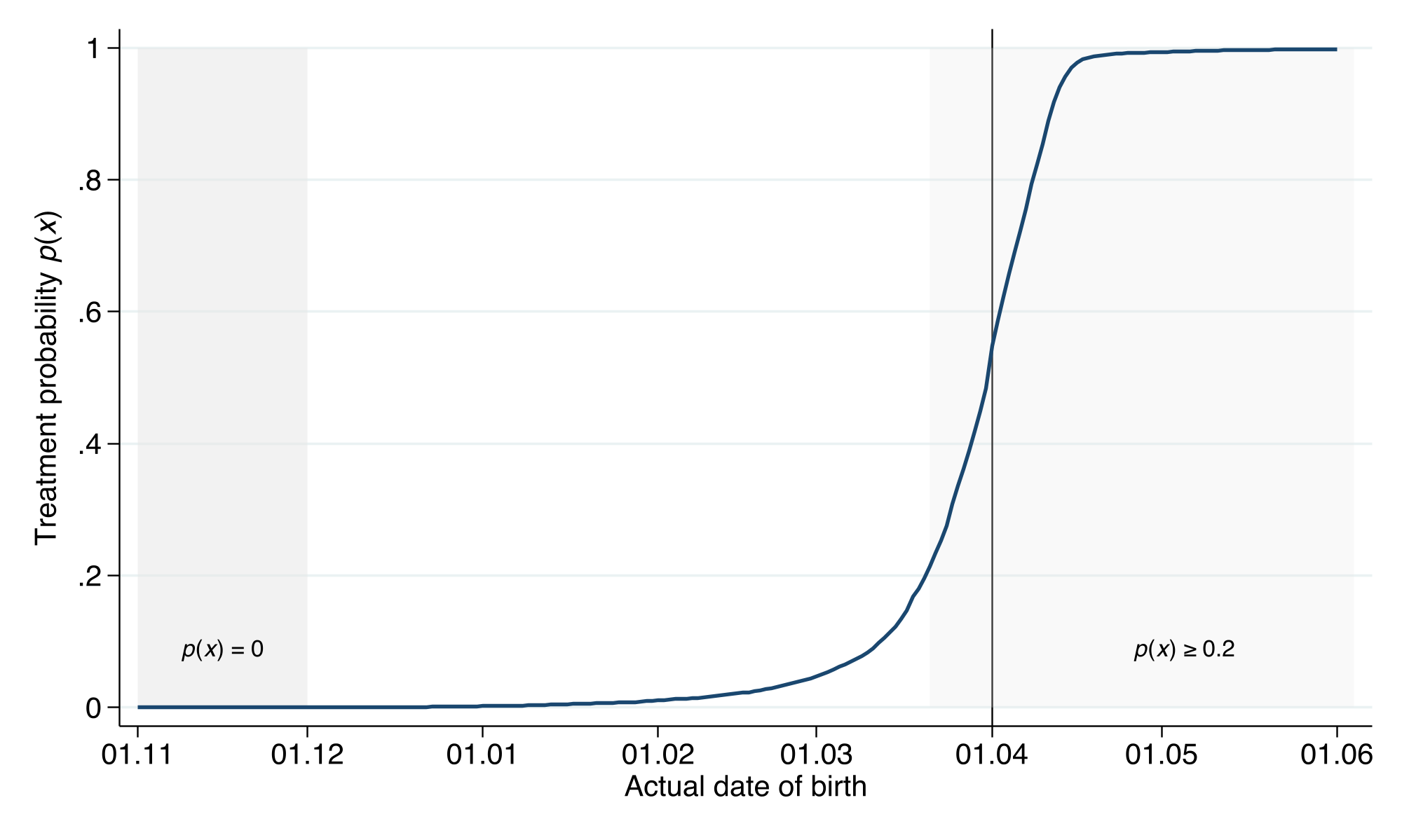} \\[0.05cm]
    {\small Panel B. Treatment probability}
  \end{minipage} \\[0.35cm]
  \refstepcounter{figure}
  \label{fig:diagnosis}
  {\small \textbf{Figure~\thefigure: Measurement error and the estimated treatment probability}} \\[0.3cm]
  \begin{minipage}{0.9\textwidth}
      \begin{footnotesize}
          \begin{spacing}{1.15}
          \textit{Notes:} Panel A shows the distribution of the difference between actual and expected dates of birth in the Millennium Cohort Study. Panel B shows the estimated treatment probability $\hat p^t(x)$ for the 1 April 2007 cutoff, obtained under the stationarity assumption discussed in \ref{sec:phat}. The left shaded area marks dates of birth before 1 December, for which $p^t(x) = 0$ and which serve to identify the reform-cohort shift $\alpha^t$ in Section~\ref{ssec:solution}; the right shaded area marks the estimation region, consisting of dates for which $\hat p^t(x) \geq 0.2$.
          \end{spacing}
      \end{footnotesize}
  \end{minipage}
\end{figure}

Second, the approaches of \citet{dong_regression_2015} and \citet{dong_when_2023} require the observed running variable to correctly classify policy-regime assignment, either in the full sample or after observations near the cutoff are excluded.\footnote{The approach of \citet{eckles2025noise} also does not apply in our setting because treatment in their framework is assigned according to the observed (error-prone) running variable $X_i^t$, so that $D_i^t=\mathbf{1}\{X_i^t\geq x_0\}$ is observed.} In our setting, however, $D_i^t$ is unobserved and generally cannot be recovered perfectly from $X_i^t$, since assignment is based on $E_i^t$ and $D_i^t\neq\mathbf{1}\{X_i^t\geq x_0\}$ for some observations. \citet{dong_when_2023} point out that correct classification can sometimes be restored in the retained sample by trimming a neighbourhood of the cutoff, in the spirit of donut RD (e.g., \citealp{almond2011role,barreca2011saving}). Under smoothness assumptions, the conditional mean functions $\mathbb E[Y_i^t(d)\mid X_i^t=x]$, $d\in\{0,1\}$, can then be extrapolated to the cutoff, permitting bias-aware inference on $\theta$ (e.g., \citealp{noack2023donut,dong_when_2023}). Because the distribution of $\varepsilon_i^t$ has wide support in our setting (see Panel~A of Figure~\ref{fig:diagnosis}), restoring correct classification would require dropping a large fraction of observations around the cutoff. The resulting inference is therefore likely to be uninformative: it would rely on substantial extrapolation and could yield wide confidence intervals that are highly sensitive to the smoothness assumptions imposed on the conditional mean functions.\footnote{Appendix Figure~\ref{fig:donutRD} also illustrates a related concern: even with linear extrapolation, conventional donut RD estimates remain highly sensitive to the choice of trimming window.}

Let
\[
p^t(x)\coloneqq \mathbb P(D_i^t=1\mid X_i^t=x)
=\mathbb P(E_i^t\geq x_0\mid X_i^t=x)
\]
denote the treatment probability at actual birth date $x$ in reform cohort $t$. For dates $x$ satisfying $0<p^t(x)<1$, the observed conditional mean can be decomposed as
\begin{align*}
\mathbb E[Y_i^t\mid X_i^t=x]
={}&p^t(x)\mathbb E[Y_i^t\mid X_i^t=x,D_i^t=1]
 +(1-p^t(x))\mathbb E[Y_i^t\mid X_i^t=x,D_i^t=0]\\*
={}&p^t(x)\mathbb E[Y_i^t(1)\mid X_i^t=x,D_i^t=1]
 +(1-p^t(x))\mathbb E[Y_i^t(0)\mid X_i^t=x,D_i^t=0]\\*
={}&p^t(x)\mathbb E[Y_i^t(1)-Y_i^t(0)\mid X_i^t=x,D_i^t=1]\\*
&+\underbrace{\begin{aligned}[t]&p^t(x)\mathbb E[Y_i^t(0)\mid X_i^t=x,D_i^t=1]\\&+(1-p^t(x))\mathbb E[Y_i^t(0)\mid X_i^t=x,D_i^t=0]\end{aligned}}_{\mathbb E[Y_i^t(0)\mid X_i^t=x]}\\*
={}&p^t(x)\mathbb E[Y_i^t(1)-Y_i^t(0)\mid X_i^t=x,D_i^t=1]
 +\mathbb E[Y_i^t(0)\mid X_i^t=x].
\end{align*}

For any date $x$ such that $p^t(x)>0$, let
\[
\mathrm{CATT}^t(x)
\coloneqq
\mathbb E[Y_i^t(1)-Y_i^t(0)\mid X_i^t=x,D_i^t=1]
\]
denote the conditional average treatment effect on the treated, and let
\[
m_0^t(x)\coloneqq\mathbb E[Y_i^t(0)\mid X_i^t=x]
\]
denote the conditional mean function of the untreated potential outcome for cohort $t$. When $p^t(x)=0$, $\mathrm{CATT}^t(x)$ is not itself defined; throughout, we adopt the convention that $p^t(x)\mathrm{CATT}^t(x)=0$. With this notation, the decomposition can be written as
\begin{align}
\mathbb E[Y_i^t\mid X_i^t=x]
=p^t(x)\mathrm{CATT}^t(x)+m_0^t(x).
\label{eq:cef}
\end{align}
At dates for which $p^t(x)=0$, equation~\eqref{eq:cef} reduces to $\mathbb E[Y_i^t\mid X_i^t=x]=m_0^t(x)$.

Notice that the representation in equation~\eqref{eq:cef} is purely definitional. It follows directly from the realised-outcome relationship $Y_i^t=Y_i^t(D_i^t)$ and requires no additional identifying assumptions. In particular, it remains valid regardless of whether the measurement error is differential or not.

Also notice that $\mathrm{CATT}^t(x)$ generally differs from the conditional average treatment effect
\[
\mathrm{CATE}^t(x)
\coloneqq
\mathbb E[Y_i^t(1)-Y_i^t(0)\mid X_i^t=x].
\]
The two estimands coincide at dates for which $p^t(x)=1$. When $0<p^t(x)<1$, they also coincide if the average treatment effect conditional on $X_i^t=x$ is the same for treated and untreated units. This condition is difficult to justify in our setting because selection into treatment conditional on $X_i^t$ is endogenous: it is determined by the measurement error $\varepsilon_i^t$, which, as discussed above, may itself affect potential outcomes. We therefore focus on identifying and estimating the conditional average treatment effect on the treated, $\mathrm{CATT}^t(x)$, taking the purely definitional representation in equation~\eqref{eq:cef} as our starting point.

\begin{figure}[t]
  \centering
  \includegraphics[width=0.7\textwidth]{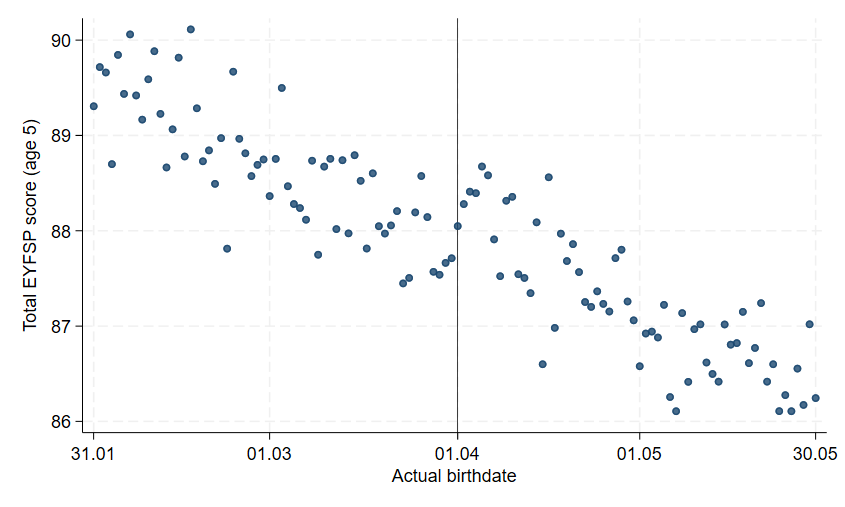} \\[0.35cm]
  \refstepcounter{figure}
  \label{fig:nodisc}
  {\small \textbf{Figure~\thefigure: $\mathbb E [Y_i^t \mid X_i^t = x ]$ for the 2007 pay reform}} \\[0.3cm]
  \begin{minipage}{0.9\textwidth}
      \begin{footnotesize}
          \begin{spacing}{1.15}
          \textit{Notes:} The figure plots average test scores at age 5 by actual date of birth for the reform cohort of the 2007 pay reform (children born in the 2006/07 academic year), for which large effects are found. The negative trend reflects the age effect: the later the child is born, the younger the child is at the time of the examination and the worse, on average, the performance.
          \end{spacing}
      \end{footnotesize}
  \end{minipage}
\end{figure}

The representation in equation~\eqref{eq:cef} also makes clear why the discontinuity on which conventional RD methods rely is absent in our setting. Because treatment is assigned according to the unobserved expected date of birth $E_i^t$, rather than the observed actual date of birth $X_i^t$, the treatment probability $p^t(x)=\Pr(E_i^t\geq x_0\mid X_i^t=x)$ changes gradually around the cutoff and, unlike in a conventional sharp RD design, does not jump from zero to one at $x_0$; see Panel~B of Figure~\ref{fig:diagnosis}. If $m_0^t(x)$ and $\mathrm{CATT}^t(x)$ are continuous at the cutoff, equation~\eqref{eq:cef} then implies that the observed conditional mean $\mathbb E[Y_i^t\mid X_i^t=x]$ is also continuous at $x_0$. Thus, conventional RD methods using $X_i^t$ as the running variable fail to detect a policy effect at the cutoff, even when $\mathrm{CATT}^t(x_0)$ is large. The 2007 reform illustrates this contrast: our approach estimates a statistically significant and sizeable effect despite the absence of a visible discontinuity in the observed conditional mean; see Figure~\ref{fig:nodisc} and Section~\ref{sec:findings} for a discussion of the findings.

Finally, notice that, even when the treatment probability $p^t(x)$ is known, the two functions $m_0^t(x)$ and $\mathrm{CATT}^t(x)$ cannot be separately identified from the single observed conditional expectation function $\mathbb E[Y_i^t\mid X_i^t=x]$ given by~\eqref{eq:cef}. Thus, identification of $\mathrm{CATT}^t(x)$ requires additional information. The next subsection shows how validation data and an untreated control cohort provide this information.

\subsection{Our solution}\label{ssec:solution}
{\noindent \bf Identification using validation data and a control cohort}

\noindent
To identify $\mathrm{CATT}^t(x)$ and construct simple sample-analogue estimators, we combine auxiliary validation data with an untreated control cohort. The validation data are used to recover the treatment probability $p^t(x)$, while the control cohort is used to impute the untreated conditional mean function $m_0^t(x)$ under a parallel-trends assumption. For ease of exposition, we initially abstract from additional covariates.

First, we use an auxiliary validation dataset, the Millennium Cohort Study (MCS), which, as described in Section~\ref{sec:contextanddata}, contains both actual and expected birth dates for a sample of 18,818 children and thus provides information on their joint distribution. Because the MCS does not cover all the reform cohorts, using this information to recover $p^t(x)$ requires a stationarity assumption that the conditional distribution of expected date of birth given actual date of birth is stable across cohorts. Under this assumption, the MCS identifies $p^t(x)$; Panel~B of Figure~\ref{fig:diagnosis} illustrates the resulting estimate for the 2007 reform, and \ref{sec:phat} describes its construction.

Second, as discussed above, knowledge of $p^t(x)$ alone is insufficient to identify $\mathrm{CATT}^t(x)$. We therefore use an untreated control cohort of mother--child pairs from the preceding academic year $t-1$ to recover $m_0^t(x)$. Because this cohort predates the reform, $D_i^{t-1}=0$ and $Y_i^{t-1}=Y_i^{t-1}(0)$, so its untreated conditional mean function is identified from observed outcomes:
\begin{align*}
    m_0^{t-1}(x)
    \coloneqq
    \mathbb E[Y_i^{t-1}(0) \mid X_i^{t-1}=x]
    =
    \mathbb E[Y_i^{t-1} \mid X_i^{t-1}=x].
\end{align*}
We link this function to the untreated conditional mean function for the reform cohort through the parallel-trends restriction
\begin{align}
    \label{eq:parallel_trends}
    m_0^t(x)=\alpha^t+m_0^{t-1}(x),
\end{align}
where $\alpha^t$ denotes an additive reform-cohort shift common across actual birth dates.

The parallel-trends restriction in \eqref{eq:parallel_trends} is the key assumption used to recover $m_0^t(x)$ and hence identify $\mathrm{CATT}^t(x)$. It allows mean untreated outcomes to differ across cohorts by the common additive term $\alpha^t$, while requiring their date-of-birth profiles to be parallel. For test-score outcomes, for example, $\alpha^t$ can capture differences in test difficulty between the reform and control cohorts. Related parallel-trends assumptions across cohorts have been used in evaluations of parental-leave reforms, typically implemented through parametric RD-DiD specifications \citep[e.g.,][]{dustmann_expansions_2012,schoenberg_expansions_2014,lalive2014parental,danzer_paid_2018}. In a nonparametric multi-cutoff RD setting, \citet{cattaneo2021extrapolating} use an analogous common-trends-type assumption across cutoff groups to recover counterfactual regression functions away from the cutoff.\footnote{The difference-in-discontinuities literature also uses related, though distinct, assumptions about the stability of confounding discontinuities across periods \citep{grembi_fiscal_2016,picchetti2024difference,leventer2024correcting}.} The restriction in \eqref{eq:parallel_trends} also has observable implications at actual birth dates for which the treatment probability is zero, $p^t(x)=0$. Remark~\ref{rem:over_id} describes how we assess these implications empirically.
{}

Given \eqref{eq:parallel_trends}, identifying $m_0^t(x)$, and hence $\mathrm{CATT}^t(x)$, requires identification of the reform-cohort shift $\alpha^t$. We identify this shift using a conservative region of actual birth dates sufficiently far before the policy cutoff that eligibility would require an exceptionally premature delivery. Since such deliveries are extremely rare among live births and survival among children born this early is very low, we take $p^t(x)$ to be zero throughout this region (see Panel~B of Figure~\ref{fig:diagnosis} for an illustration). Let $\tilde{x}$ denote any date in this region, so that $p^t(\tilde{x})=0$. At such a date, \eqref{eq:cef} implies that
\begin{align*}
    \mathbb E[Y_i^t \mid X_i^t=\tilde{x}]=m_0^t(\tilde{x}).
\end{align*}
Combining this equality with \eqref{eq:parallel_trends} identifies $\alpha^t$ as
\begin{align}
    \label{eq:alpha_ID}
    \alpha^t
    =
    \mathbb E[Y_i^t \mid X_i^t=\tilde{x}]
    -
    \mathbb E[Y_i^{t-1} \mid X_i^{t-1}=\tilde{x}].
\end{align}
This, in turn, identifies $\mathrm{CATT}^t(x)$ at every actual birth date $x$ for which $p^t(x)>0$. The following proposition summarises this result.

\begin{proposition}
    \label{prop:NPID}
    Suppose that $p^t(\cdot)$ is identified from the validation data, the parallel-trends restriction in \eqref{eq:parallel_trends} holds, and $p^t(\tilde{x})=0$ for some actual birth date $\tilde{x}$. Then, for every $x$ such that $p^t(x)>0$, $\mathrm{CATT}^t(x)$ is nonparametrically identified as
    \begin{align*}
        \mathrm{CATT}^t(x)
        =
        \frac{
        \mathbb E[Y_i^t \mid X_i^t=x]
        -
        \mathbb E[Y_i^{t-1} \mid X_i^{t-1}=x]
        -
        \alpha^t
        }{p^t(x)},
    \end{align*}
    where $\alpha^t$ is given by \eqref{eq:alpha_ID}.
\end{proposition}

Proposition~\ref{prop:NPID} establishes identification of $\mathrm{CATT}^t(x)$ for all actual birth dates $x$ with $p^t(x)>0$, both at and away from the policy cutoff. Importantly, this result is nonparametric: we do not impose a particular functional form or any smoothness assumptions on the untreated conditional mean function $m_0^t(x)$, and we do not restrict treatment-effect heterogeneity across birth dates, allowing $\mathrm{CATT}^t(x)$ to vary freely with $x$. Thus, unlike donut RD approaches, our result does not rely on extrapolation of the outcome regression functions. Most importantly for our setting, our approach allows the measurement error $\varepsilon_i^t=X_i^t-E_i^t$ to affect potential outcomes, whereas many existing approaches to RD with measurement error require it to be non-differential.

\begin{remark}[Connection to fuzzy difference-in-differences]
    \label{rem:fuzzy_did}
    The estimand identified in Proposition~\ref{prop:NPID} can be written in a form closely related to fuzzy difference-in-differences.\footnote{We thank our discussant Yingying Dong for pointing out this connection.} Specifically, substituting \eqref{eq:alpha_ID} into the identifying expression gives
    \begin{align}
        \label{eq:fuzzy_did}
        \mathrm{CATT}^t(x) = \frac{1}{p^t(x)}\Big\{\big[\mathbb E(Y_i^t \mid X_i^t=x)-\mathbb E(Y_i^{t-1} \mid X_i^{t-1}=x)\big] \notag\\
        \qquad\qquad -\big[\mathbb E(Y_i^t \mid X_i^t=\tilde{x})-\mathbb E(Y_i^{t-1} \mid X_i^{t-1}=\tilde{x})\big]\Big\}.
    \end{align}
    For the control cohort, define $p^{t-1}(z)=\mathbb P(D_i^{t-1}=1 \mid X_i^{t-1}=z)$. Because this cohort is untreated, $p^{t-1}(z)=0$ for every $z$. Using this fact together with $p^t(\tilde{x})=0$, the denominator in \eqref{eq:fuzzy_did} can equivalently be written as
    \begin{align*}
        p^t(x)
        =
        \left[p^t(x)-p^{t-1}(x)\right]
        -
        \left[p^t(\tilde{x})-p^{t-1}(\tilde{x})\right].
    \end{align*}
    The numerator in \eqref{eq:fuzzy_did} is a difference-in-differences in outcomes, while the denominator is the corresponding difference-in-differences in treatment probabilities. The resulting ratio has the same algebraic form as the Wald-DiD estimand familiar from fuzzy difference-in-differences \citep{dechaisemartin_fuzzy_2018}. Specifically, in our setting, actual birth dates $x$ and $\tilde{x}$ index the two groups, while cohorts $t$ and $t-1$ provide the reform- and pre-reform-cohort observations within each group. Because the treatment probability changes only for the group indexed by $x$ in the reform cohort, the treated mother--child pairs at date $x$ in that cohort correspond to the switchers in fuzzy-DiD terminology, and the average treatment effect among these switchers coincides with $\mathrm{CATT}^t(x)$. However, unlike in a conventional fuzzy-DiD application, individual policy-regime assignment is not observed in our outcome data. Instead, the treatment probability entering the denominator is recovered from separate validation data.
\end{remark}

\begin{remark}[Overidentification and observable implications]
    \label{rem:over_id}
    Because \eqref{eq:alpha_ID} holds for every actual birth date $x$ in the zero-treatment region, the reform-cohort shift $\alpha^t$ is overidentified whenever this region contains multiple dates. In particular, the parallel-trends restriction implies that
    \[
        \mathbb E[Y_i^t\mid X_i^t=x]
        -
        \mathbb E[Y_i^{t-1}\mid X_i^{t-1}=x]
        =
        \alpha^t
    \]
    is constant across all such dates. This observable implication provides a specification check within the zero-treatment region, although it does not directly test the parallel-trends restriction at dates for which $p^t(x)>0$. We assess this implication separately for each reform and outcome and find it consistent with the data. Figure~\ref{fig:implementation} illustrates the assessment for the main outcome of the 2007 pay reform, while \ref{sec:paralleltrendstest} presents the full set of graphical diagnostics.
\end{remark}

\bigskip

{\noindent \bf Nonparametric estimation}

\noindent
Let $\mu^\tau(x)\coloneqq\mathbb E[Y_i^\tau\mid X_i^\tau=x]$ for $\tau\in\{t-1,t\}$. Proposition~\ref{prop:NPID} suggests the sample-analogue estimator
\begin{align}
    \widehat{\mathrm{CATT}}^t(x)
    =
    \frac{
        \widehat\mu^t(x)
        -
        \widehat\mu^{t-1}(x)
        -
        \widehat\alpha^t
    }{
        \widehat p^t(x)
    },
    \label{eq:np_estimator}
\end{align}
for values of $x$ such that $p^t(x)>0$. Here, $\widehat\mu^\tau(x)$ denotes a nonparametric estimator of $\mu^\tau(x)$, $\widehat\alpha^t$ is obtained from the sample counterpart of \eqref{eq:alpha_ID} using observations in the zero-treatment region, and $\widehat p^t(x)$ is estimated from the validation data.
Depending on the support of $X_i^\tau$, $\widehat\mu^\tau(x)$ may be constructed using cell averages when $X_i^\tau$ is discrete or standard nonparametric regression methods otherwise. Section~\ref{ssec:implementation} describes the specific implementation choices made in our application.

\bigskip

{\noindent \bf Regression-based estimation}

\noindent
Under the parallel-trends restriction in \eqref{eq:parallel_trends}, the conditional mean for the reform and control cohorts can be written as
\begin{align*}
    \mathbb E[Y_i^\tau\mid X_i^\tau=x]
    =
    \alpha^t\mathbf 1\{\tau=t\}
    +
    m_0^{t-1}(x)
    +
    p^\tau(x)\mathrm{CATT}^t(x),
    \qquad \tau\in\{t-1,t\},
\end{align*}
where $p^{t-1}(x)=0$ for the untreated control cohort. Hence, the unknown functions $m_0^{t-1}(x)$ and $\mathrm{CATT}^t(x)$ can be estimated from the regression
\begin{align*}
    Y_i^\tau
    =
    \alpha^t\mathbf 1\{\tau=t\}
    +
    m_0^{t-1}(X_i^\tau)
    +
    p^\tau(X_i^\tau)\mathrm{CATT}^t(X_i^\tau)
    +
    u_i^\tau,
    \qquad \tau\in\{t-1,t\},
\end{align*}
where observations from both cohorts are pooled. A feasible regression replaces $p^\tau(\cdot)$ with its estimate from the validation data. The functions $m_0^{t-1}(x)$ and $\mathrm{CATT}^t(x)$ can be approximated flexibly using basis functions, such as polynomials or splines, or specified parametrically when fully nonparametric estimation is not feasible with the available data. The regression framework also makes it straightforward to include additional predetermined covariates as controls, as we do for mothers' outcomes in Section~\ref{ssec:implementation}.

\subsection{Empirical implementation}\label{ssec:implementation}
\begin{figure}[p]
  \centering
  \textbf{Step 1}: Estimating the reform-cohort shift and visualising the parallel-trends diagnostic\\[0.1cm]
  \begin{adjustbox}{max width=\textwidth}
  \begin{tabular}{cc}
    \includegraphics[width=0.42\textwidth]{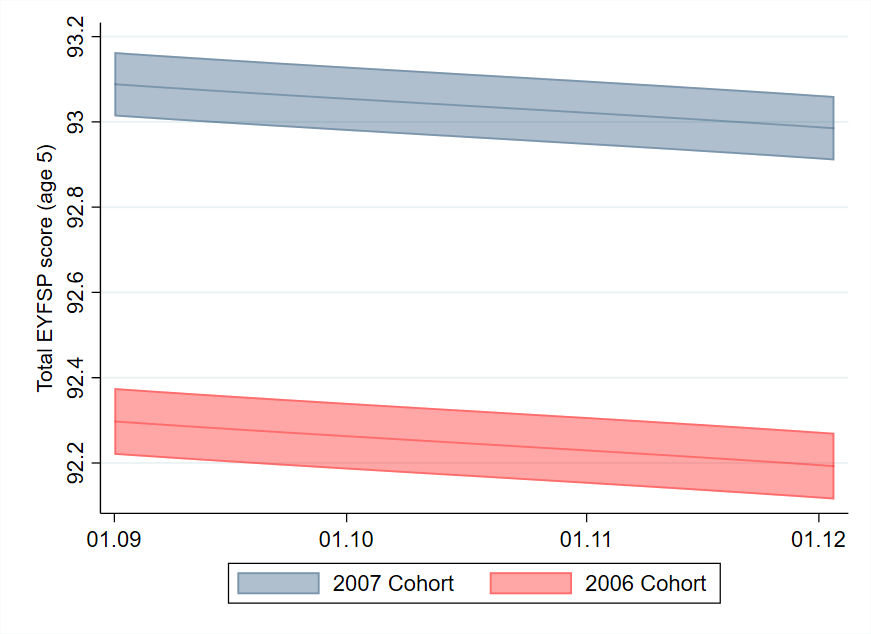} &
    \includegraphics[width=0.42\textwidth]{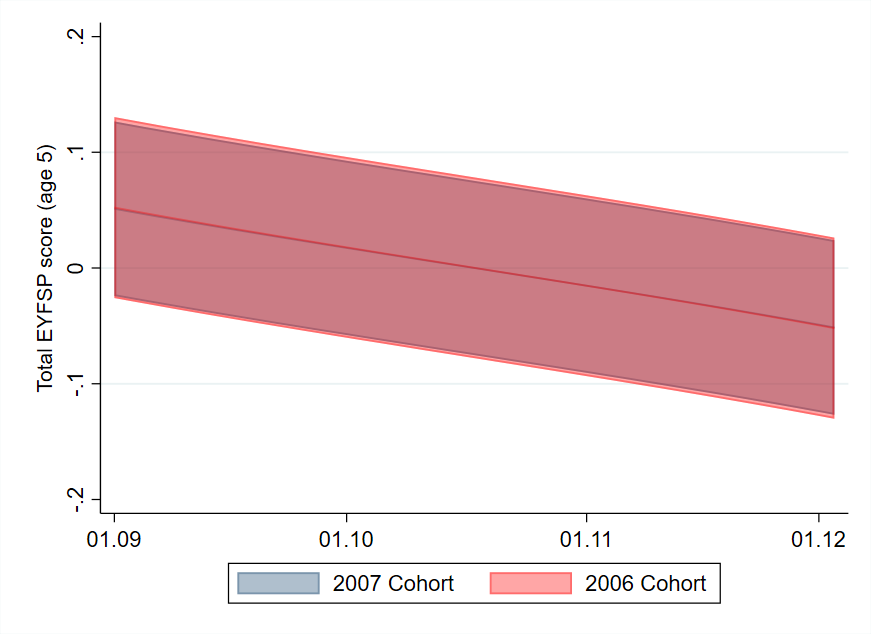} \\
    {\small Panel A. Raw outcomes} & {\small Panel B. Demeaned outcomes} \\
  \end{tabular}
  \end{adjustbox}\\[0.15cm]

  \vspace{0.25cm}

  \textbf{Step 2}: Estimating daily CATTs nonparametrically\\[0.1cm]
  \includegraphics[width=0.44\textwidth]{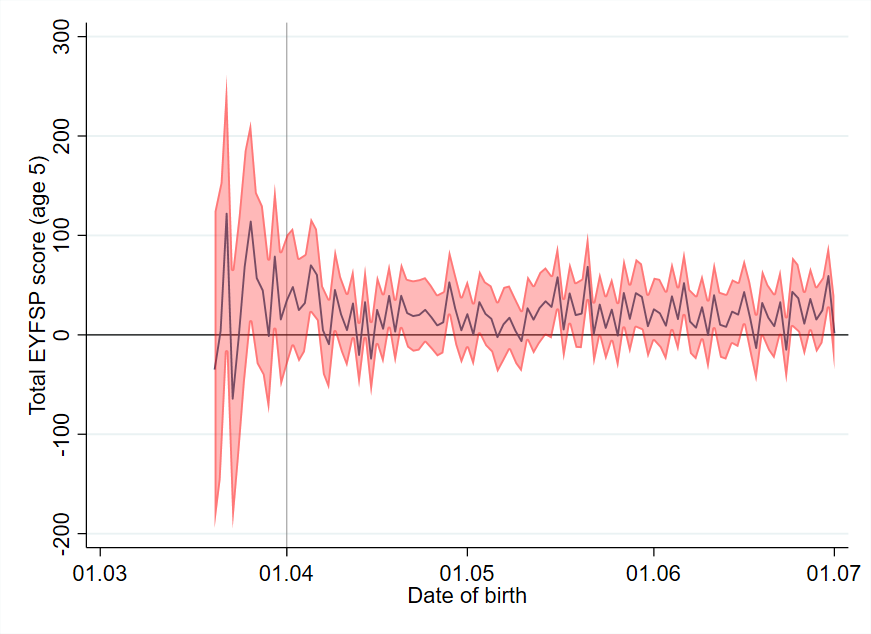}\\[0.15cm]

  \vspace{0.25cm}

  \textbf{Step 3}: Constructing weekly and monthly averages\\[0.1cm]
  \begin{adjustbox}{max width=\textwidth}
  \begin{tabular}{cc}
    \includegraphics[width=0.42\textwidth]{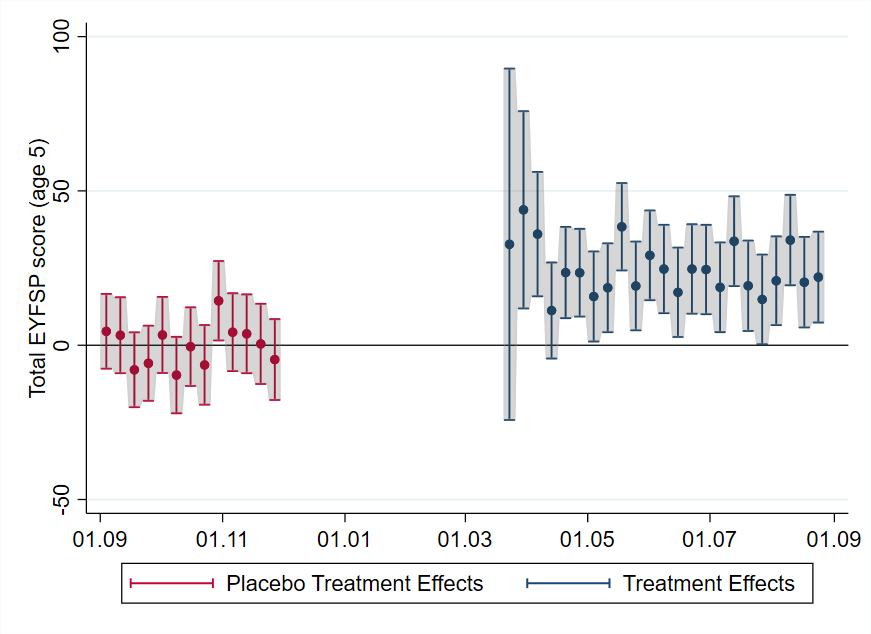} &
    \includegraphics[width=0.42\textwidth]{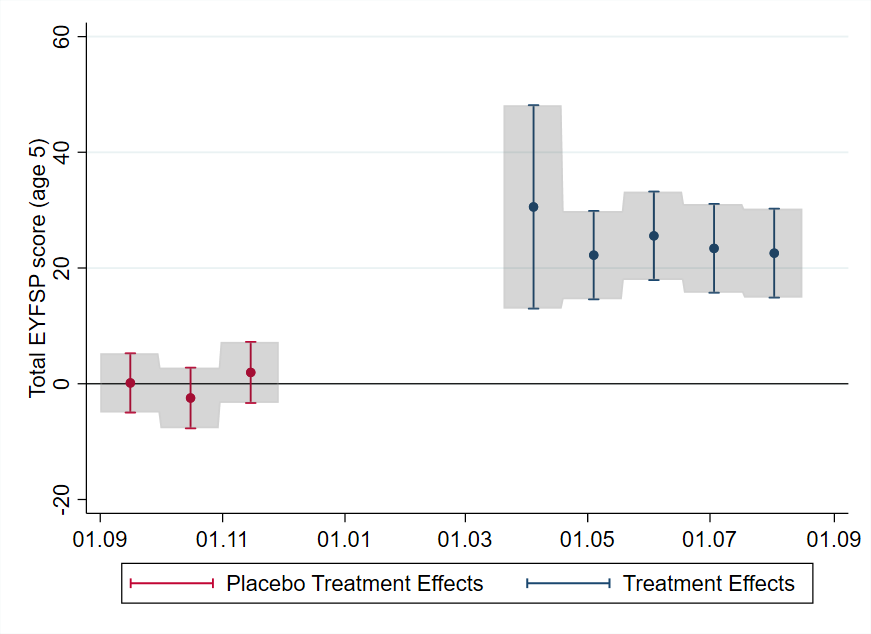} \\
    {\small Panel A. Weekly averages} & {\small Panel B. Monthly averages} \\
  \end{tabular}
  \end{adjustbox}\\[0.15cm]
  \refstepcounter{figure}
  \label{fig:implementation}
  {\small \textbf{Figure~\thefigure: Empirical implementation: 2007 reform and total test score at age 5}}\\[0.3cm]
  \begin{minipage}{0.9\textwidth}\begin{footnotesize}\begin{spacing}{1.15}
    \textit{Notes:} Step 1: Panel A shows raw cohort profiles in the zero-treatment region. Panel B demeans each by its equal-date mean over the same dates; their reform--control mean difference is $\widehat\alpha^t$. {} Step 2: The panel reports daily CATT estimates and pointwise 95\% confidence intervals where $\widehat p^t(x)\geq0.2$, expressed in days of age at testing. Step 3: Panel A reports equal-date weekly CATT and placebo estimates; Panel B reports monthly estimates. Placebos use zero-treatment windows; bars and shaded bands show pointwise 95\% confidence intervals. Estimates are expressed in days of age at testing. Horizontal-axis tick labels give the date of birth in day.month format.
  \end{spacing}\end{footnotesize}\end{minipage}
\end{figure}

For children's outcomes, the large administrative samples allow us to implement the nonparametric estimator in \eqref{eq:np_estimator} using daily date-of-birth cell averages. Figure~\ref{fig:implementation} illustrates the three steps of this procedure using the 2007 reform and children's total test score at age 5. For mothers' outcomes, the smaller QLFS samples make daily nonparametric estimation impractical, so we instead use the regression-based approach described below.

First, we use the zero-treatment region to estimate the reform-cohort shift $\alpha^t$ and construct the graphical diagnostic discussed in Remark~\ref{rem:over_id}.\footnote{For the 2007 reform, we define this region as actual births before 1 December, four months before the 1 April policy cutoff, and take $p^t(x)=0$ throughout; see Panel~B of Figure~\ref{fig:diagnosis}.} We compute an equal-date average of the daily cell means for each cohort over the same dates; the reform--control difference between these averages is $\widehat\alpha^t$. We then demean each cohort's outcome profile by its own average. If the cohort difference is constant across dates, the demeaned profiles should coincide up to sampling variation.

Second, we construct daily estimates $\widehat{\mathrm{CATT}}^t(x)$ by applying \eqref{eq:np_estimator} to the daily cell means and the estimated treatment probabilities $\widehat p^t(x)$ obtained from the validation data. \ref{sec:phat} describes the construction of $\widehat p^t(x)$. Because division by a small treatment probability amplifies sampling noise, we report daily CATT estimates only for dates satisfying $\widehat p^t(x)\geq0.2$, corresponding to the right shaded region in Panel~B of Figure~\ref{fig:diagnosis}. To make estimates comparable across outcomes and test ages, we express all daily CATT estimates---and the placebo estimates and averages derived from them below---in units of days of age at testing: each estimate is divided by the effect of one additional day of age on the outcome, estimated from the variation in daily cell means across dates of birth in the zero-treatment region.{}
{}

Third, to improve precision, we aggregate the daily CATT estimates into weekly and monthly averages.\footnote{For simplicity, we report averages that weight each date equally. Weighting dates in proportion to the number of births on each date does not materially change the results. {}} We also partition the zero-treatment region into weekly and monthly windows and construct a placebo estimate for each window by averaging the daily reform--control differences in mean outcomes within the window and subtracting $\widehat\alpha^t$. We then examine whether these window-specific placebo estimates are statistically distinguishable from zero, providing a diagnostic of the observable implications of parallel trends within the zero-treatment region.

For mothers' outcomes, we implement the regression-based approach described in Section~\ref{ssec:solution}. We specify the birth-date component of the untreated conditional mean as linear in actual date of birth $x$ and assume a constant treatment effect across birth dates, $\mathrm{CATT}^t(x)=\beta$. To improve precision, we also include additional covariates additively.

As detailed in Section~\ref{sec:Data}, the mothers' data are structured to mirror the cohort structure of the children's data. For each reform, we pool mothers whose children were born in the reform academic year, $t$, and the preceding academic year, $t-1$.\footnote{We observe different populations of mothers at different survey horizons, just as we analyse children's test scores at different ages.}

Specifically, we estimate
\begin{align}
    Y_i^\tau
    =
    \gamma_0
    +\alpha^t\mathbf 1\{\tau=t\}
    +\lambda X_i^\tau
    +\beta\,\widehat p^\tau(X_i^\tau)
    +W_i^{\tau\prime}\delta
    +u_i^\tau,
    \qquad \tau\in\{t-1,t\}.
    \label{eq:semiparametric_regression_mothers}
\end{align}
Here, $\alpha^t$ captures the reform-cohort shift, $\gamma_0+\lambda x$ is the linear birth-date component, and $\beta$ is the constant CATT. As before, we set $\widehat p^{t-1}(x)=0$ for the untreated control cohort.

The vector of controls $W_i^\tau$ includes the number of children, maternal age, marital status, and years of schooling. These controls enter the untreated conditional mean additively, with coefficients common across cohorts, and we assume that the treatment component remains $p^\tau(x)\beta$ after conditioning on them.\footnote{A sufficient condition is that, conditional on actual date of birth, the treatment probability does not vary with $W_i^t$ and the constant-CATT restriction continues to hold after conditioning on $W_i^t$.} Importantly, we measure the controls at the reform date, or at the corresponding date for the control cohort, since some of them, such as the number of children, may themselves be affected by the reform.

\ref{sec:standarderrors} derives standard errors for both procedures. For the nonparametric estimates, the formulas account for the common estimation of the reform-cohort shift and treatment probabilities across birth dates; for the regression estimates, they account for the use of estimated treatment probabilities as generated regressors.

\pagebreak
\section{Findings}
\label{sec:findings}
\subsection{Children's outcomes}
\begin{figure}[!htbp]
    \centering
    \includegraphics[width=0.6\textwidth]{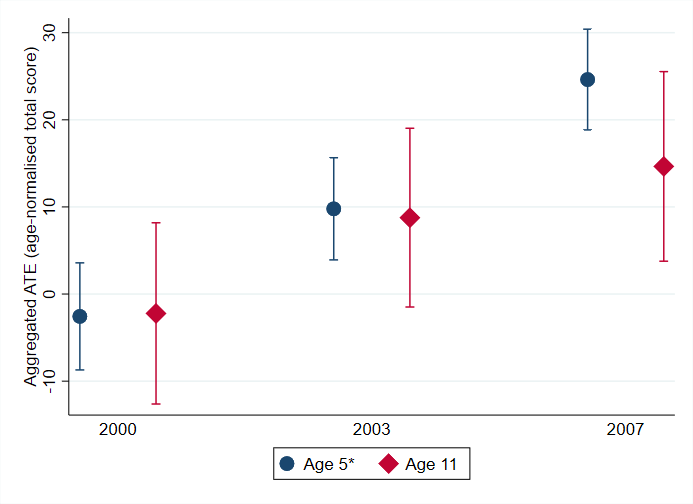}\\
    \refstepcounter{figure}
    \label{fig:coefplot}
    \vspace{0.35cm}
    \begin{small}
        \textbf{Figure~\thefigure: Aggregated average treatment effects on the treated}
    \end{small}  \\
    \vspace{0.3cm}
    \begin{minipage}{0.9\textwidth}
        \begin{footnotesize}
            \begin{spacing}{1.15}
               \textit{Notes:} The figure reports average conditional average treatment effects on the treated (CATTs), aggregated over all dates in the analysis period with predicted treatment probability $\hat p^t(x) \in [0.2,1]$. Estimates are normalised by the effect of an additional day of age on the corresponding test score, estimated from the variation in average outcomes across dates of birth in the region with $p^t(x) = 0$. The outcomes are the total EYFSP score at age 5, the total KS1 score at age 7, and the average KS2 score at age 11. Blue circles (labelled ``Age 5*'' in the legend) correspond to the earliest available outcomes: age 5 (EYFSP) for the 2003 and 2007 reforms and, as flagged by the asterisk, age 7 (KS1) for the 2000 reform. Red diamonds correspond to age 11 (KS2) outcomes. Vertical lines indicate pointwise 95\% confidence intervals. {}
            \end{spacing}
        \end{footnotesize}
    \end{minipage}
\end{figure}

Figure~\ref{fig:coefplot} provides an overview of the results. It shows the average of all estimated conditional average treatment effects on the treated (CATTs) over the analysis period,\footnote{That is, for all days in the academic year for which the predicted treatment probability ($\hat p^t(x)$) lies between 0.2 and 1---from slightly before the start of the reform until the end of the academic year, 31~August of the year of the respective reform.} for the earliest available test score (age 7 for the 2000 reform, for which no age-5 scores are available in our data, and age 5 for the 2003 and 2007 reforms) and the latest available test score (age 11). It shows that the effects differ across reform types: the 2000 job-protection reform has no measurable effect; the 2003 bundle reform shows small effects at age 5 that are not statistically distinguishable from zero at age 11; and the 2007 pay reform yields a comparatively large effect on test scores---roughly equivalent to that of being 25 days older at the time of testing (0.045 standard deviations).

Below, we examine the consistency and robustness of the effects shown in Figure~\ref{fig:coefplot} across outcomes and samples, and compare them to placebo estimates for the zero-treatment region (Section~\ref{ssec:implementation}).\footnote{The definition of the placebo estimates and the derivation of their standard errors are provided in \ref{sec:standarderrors}.} The reforms are discussed in chronological order, starting with the 2000 job-protection reform, followed by the 2003 bundle reform, and concluding with the 2007 pay reform.

\begin{figure}[!htbp]
  \centering
  \begin{tabular}{cc}
    \includegraphics[width=0.45\textwidth]{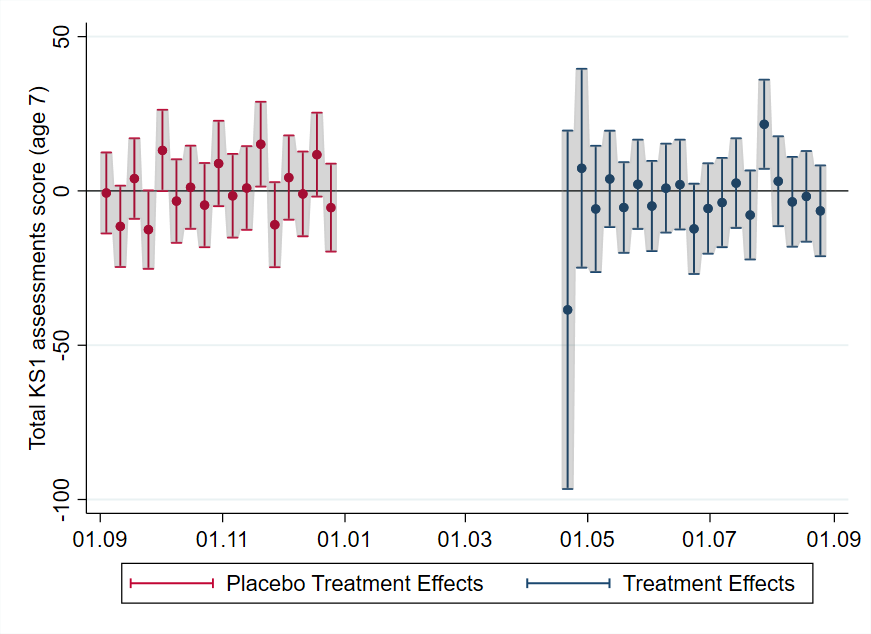}  &
    \includegraphics[width=0.45\textwidth]{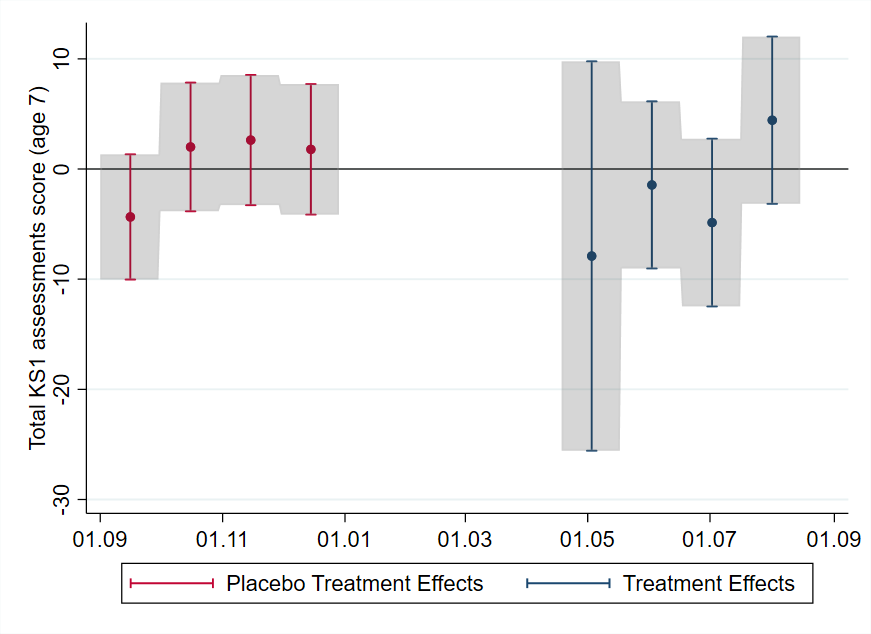} \\
    {\small Panel A. Week aggregation} & {\small Panel B. Month aggregation} \\
  \end{tabular}\\
  \refstepcounter{figure}  %
  \label{fig:fullsample2000}  %
    \vspace{0.35cm}
    \begin{small}
   \textbf{Figure~\thefigure: Estimated effects of the 2000 job-protection reform}
    \end{small}

    \vspace{0.3cm}
    \begin{minipage}{0.9\textwidth}
        \begin{footnotesize}
            \begin{spacing}{1.15}
                \textit{Notes:} The figure plots conditional average treatment effects on the treated for birth dates with $\hat p^t(x) \geq 0.2$, aggregated at the weekly level in the left panel and at the monthly level in the right panel. Placebo estimates are computed for pre-reform birth dates with $p^t(x) = 0$ as the difference between reform- and control-cohort outcomes net of the estimated reform-cohort shift $\hat \alpha^t$; under the parallel-trends assumption, they should be indistinguishable from zero. The outcome variable considered is the total score at age 7 (KS1). Estimates are expressed in days of age (see the notes to Figure~\ref{fig:coefplot}); vertical lines and shaded areas indicate pointwise 95\% confidence intervals. Horizontal-axis tick labels give the date of birth in day.month format.
            \end{spacing}
        \end{footnotesize}
    \end{minipage}
\end{figure}

The 2000 job-protection reform does not appear to have any measurable effect on children's school performance. The aggregated estimate in Figure~\ref{fig:coefplot} is close to zero, and the upper bound of its 95\% confidence interval rules out positive effects larger than the equivalent of about four days of age at testing. This result is confirmed by the disaggregated weekly and monthly estimates in Figure~\ref{fig:fullsample2000}, by analyses of maths and language scores at the same age (results not reported), and by the age-11 estimate in Figure~\ref{fig:coefplot}, which is also close to zero. We also find no clear evidence of heterogeneity across the subgroups examined in Appendix Figure~\ref{fig:geo2000}.

\begin{figure}[!htbp]
    \centering
    \begin{tabular}{cc}
        \includegraphics[width=0.45\textwidth]{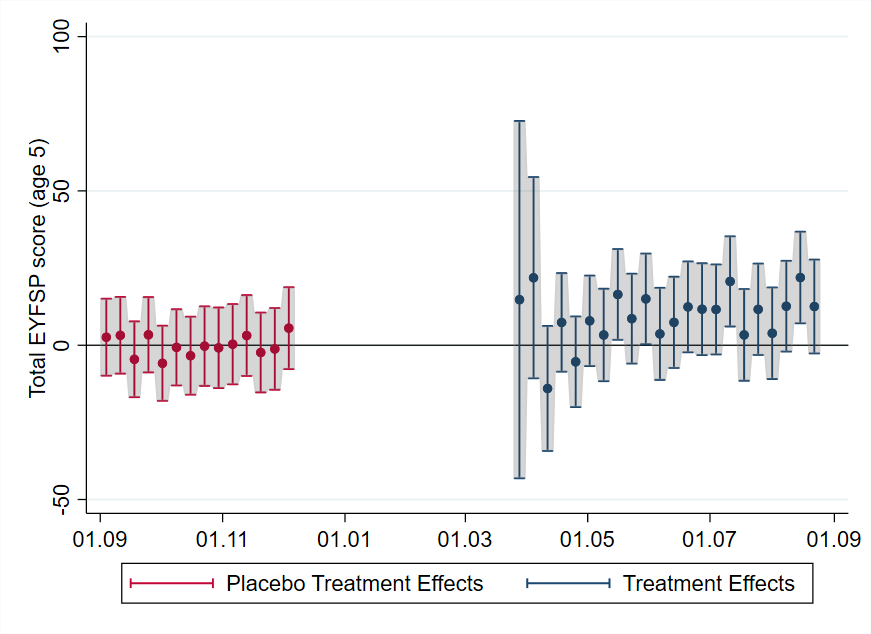} &
        \includegraphics[width=0.45\textwidth]{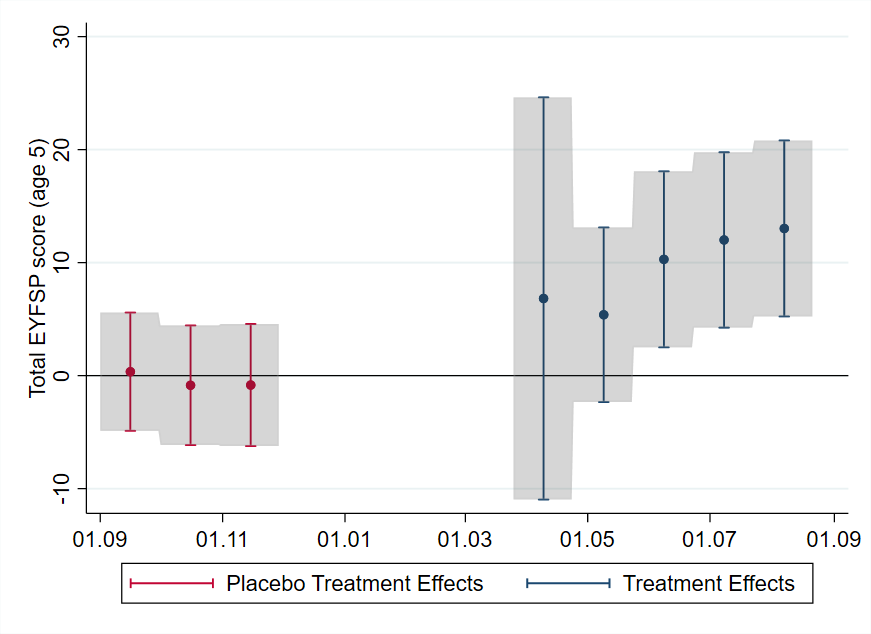} \\
        {\small Panel A. Week aggregation} & {\small Panel B. Month aggregation} \\
    \end{tabular}

    \refstepcounter{figure}  %
    \label{fig:fullsample2003}  %

    \vspace{0.35cm}
    \begin{small}
        \textbf{Figure~\thefigure: Estimated effects of the 2003 bundle reform}
    \end{small}

    \vspace{0.3cm}
    \begin{minipage}{0.9\textwidth}
        \begin{footnotesize}
            \begin{spacing}{1.15}
                \textit{Notes:} The figure plots conditional average treatment effects on the treated for birth dates with $\hat p^t(x) \geq 0.2$, aggregated at the weekly level in the left panel and at the monthly level in the right panel. Placebo estimates are computed for pre-reform birth dates with $p^t(x) = 0$ as the difference between reform- and control-cohort outcomes net of the estimated reform-cohort shift $\hat \alpha^t$; under the parallel-trends assumption, they should be indistinguishable from zero. The outcome variable considered is the total score at age 5 (EYFSP). Estimates are expressed in days of age (see the notes to Figure~\ref{fig:coefplot}); vertical lines and shaded areas indicate pointwise 95\% confidence intervals. Horizontal-axis tick labels give the date of birth in day.month format.
            \end{spacing}
        \end{footnotesize}
    \end{minipage}
\end{figure}

For the 2003 reform, the evidence is less conclusive. At the earliest test age (age 5), the estimated effects are positive and reach statistical significance at the monthly aggregation level, with an effect size equivalent to about ten days of age (Figure~\ref{fig:fullsample2003}). These effects are robust across subjects: Appendix Figure~\ref{fig:robustness2003} shows that the effects on maths and language scores are similar in magnitude and significance. However, three caveats apply. First, the estimates are less precise at later ages: the age-11 effect is not statistically significant, although its point estimate is of similar size to the age-5 effect (Figure~\ref{fig:coefplot}; see also Appendix Figure~\ref{fig:robustness2003}). Second, the monthly point estimates vary across birth months.\footnote{If the new maternity benefits affect mothers of May-born and June-born children similarly, one would expect their average treatment effects---aggregated at the month level---to be comparable.} Third, the heterogeneity patterns are not straightforward: while the effects seem larger in rural areas, no clear differences emerge by child poverty status (Appendix Figure~\ref{fig:geo2003}), which makes it difficult to attribute the full-sample effects in Figure~\ref{fig:fullsample2003} to either component of the bundle.

\begin{figure}[!htbp]
    \centering
    \begin{tabular}{cc}
        \includegraphics[width=0.45\textwidth]{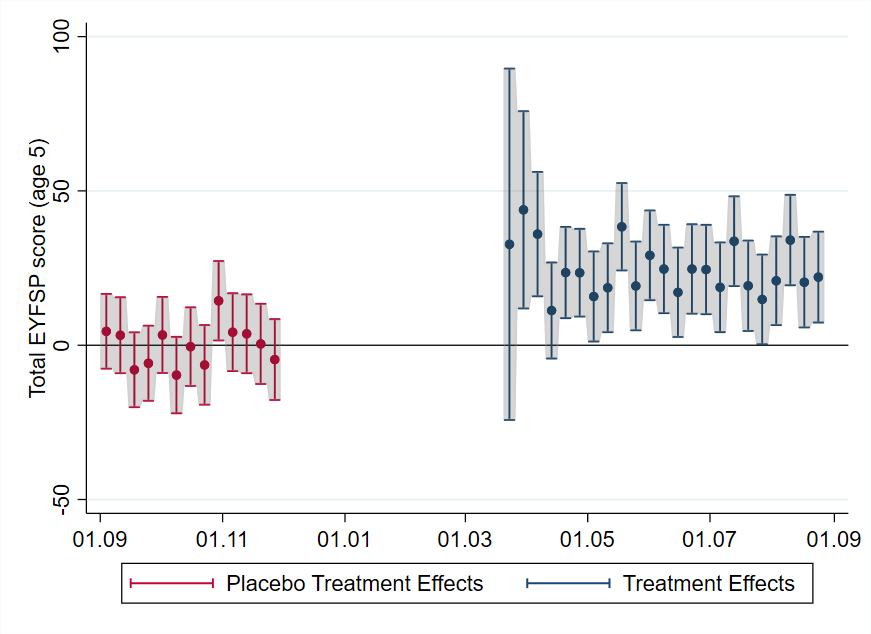}  &
        \includegraphics[width=0.45\textwidth]{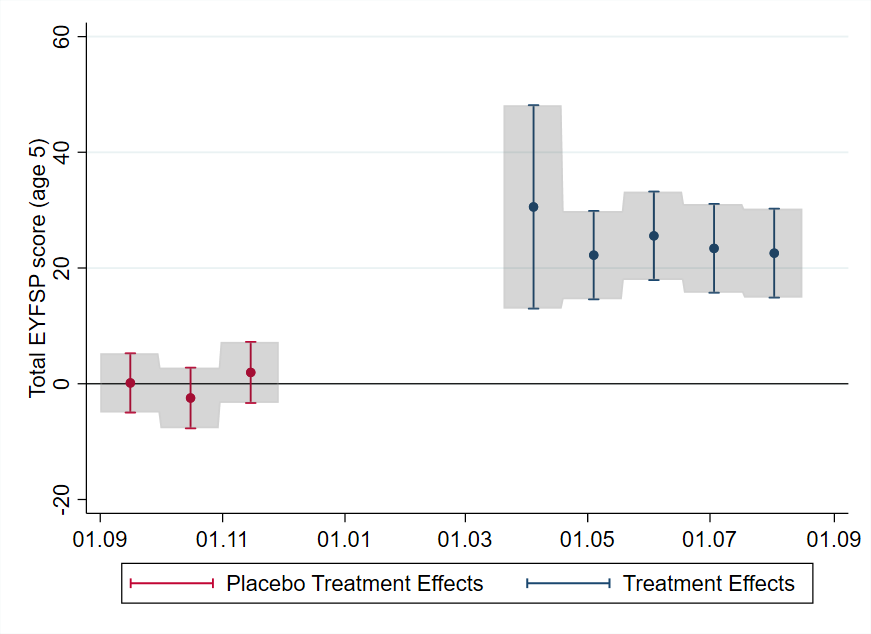} \\
        {\small Panel A. Week aggregation} & {\small Panel B. Month aggregation} \\
    \end{tabular}

    \refstepcounter{figure}
    \label{fig:fullsample2007}

    \vspace{0.35cm}
    \begin{small}
        \textbf{Figure~\thefigure: Estimated effects of the 2007 pay reform}
    \end{small}

    \vspace{0.3cm}
    \begin{minipage}{0.9\textwidth}
        \begin{footnotesize}
            \begin{spacing}{1.15}
                \textit{Notes:} The figure plots conditional average treatment effects on the treated for birth dates with $\hat p^t(x) \geq 0.2$, aggregated at the weekly level in the left panel and at the monthly level in the right panel. Placebo estimates are computed for pre-reform birth dates with $p^t(x) = 0$ as the difference between reform- and control-cohort outcomes net of the estimated reform-cohort shift $\hat \alpha^t$; under the parallel-trends assumption, they should be indistinguishable from zero. The outcome variable considered is the total score at age 5 (EYFSP). Estimates are expressed in days of age (see the notes to Figure~\ref{fig:coefplot}); vertical lines and shaded areas indicate pointwise 95\% confidence intervals. Horizontal-axis tick labels give the date of birth in day.month format.
            \end{spacing}
        \end{footnotesize}
    \end{minipage}
\end{figure}

For the 2007 pay reform, the evidence of positive effects is clearer. At the aggregate level, the estimated effects are roughly two and a half times as large as those for the 2003 reform, corresponding to about 25 days of age or 0.045 standard deviations. The weekly estimates are statistically significant for almost every birth week (Figure~\ref{fig:fullsample2007}).
An effect of 0.045 standard deviations is small relative to those of intensive, targeted early-childhood programmes such as Head Start, whose short-run test-score effects are around 0.15 standard deviations at ages 5--6 \parencite{deming_early_2009}, but comparable to that of Sure Start, the area-based early-years programme rolled out in England over the same period: access to a Sure Start centre from birth to age 5 raised test scores by 0.018 standard deviations at age 7 and by 0.042 standard deviations at age 11 \parencite{carneiro_short-_2024}.

\begin{figure}[!htbp]
    \centering
    \begin{tabular}{cc}
        \includegraphics[width=0.45\textwidth]{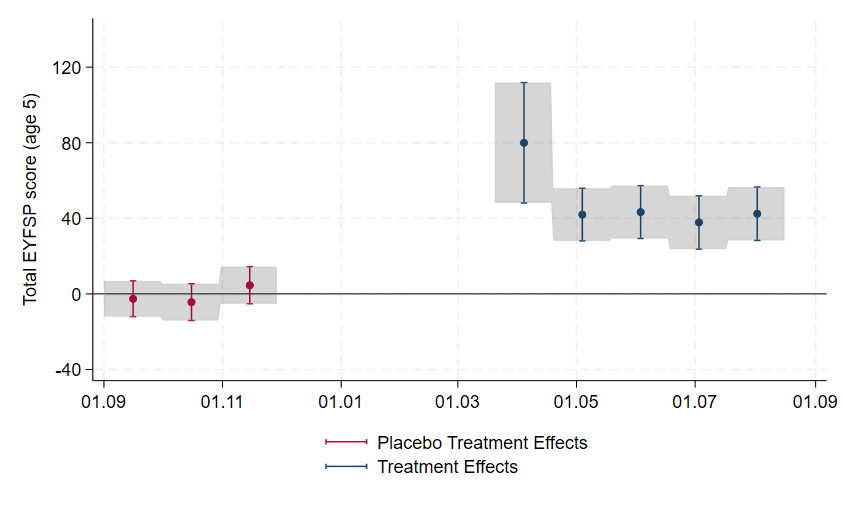}  &
        \includegraphics[width=0.45\textwidth]{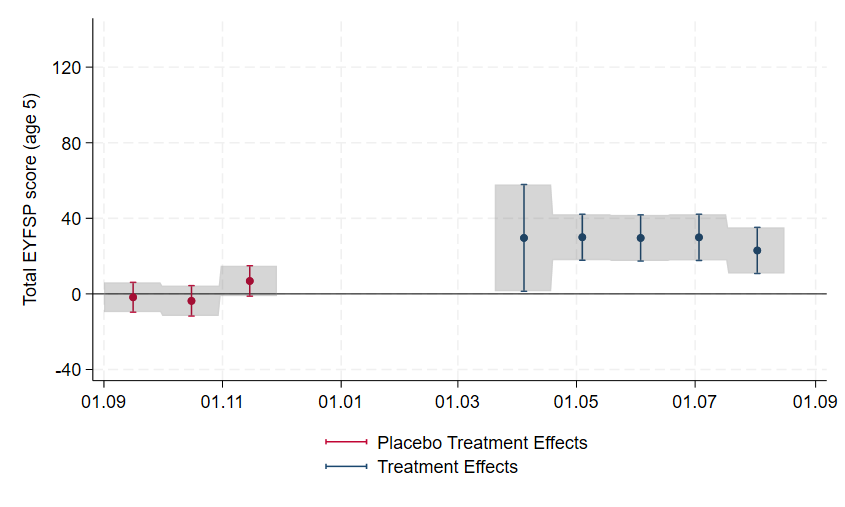} \\
        {\small Panel A. Rural regions} & {\small Panel B. Urban regions} \\
        \includegraphics[width=0.45\textwidth]{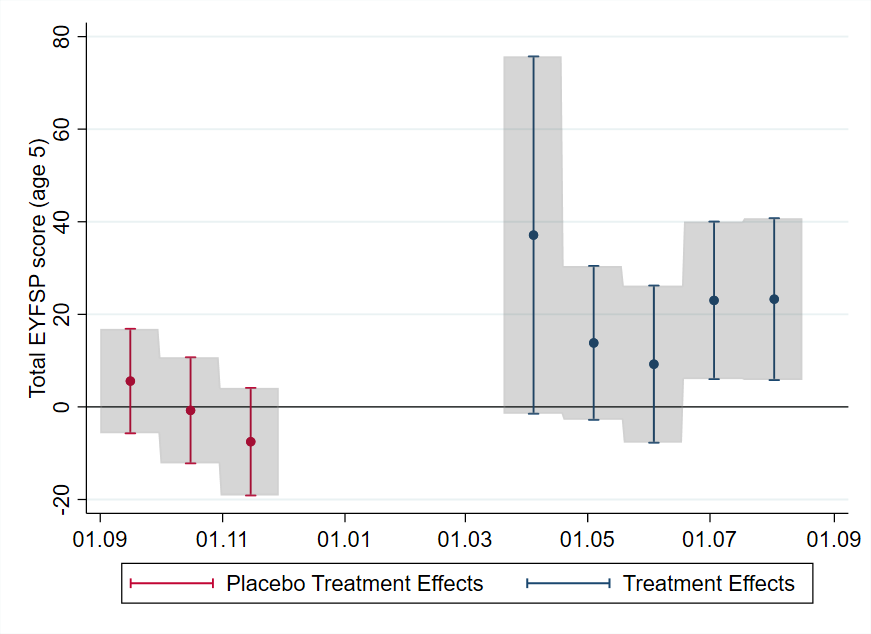}  &
        \includegraphics[width=0.45\textwidth]{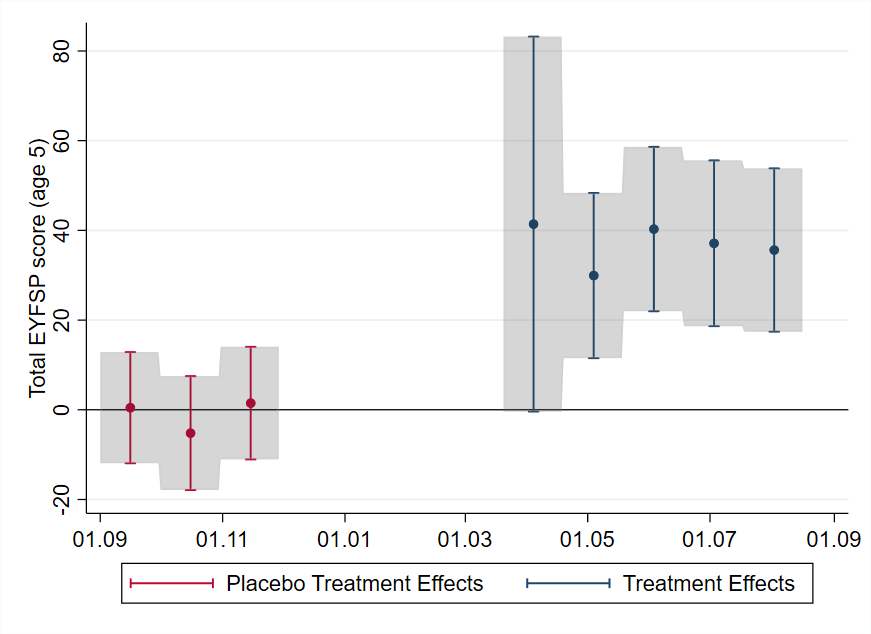} \\
        {\small Panel C. Bottom $20\%$ child poverty} & {\small Panel D. Top $20\%$ child poverty} \\
    \end{tabular}

    \refstepcounter{figure}
    \label{fig:geo2007}

    \vspace{0.35cm}
    \begin{small}
\textbf{Figure~\thefigure: Geographic heterogeneity in the effects of the 2007 pay reform}
    \end{small}

    \vspace{0.3cm}
    \begin{minipage}{0.9\textwidth}
        \begin{footnotesize}
            \begin{spacing}{1.15}
                \textit{Notes:} The figure plots conditional average treatment effects on the treated, aggregated at the monthly level, for different subsamples. The top row restricts the sample to rural (left) and urban (right) regions, while the bottom row restricts the sample to areas with low and high child poverty, defined as the bottom and top quintiles of the IDACI distribution, respectively. The outcome variable considered is the total score at age 5 (EYFSP). Placebo estimates for pre-reform birth dates with $p^t(x) = 0$ are shown in red. Estimates are expressed in days of age (see the notes to Figure~\ref{fig:coefplot}); vertical lines and shaded areas indicate pointwise 95\% confidence intervals. The vertical scales differ across rows. Horizontal-axis tick labels give the date of birth in day.month format. {}
            \end{spacing}
        \end{footnotesize}
    \end{minipage}
\end{figure}

Heterogeneity analyses suggest that these effects are larger in rural regions and in areas in the top quintile of the child poverty index. The point estimates are roughly one and a half times as large in rural as in urban areas, and about twice as large in the most deprived quintile as in the least deprived quintile (Figure~\ref{fig:geo2007}).\footnote{The subgroup estimates use the full-sample treatment probability $\hat p^t(x)$, which is appropriate if the gap between actual and expected dates of birth is distributed similarly across subgroups; the MCS could be used to check this.} This is consistent with expectations: maternity pay is likely to have stronger effects in low-income households, and maternity leave may matter more in areas with limited childcare infrastructure. No systematic heterogeneity is found by ethnicity, first language, or free-school-meal eligibility (Appendix Figure~\ref{fig:heterogeneity2007}); the last of these is not at odds with the income mechanism, since free-school-meal eligibility identifies the poorest households, whose mothers are the least likely to be in the employment that qualifies for Statutory Maternity Pay.

\begin{figure}[!htbp]
    \centering
    \begin{tabular}{cc}
        \includegraphics[width=0.45\textwidth]{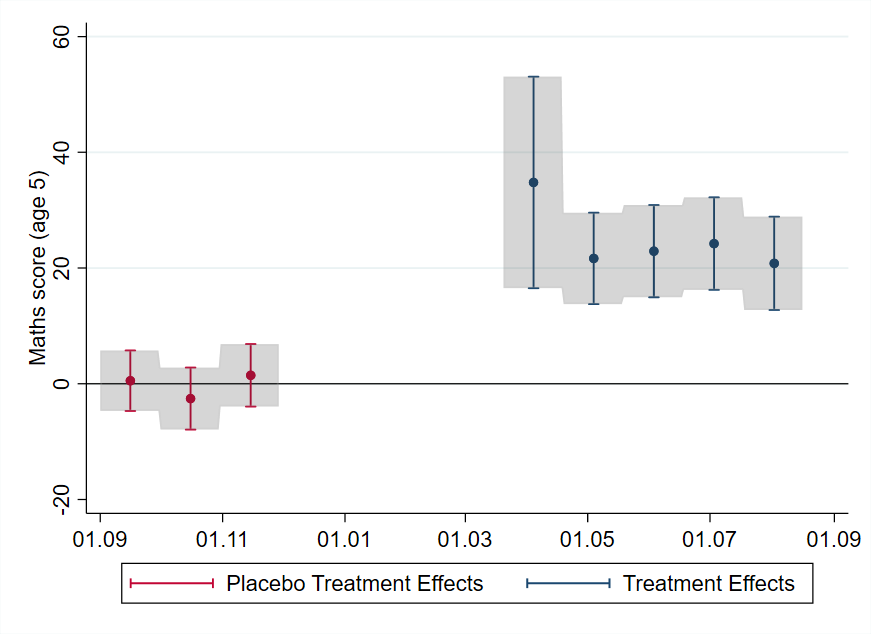}  &
        \includegraphics[width=0.45\textwidth]{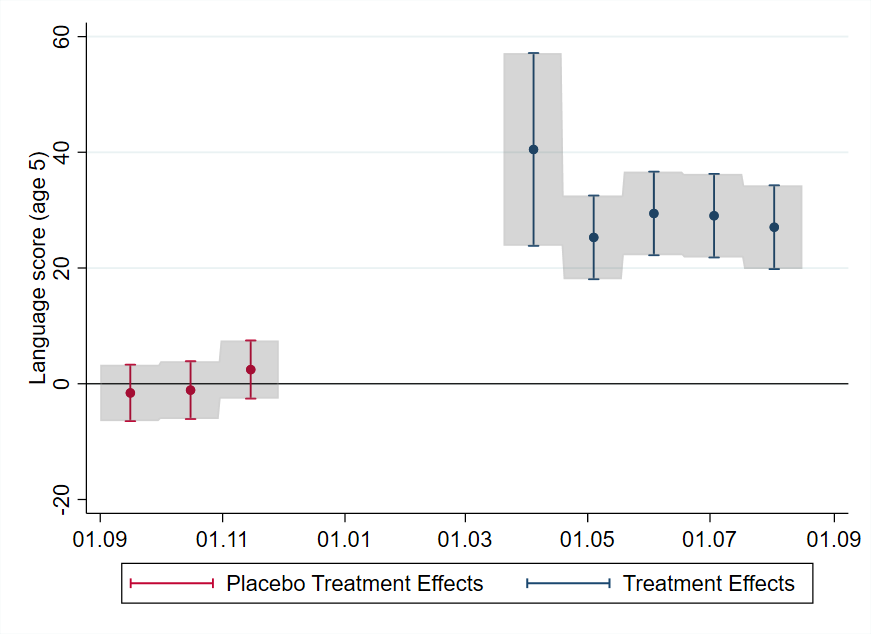} \\
        {\small Panel A. Maths score (age 5)} & {\small Panel B. Language score (age 5)} \\
        \includegraphics[width=0.45\textwidth]{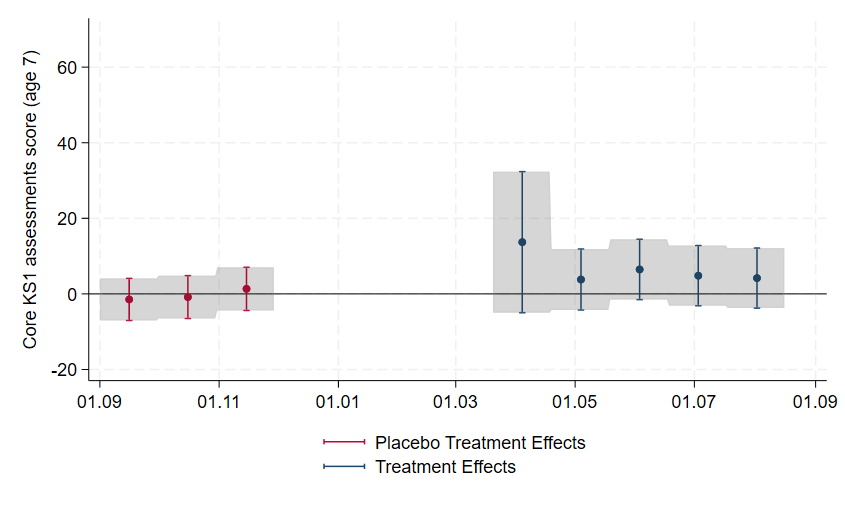}  &
        \includegraphics[width=0.45\textwidth]{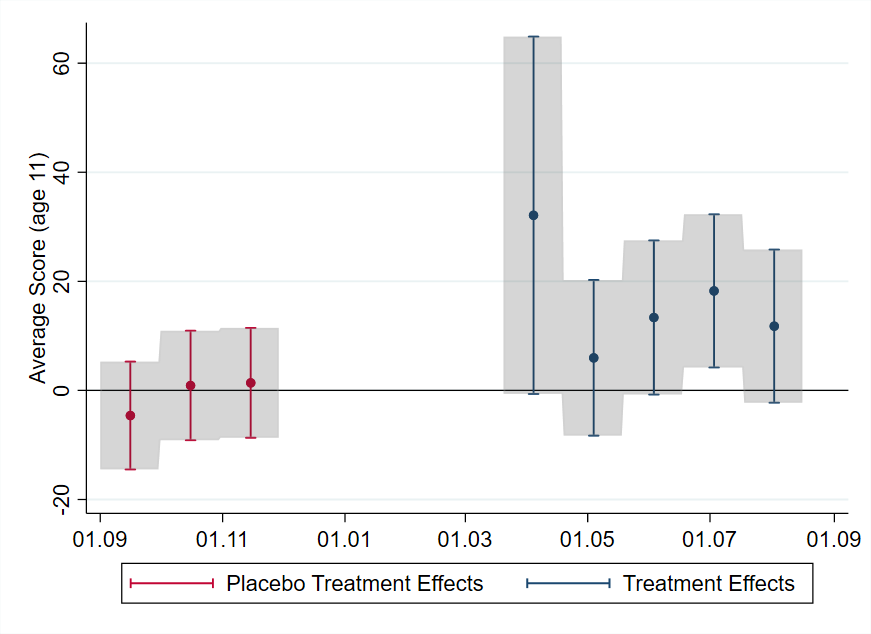} \\
        {\small Panel C. Total score (age 7)} & {\small Panel D. Average score (age 11)} \\
    \end{tabular}

    \refstepcounter{figure}
    \label{fig:robustness2007}

    \vspace{0.35cm}
    \begin{small}
      \textbf{Figure~\thefigure: Outcome and age robustness for the 2007 pay reform}
    \end{small}

    \vspace{0.3cm}
    \begin{minipage}{0.9\textwidth}
        \begin{footnotesize}
            \begin{spacing}{1.15}
                \textit{Notes:} This figure examines the robustness of the results for the 2007 pay reform across outcome measures and test ages. The top row reports conditional average treatment effects on the treated using maths and language scores at age 5, instead of the total score shown in the previous figures. The bottom row shows effects for the total Key Stage 1 score at age 7 and the average Key Stage 2 score at age 11. Estimates are aggregated at the monthly level and expressed in days of age; placebo estimates for pre-reform birth dates with $p^t(x) = 0$, constructed as described in the notes to Figure~\ref{fig:fullsample2000}, are shown in red. Vertical lines and shaded areas indicate pointwise 95\% confidence intervals. The vertical scales differ across panels. Horizontal-axis tick labels give the date of birth in day.month format. {}
            \end{spacing}
        \end{footnotesize}
    \end{minipage}
\end{figure}

Figure~\ref{fig:robustness2007} further shows that the results are robust to the choice of outcome variable. The total score effect is of a similar magnitude when analysing the individual subjects, maths and language, separately. Point estimates decline and standard errors increase at later ages. At age 7 the monthly estimates are positive but imprecise and not statistically significant, whereas at age 11 the aggregated effect is again positive and statistically significant, equivalent to about 15 days of age at testing (Figure~\ref{fig:coefplot}).

Figure~\ref{fig:robustness2007} already provides two types of robustness checks: placebo estimates for the pre-reform period and the analysis of alternative outcomes (maths and language scores). An additional and potentially more important test of the estimation strategy is its sensitivity to the choice of control cohort. \ref{sec:robustness} (Figures~\ref{fig:cohortrobustness2000}--\ref{fig:robustness2007robustness}) shows that the estimated effects for the 2007 reform, as well as those for the 2000 and 2003 reforms, remain similar when, instead of using the immediately preceding cohort as the control, the second preceding cohort is used.

\Needspace*{8\baselineskip}
\subsection{Maternal outcomes and mechanisms}

\begin{table}[!tb]
\centering
\caption{Estimated effects of the 2000, 2003, and 2007 reforms on hours worked over time}
\label{tab:mothers_hours}
\begin{threeparttable}
\begin{tabular}{l
>{\centering\arraybackslash}p{2.8cm}
>{\centering\arraybackslash}p{2.8cm}
>{\centering\arraybackslash}p{2.8cm}}\toprule
 & (1) & (2) & (3) \\
Horizon & 2000 reform & 2003 reform & 2007 reform \\
\midrule
One year after
& -2.148*** & -2.461*** & -2.568*** \\
& (0.629)   & (0.638)   & (0.657)   \\

Two years after
& -1.864*** & -2.108*** & -2.636*** \\
& (0.649)   & (0.615)   & (0.647)   \\

Three years after
& -1.756*** & -2.121*** & -1.618**  \\
& (0.659)   & (0.658)   & (0.695)   \\

Four years after
& -1.174    & -3.118*** & -1.894*** \\
& (0.735)   & (0.636)   & (0.715)   \\

Five years after
& -1.443**  & -1.431**  & -2.151*** \\
& (0.698)   & (0.671)   & (0.674)   \\

Six years after
& -1.998*** & -1.416**  & -1.199*   \\
& (0.711)   & (0.695)   & (0.689)   \\

\midrule
Controls
& \multicolumn{3}{p{\dimexpr 8.4cm+4\tabcolsep\relax}}{Number of children, age, years of schooling and marital status, all measured at the time of the reform; a linear date-of-birth trend; and cohort fixed effects} \\
\bottomrule
\end{tabular}

\begin{tablenotes}[flushleft]
\item \textit{Notes:} This table reports the estimated coefficients $\beta$ from the regression-based specification in equation~\eqref{eq:semiparametric_regression_mothers} of Section~\ref{sec:methodology}. Each coefficient measures the effect of the corresponding reform on hours worked at the indicated horizon relative to the child's date of birth. The outcome is the number of hours worked in the reference week, with mothers who are not employed recorded as zero, so the estimates combine extensive- and intensive-margin responses. Standard errors, reported in parentheses, are adjusted for the estimation of the treatment probability (see \ref{sec:standarderrors}). {} Sample sizes by horizon are as follows.
2000 reform: 4,887 (year 1), 4,861 (year 2), 4,919 (year 3), 4,642 (year 4), 4,425 (year 5), 4,350 (year 6).
2003 reform: 4,315 (year 1), 4,211 (year 2), 4,151 (year 3), 4,239 (year 4), 4,167 (year 5), 3,911 (year 6).
2007 reform: 4,389 (year 1), 4,239 (year 2), 3,995 (year 3), 3,988 (year 4), 4,079 (year 5), 3,935 (year 6). \\
*** $p<0.01$, ** $p<0.05$, * $p<0.1$
\end{tablenotes}
\end{threeparttable}
\end{table}

\begin{table}[!tb]
\centering
\caption{Estimated effects of the 2000, 2003, and 2007 reforms on the probability of being employed over time}
\label{tab:mothers_employment}
\begin{threeparttable}
\begin{tabular}{l
>{\centering\arraybackslash}p{2.8cm}
>{\centering\arraybackslash}p{2.8cm}
>{\centering\arraybackslash}p{2.8cm}}\toprule
 & (1) & (2) & (3) \\
Horizon & 2000 reform & 2003 reform & 2007 reform \\
\midrule
One year after
& -0.093*** & -0.091*** & -0.098*** \\
& (0.022)   & (0.023)   & (0.021)   \\

Two years after
& -0.089*** & -0.077*** & -0.117*** \\
& (0.022)   & (0.022)   & (0.021)   \\

Three years after
& -0.072*** & -0.108*** & -0.085*** \\
& (0.022)   & (0.022)   & (0.023)   \\

Four years after
& -0.054**  & -0.098*** & -0.091*** \\
& (0.024)   & (0.022)   & (0.023)   \\

Five years after
& -0.051**  & -0.046**  & -0.062*** \\
& (0.023)   & (0.023)   & (0.022)   \\

Six years after
& -0.056**  & -0.066*** & -0.071*** \\
& (0.023)   & (0.023)   & (0.023)   \\

\midrule
Controls
& \multicolumn{3}{p{\dimexpr 8.4cm+4\tabcolsep\relax}}{Number of children, age, years of schooling and marital status, all measured at the time of the reform; a linear date-of-birth trend; and cohort fixed effects} \\
\bottomrule
\end{tabular}

\begin{tablenotes}[flushleft]
\item \textit{Notes:} This table reports the estimated coefficients $\beta$ from the regression-based specification in equation~\eqref{eq:semiparametric_regression_mothers} of Section~\ref{sec:methodology}. Each coefficient measures the effect of the corresponding reform on the probability of being employed at the indicated horizon relative to the child's date of birth. Standard errors, reported in parentheses, are adjusted for the estimation of the treatment probability (see \ref{sec:standarderrors}). Sample sizes by horizon are identical to those in Table~\ref{tab:mothers_hours}. \\
*** $p<0.01$, ** $p<0.05$, * $p<0.1$
\end{tablenotes}
\end{threeparttable}
\end{table}

While the question of whether the reforms increased maternal labour-market attachment is interesting in its own right, it is also relevant for understanding the mechanisms behind the 2007 reform's impact on children. Potential channels include changes in mothers' time at home and family income.

Given the smaller QLFS sample, we analyse maternal outcomes using the regression-based implementation of our approach described in Section~\ref{sec:methodology}. This uses the same identification strategy as for children's outcomes, but assumes a linear birth-date trend in outcomes absent the reform and a treatment effect that is constant across birth dates for each reform and follow-up horizon. Tables~\ref{tab:mothers_hours} and~\ref{tab:mothers_employment} report the resulting estimates for hours worked and employment.

The patterns are consistent across reforms and horizons.{} Despite targeting different aspects of maternity leave, all three reforms reduced maternal labour supply in the short run: the probability of being employed falls by about 9 to 10 percentage points in the year after childbirth (Table~\ref{tab:mothers_employment}), and weekly hours worked (measured over all mothers in the sample, including those not employed) fall by more than two hours (Table~\ref{tab:mothers_hours}). These first-year effects likely reflect benefit take-up.

Over longer horizons, the analysis does not support the view that maternity leave increases maternal labour-market attachment by helping women stay employed in the long run. Across all three reforms, we find that the negative effect on labour supply persists: the point estimates tend to be somewhat smaller at longer horizons, particularly for employment, but the pattern is not monotone and the effect does not reverse. Six years after childbirth, the effects on hours worked remain negative for all three reforms (statistically significant at the 1\% level for the 2000 reform, at the 5\% level for the 2003 reform, and at the 10\% level for the 2007 reform), and the effects on the probability of being employed remain negative and statistically significant at the 5\% level or better for all three reforms.

Finally, the comparison across reforms provides suggestive evidence about the mechanisms behind the 2007 reform's effects on children. Although the three reforms differ in design and in their effects on children, the point estimates suggest broadly comparable reductions in maternal hours and employment. In particular, the 2000 job-protection reform reduced maternal labour supply without detectable gains in children's outcomes. This pattern suggests that additional time at home alone may not be sufficient to improve children's outcomes and is consistent with a role for income support during leave, potentially through reduced financial stress complementing maternal time investments. Two caveats apply: the 2007 reform also extended job protection to mothers with short tenure (who, as noted in Section~\ref{sec:contextanddata}, mostly did not qualify for Statutory Maternity Pay), and employment one year after childbirth is a coarse proxy for time spent at home.

\section{Conclusion}
\label{sec:conclusion}
Most evaluations of maternity leave find no average effect on children's outcomes, a conclusion drawn largely from countries with generous provision. In the UK, where provision was initially low, we find that maternity-leave policy had a small but robust and lasting positive effect on children's test scores. The 2007 reform, primarily an extension of pay, raised scores at age 5 by 0.045 standard deviations, equivalent to about 25 additional days of age at testing. At age 11, the estimated effect remains statistically significant and is equivalent to about 15 additional days of age at testing. The estimates are stable across outcome measures, birth weeks, and control cohorts. The estimated age-5 effect is about twice as large in the most deprived fifth of areas, where the additional income is likely to matter most, as in the least deprived. The 2003 reform, which extended both pay and job protection, had an estimated age-5 effect less than half that of the 2007 reform, and the 2000 extension of job protection alone had no measurable effect. All three reforms reduced maternal employment and hours worked up to six years after childbirth, which does not support the premise that leave expansions can strengthen mothers' long-run labour-market attachment while improving children's outcomes.

None of these effects would have been detected using a conventional RD design. Eligibility for the reforms depended on the expected date of birth, which neither of our outcome datasets records. The actual date of birth is therefore only an imperfect measure of the true running variable, and the resulting measurement error effectively eliminates the discontinuity on which conventional RD relies. Existing corrections are of limited use in this setting, because the error has wide support and may itself be related to potential outcomes. Our imputation-based approach uses an untreated pre-reform cohort to impute average outcomes absent the reform under a parallel-trends assumption with testable implications. We combine these imputed outcomes with the treated share at each actual birth date, recovered from validation data on actual and expected birth dates. Dividing the difference between observed and imputed average outcomes by that share identifies the effect on children who were eligible at each birth date with a positive treated share, rather than at the cutoff alone as in conventional RD. Unlike many existing correction methods, our approach remains valid when the gap between actual and expected birth dates is related to potential outcomes. This is crucial in our application because premature delivery, for example, may itself affect children's subsequent development.

Beyond maternity leave, the approach may be useful in other RD settings in which true treatment status is unobserved because the relevant running variable is latent or measured imperfectly. It requires a credible way of imputing counterfactual outcomes, here a comparable untreated cohort, and auxiliary data that recover the treatment probability conditional on the observed running variable. Pension reforms that combine age cutoffs with imperfectly observed contribution histories, family-policy reforms for which eligibility depends on an imperfectly recorded employment history, and fetal-origins designs in which exposure to a shock is inferred from the date of birth are possible examples. In these applications, as in ours, measurement error may have economic content of its own and be related to potential outcomes. Our approach accommodates this possibility, making it a promising option when both conventional RD and existing corrections requiring measurement error to be conditionally independent of potential outcomes are inapplicable.

\clearpage
\printbibliography[heading=bibintoc]

@article{grembi_fiscal_2016,
    ids = {grembi2016fiscal},
    publisher = {JSTOR},
    title = {Do {Fiscal} {Rules} {Matter}?},
    volume = {8},
    issn = {1945-7782},
    url = {https://www.aeaweb.org/articles?id=10.1257/app.20150076},
    doi = {10.1257/app.20150076},
    language = {en},
    number = {3},
    urldate = {2021-08-11},
    journal = {American Economic Journal: Applied Economics},
    author = {Grembi, Veronica and Nannicini, Tommaso and Troiano, Ugo},
    month = jul,
    year = {2016},
    pages = {1--30},
}

@article{carneiro_flying_2015,
    title = {A {Flying} {Start}? {Maternity} {Leave} {Benefits} and {Long}-{Run} {Outcomes} of {Children}},
    volume = {123},
    number = {2},
    journal = {Journal of Political Economy},
    author = {Carneiro, Pedro and L{\o}ken, Katrine V. and Salvanes, Kjell G.},
    year = {2015},
    pages = {365--412},
}

@techreport{koslowski_17th_2021,
    title = {17th {International} {Review} of {Leave} {Policies} and {Related} {Research} 2021},
    type = {Report},
    institution = {{International Network on Leave Policies and Research}},
    url = {https://www.leavenetwork.org/annual-review-reports/review-2021/},
    author = {Koslowski, Alison and Blum, Sonja and Dobroti{\'c}, Ivana and Kaufman, Gayle and Moss, Peter},
    year = {2021},
}

@online{lee_honey_2021,
    title = {Honey, {I} shrunk the {Shared} {Parental} {Leave} take-up figures},
    organization = {Maternity Action},
    author = {Dunstan, Richard},
    url = {https://maternityaction.org.uk/2021/02/honey-i-shrunk-the-shared-parental-leave-take-up-figures/},
    date = {2021-02-15},
}

@article{baker_maternal_2008,
    title = {Maternal Employment, Breastfeeding, and Health: {Evidence} from Maternity Leave Mandates},
    volume = {27},
    number = {4},
    journal = {Journal of Health Economics},
    author = {Baker, Michael and Milligan, Kevin},
    year = {2008},
    pages = {871--887},
}

@article{berger_maternity_2005,
    title = {Maternity {Leave}, {Early} {Maternal} {Employment} and {Child} {Health} and {Development} in the {US}},
    volume = {115},
    number = {501},
    journal = {The Economic Journal},
    author = {Berger, Lawrence M. and Hill, Jennifer and Waldfogel, Jane},
    year = {2005},
    pages = {F29--F47},
}

@article{dahl_what_2016,
    ids = {dahl_what_2016-1},
    title = {What {Is} the {Case} for {Paid} {Maternity} {Leave}?},
    volume = {98},
    issn = {0034-6535, 1530-9142},
    url = {https://direct.mit.edu/rest/article/98/4/655-670/58344},
    doi = {10.1162/REST_a_00602},
    language = {en},
    number = {4},
    urldate = {2022-09-07},
    journal = {Review of Economics and Statistics},
    author = {Dahl, Gordon B. and L{\o}ken, Katrine V. and Mogstad, Magne and Salvanes, Kari Vea},
    month = oct,
    year = {2016},
    pages = {655--670},
}

@article{danzer_paid_2018,
    title = {Paid {Parental} {Leave} and {Children}'s {Schooling} {Outcomes}},
    volume = {128},
    number = {608},
    journal = {The Economic Journal},
    author = {Danzer, Natalia and Lavy, Victor},
    year = {2018},
    pages = {81--117},
}

@article{dustmann_expansions_2012,
    ids = {dustmann_expansions_2012-1},
    doi = {10.1257/app.4.3.190},
    url = {https://pubs.aeaweb.org/doi/10.1257/app.4.3.190},
    month = jul,
    title = {Expansions in {Maternity} {Leave} {Coverage} and {Children}'s {Long}-{Term} {Outcomes}},
    volume = {4},
    number = {3},
    journal = {American Economic Journal: Applied Economics},
    author = {Dustmann, Christian and Sch{\"o}nberg, Uta},
    year = {2012},
    pages = {190--224},
}

@article{lalive_how_2009,
    title = {How {Does} {Parental} {Leave} {Affect} {Fertility} and {Return} to {Work}? {Evidence} from {Two} {Natural} {Experiments}},
    volume = {124},
    number = {3},
    journal = {The Quarterly Journal of Economics},
    author = {Lalive, Rafael and Zweim{\"u}ller, Josef},
    year = {2009},
    pages = {1363--1402},
}

@article{rossin_effects_2011,
    title = {The Effects of Maternity Leave on Children's Birth and Infant Health Outcomes in the {United} {States}},
    volume = {30},
    number = {2},
    journal = {Journal of Health Economics},
    author = {Rossin, M.},
    year = {2011},
    pages = {221--239},
}

@article{rossin-slater_effects_2013,
    title = {The {Effects} of {California}'s {Paid} {Family} {Leave} {Program} on {Mothers}' {Leave}-{Taking} and {Subsequent} {Labor} {Market} {Outcomes}},
    author = {Rossin-Slater, Maya and Ruhm, Christopher J. and Waldfogel, Jane},
    journal = {Journal of Policy Analysis and Management},
    volume = {32},
    number = {2},
    pages = {224--245},
    doi = {10.1002/pam.21676},
    year = {2013},
}

@article{schoenberg_expansions_2014,
    ids = {schonberg_expansions_2014},
    title = {Expansions in {Maternity} {Leave} {Coverage} and {Mothers}' {Labor} {Market} {Outcomes} after {Childbirth}},
    volume = {32},
    language = {en},
    number = {3},
    journal = {Journal of Labor Economics},
    author = {Sch{\"o}nberg, Uta and Ludsteck, Johannes},
    month = jul,
    year = {2014},
    pages = {469--505},
}

@article{waldfogel_effects_2002,
    title = {The Effects of Early Maternal Employment on Child Cognitive Development},
    volume = {39},
    number = {2},
    journal = {Demography},
    author = {Waldfogel, Jane and Han, Wen-Jui and Brooks-Gunn, Jeanne},
    year = {2002},
    pages = {369--392},
}

@article{rasmussen_increasing_2010,
    title = {Increasing the Length of Parents' Birth-related Leave: {The} Effect on Children's Long-term Educational Outcomes},
    volume = {17},
    number = {1},
    journal = {Labour Economics},
    author = {Rasmussen, Astrid W{\"u}rtz},
    year = {2010},
    pages = {91--100},
}

@article{zabel_eligibility_2009,
    title = {Eligibility for {Maternity} {Leave} and {First} {Birth} {Timing} in {Great} {Britain}},
    volume = {28},
    number = {3},
    journal = {Population Research and Policy Review},
    author = {Zabel, C.},
    month = jun,
    year = {2009},
    pages = {251--270},
}

@article{gregg_employment_2007,
    title = {The {Employment} of {Married} {Mothers} in {Great} {Britain}, 1974–2000},
    volume = {74},
    number = {296},
    journal = {Economica},
    author = {Gregg, Paul and Guti{\'e}rrez-Dom{\`e}nech, Maria and Waldfogel, Jane},
    year = {2007},
    pages = {842--864},
}

@article{davezies_regression_2017,
    title = {Regression Discontinuity Design with Continuous Measurement Error in the Running Variable},
    volume = {200},
    issn = {03044076},
    url = {https://linkinghub.elsevier.com/retrieve/pii/S0304407617300969},
    doi = {10.1016/j.jeconom.2017.06.010},
    language = {en},
    number = {2},
    urldate = {2022-09-07},
    journal = {Journal of Econometrics},
    author = {Davezies, Laurent and Le Barbanchon, Thomas},
    month = oct,
    year = {2017},
    pages = {260--281},
}

@article{kleven_children_2019,
    title = {Children and {Gender} {Inequality}: {Evidence} from {Denmark}},
    volume = {11},
    number = {4},
    journal = {American Economic Journal: Applied Economics},
    author = {Kleven, Henrik and Landais, Camille and S{\o}gaard, Jakob Egholt},
    year = {2019},
    pages = {181--209},
}

@article{kleven_child_2019,
    title = {Child {Penalties} across {Countries}: {Evidence} and {Explanations}},
    volume = {109},
    journal = {AEA Papers and Proceedings},
    author = {Kleven, Henrik and Landais, Camille and Posch, Johanna and Steinhauer, Andreas and Zweim{\"u}ller, Josef},
    year = {2019},
    pages = {122--126},
}

@article{liu_duration_2010,
    title = {The {Duration} of {Paid} {Parental} {Leave} and {Children}'s {Scholastic} {Performance}},
    volume = {10},
    number = {1},
    journal = {The B.E. Journal of Economic Analysis \& Policy},
    author = {Liu, Qian and {Nordstr{\"o}m Skans}, Oskar},
    year = {2010},
}

@article{heckman_skill_2006,
    title = {Skill {Formation} and the {Economics} of {Investing} in {Disadvantaged} {Children}},
    volume = {312},
    number = {5782},
    journal = {Science},
    author = {Heckman, James},
    year = {2006},
    pages = {1900--1902},
}

@article{troller-renfree_impact_2022,
    title = {The Impact of a Poverty Reduction Intervention on Infant Brain Activity},
    volume = {119},
    number = {5},
    journal = {Proceedings of the National Academy of Sciences},
    author = {Troller-Renfree, S. and Costanzo, M. and Duncan, G. and Magnuson, K. and Gennetian, L. and Yoshikawa, H. and Halpern-Meekin, S. and Fox, N. and Noble, K.},
    year = {2022},
    eid = {e2115649119},
    doi = {10.1073/pnas.2115649119},
}

@techreport{chzhen_are_2019,
    title = {Are the World's Richest Countries Family Friendly? {Policy} in the {OECD} and {EU}},
    type = {Report},
    institution = {{UNICEF Office of Research -- Innocenti}},
    address = {Florence},
    author = {Chzhen, Yekaterina and Gromada, Anna and Rees, Gwyther},
    month = jun,
    year = {2019},
}

@techreport{lourie_family_1999,
    title = {Family {Leave}},
    type = {Research Paper},
    number = {99/89},
    institution = {House of Commons Library},
    address = {London},
    author = {Lourie, Julia},
    month = nov,
    year = {1999},
}

@article{boelmann_wind_2021,
    title = {Wind of {Change}? {Cultural} {Determinants} of {Maternal} {Labor} {Supply}},
    volume = {17},
    number = {2},
    journal = {American Economic Journal: Applied Economics},
    author = {Boelmann, Barbara and Raute, Anna and Sch{\"o}nberg, Uta},
    year = {2025},
    pages = {41--74},
    doi = {10.1257/app.20220567},
    url = {https://doi.org/10.1257/app.20220567},
}

@article{dong_when_2023,
    title = {When {Can} {We} {Ignore} {Measurement} {Error} in the {Running} {Variable}?},
    volume = {38},
    issn = {0883-7252, 1099-1255},
    url = {http://arxiv.org/abs/2111.07388},
    doi = {10.1002/JAE.2974},
    language = {en},
    number = {5},
    urldate = {2023-10-09},
    journal = {Journal of Applied Econometrics},
    author = {Dong, Yingying and Kolesár, Michal},
    year = {2023},
    pages = {735--750},
}

@article{dong_regression_2015,
    title = {Regression {Discontinuity} {Applications} with {Rounding} {Errors} in the {Running} {Variable}},
    volume = {30},
    issn = {0883-7252, 1099-1255},
    url = {https://onlinelibrary.wiley.com/doi/10.1002/jae.2369},
    doi = {10.1002/jae.2369},
    language = {en},
    number = {3},
    urldate = {2023-10-09},
    journal = {Journal of Applied Econometrics},
    author = {Dong, Yingying},
    year = {2015},
    pages = {422--446},
}

@incollection{pei_devil_2017,
    title = {The {Devil} is in the {Tails}: {Regression} {Discontinuity} {Design} with {Measurement} {Error} in the {Assignment} {Variable}},
    booktitle = {Regression {Discontinuity} {Designs}: {Theory} and {Applications}},
    series = {Advances in {Econometrics}},
    volume = {38},
    editor = {Cattaneo, Matias D. and Escanciano, Juan Carlos},
    publisher = {Emerald Publishing Limited},
    author = {Pei, Zhuan and Shen, Yi},
    year = {2017},
    pages = {455--502},
}

@article{markussen_not_2023,
    title = {Not a {Flying} {Start} after {All}? {A} {Comment}},
    volume = {132},
    number = {12},
    journal = {Journal of Political Economy},
    author = {Lilleb{\o}, Otto Sevaldson and Markussen, Simen and R{\o}ed, Knut and Zhao, Yuejun},
    month = dec,
    year = {2024},
    pages = {4205--4212},
    doi = {10.1086/732218},
    url = {https://www.journals.uchicago.edu/doi/10.1086/732218},
}

@article{kleven_family_2024,
    title = {Do {Family} {Policies} {Reduce} {Gender} {Inequality}? {Evidence} from 60 {Years} of {Policy} {Experimentation}},
    volume = {16},
    number = {2},
    journal = {American Economic Journal: Economic Policy},
    author = {Kleven, Henrik and Landais, Camille and Posch, Johanna and Steinhauer, Andreas and Zweim{\"u}ller, Josef},
    year = {2024},
    pages = {110--149},
}

@article{raute_can_2019,
    title = {Can Financial Incentives Reduce the Baby Gap? {Evidence} from a Reform in Maternity Leave Benefits},
    volume = {169},
    journal = {Journal of Public Economics},
    author = {Raute, A.},
    year = {2019},
    pages = {203--222},
}

@techreport{seidlitz_effects_2021,
    title = {Effects of {Motherhood} and {Selection} into {Motherhood}: {Revisiting} the 2007 {Parental} {Benefits} {Reform} in {Germany}},
    type = {Beitr{\"a}ge zur Jahrestagung des Vereins f{\"u}r Socialpolitik 2021},
    institution = {{ZBW -- Leibniz Information Centre for Economics}},
    url = {https://hdl.handle.net/10419/242435},
    author = {Seidlitz, Arnim and Fitzenberger, Bernd},
    year = {2021},
}

@article{bailey_long-run_2025,
    title = {The {Long}-{Run} {Effects} of {California}'s {Paid} {Family} {Leave} {Act} on {Women}'s {Careers} and {Childbearing}: {New} {Evidence} from a {Regression} {Discontinuity} {Design} and {U}.{S}. {Tax} {Data}},
    volume = {17},
    number = {1},
    pages = {401--431},
    doi = {10.1257/pol.20200277},
    journal = {American Economic Journal: Economic Policy},
    author = {Bailey, Martha J. and Byker, Tanya S. and Patel, Elena and Ramnath, Shanthi},
    year = {2025},
    ids = {bailey_long-run_nodate},
}

@article{almond2011role,
    title = {{The Role of Hospital Heterogeneity in Measuring Marginal Returns to Medical Care: A Reply to Barreca, Guldi, Lindo, and Waddell}},
    author = {Almond, Douglas and Doyle, Jr., Joseph J. and Kowalski, Amanda E. and Williams, Heidi},
    journal = {The Quarterly Journal of Economics},
    volume = {126},
    number = {4},
    pages = {2125--2131},
    year = {2011},
    publisher = {Oxford University Press},
}

@article{barreca2011saving,
    ids = {barreca_saving_2011},
    month = nov,
    title = {{Saving Babies? Revisiting the Effect of Very Low Birth Weight Classification}},
    author = {Barreca, Alan I and Guldi, Melanie and Lindo, Jason M and Waddell, Glen R},
    journal = {The Quarterly Journal of Economics},
    volume = {126},
    number = {4},
    pages = {2117--2123},
    year = {2011},
    publisher = {Oxford University Press},
}

@misc{noack2023donut,
    title = {{Donut Regression Discontinuity Designs}},
    author = {Noack, Claudia and Rothe, Christoph},
    eprinttype = {arxiv},
    eprint = {2308.14464},
    year = {2023},
}

@article{battistin2009retirement,
    title = {{The Retirement Consumption Puzzle: Evidence from a Regression Discontinuity Approach}},
    author = {Battistin, Erich and Brugiavini, Agar and Rettore, Enrico and Weber, Guglielmo},
    journal = {American Economic Review},
    volume = {99},
    number = {5},
    pages = {2209--2226},
    year = {2009},
    publisher = {American Economic Association},
}

@article{hullegie2010effect,
    title = {{The Effect of Private Health Insurance on Medical Care Utilization and Self-assessed Health in Germany}},
    author = {Hullegie, Patrick and Klein, Tobias J},
    journal = {Health Economics},
    volume = {19},
    number = {9},
    pages = {1048--1062},
    year = {2010},
    publisher = {Wiley Online Library},
}

@article{bartalotti2021correction,
    ids = {bartalotti_correction_2021},
    title = {{A Correction for Regression Discontinuity Designs with Group-Specific Mismeasurement of the Running Variable}},
    author = {Bartalotti, Ot{\'a}vio and Brummet, Quentin and Dieterle, Steven},
    journal = {Journal of Business \& Economic Statistics},
    volume = {39},
    number = {3},
    pages = {833--848},
    year = {2021},
    publisher = {Taylor \& Francis},
}

@article{cattaneo2021extrapolating,
    title = {{Extrapolating Treatment Effects in Multi-Cutoff Regression Discontinuity Designs}},
    author = {Cattaneo, Matias D and Keele, Luke and Titiunik, Roc{\'\i}o and Vazquez-Bare, Gonzalo},
    journal = {Journal of the American Statistical Association},
    volume = {116},
    number = {536},
    pages = {1941--1952},
    year = {2021},
    publisher = {Taylor \& Francis},
}

@misc{leventer2024correcting,
    title = {{Correcting Invalid Regression Discontinuity Designs with Multiple Time Period Data}},
    author = {Leventer, Dor and Nevo, Daniel},
    eprinttype = {arxiv},
    eprint = {2408.05847},
    year = {2024},
}

@misc{picchetti2024difference,
    title = {{Difference-in-Discontinuities: Estimation, Inference and Validity Tests}},
    author = {Picchetti, Pedro and Pinto, Cristine CX and Shinoki, Stephanie T},
    eprinttype = {arxiv},
    eprint = {2405.18531},
    year = {2024},
}

@article{lalive2014parental,
    ids = {lalive_parental_2014},
    title = {{Parental Leave and Mothers' Careers: The Relative Importance of Job Protection and Cash Benefits}},
    author = {Lalive, Rafael and Schlosser, Anal{\'\i}a and Steinhauer, Andreas and Zweim{\"u}ller, Josef},
    journal = {Review of Economic Studies},
    volume = {81},
    number = {1},
    pages = {219--265},
    year = {2014},
    publisher = {Oxford University Press},
}

@article{almond_influenza_2006,
    title = {Is the 1918 Influenza Pandemic Over? Long-Term Effects of In Utero Influenza Exposure in the Post-1940 U.S. Population},
    author = {Almond, Douglas},
    journal = {Journal of Political Economy},
    volume = {114},
    number = {4},
    pages = {672--712},
    year = {2006},
    doi = {10.1086/507154},
}

@article{almond_chernobyl_2009,
    title = {Chernobyl's Subclinical Legacy: Prenatal Exposure to Radioactive Fallout and School Outcomes in Sweden},
    author = {Almond, Douglas and Edlund, Lena and Palme, M{\aa}rten},
    journal = {The Quarterly Journal of Economics},
    volume = {124},
    number = {4},
    pages = {1729--1772},
    year = {2009},
    doi = {10.1162/qjec.2009.124.4.1729},
}

@article{dechaisemartin_fuzzy_2018,
    title = {Fuzzy Differences-in-Differences},
    author = {{de Chaisemartin}, Cl{\'e}ment and D'Haultf{\oe}uille, Xavier},
    journal = {The Review of Economic Studies},
    volume = {85},
    number = {2},
    pages = {999--1028},
    year = {2018},
    doi = {10.1093/restud/rdx049},
}

@article{deming_early_2009,
    title = {Early {Childhood} {Intervention} and {Life}-{Cycle} {Skill} {Development}: {Evidence} from {Head} {Start}},
    volume = {1},
    number = {3},
    journal = {American Economic Journal: Applied Economics},
    author = {Deming, David},
    year = {2009},
    pages = {111--134},
    doi = {10.1257/app.1.3.111},
}

@report{carneiro_short-_2024,
    title = {The {Short}- and {Medium}-{Term} {Impacts} of {Sure} {Start} on {Educational} {Outcomes}},
    type = {IFS Report},
    number = {R307},
    institution = {Institute for Fiscal Studies},
    author = {Carneiro, Pedro and Cattan, Sarah and Ridpath, Nick},
    year = {2024},
    address = {London},
    doi = {10.1920/re.ifs.2024.0307},
}

@article{eckles2025noise,
    title = {Noise-induced Randomization in Regression Discontinuity Designs},
    author = {Eckles, Dean and Ignatiadis, Nikolaos and Wager, Stefan and Wu, Han},
    journal = {Biometrika},
    volume = {112},
    number = {2},
    pages = {asaf003},
    year = {2025},
    publisher = {Oxford University Press},
}
\newpage
\renewcommand{\thesubsection}{Appendix~\Alph{subsection}}
\renewcommand{\thesubsubsection}{\Alph{subsection}.\arabic{subsubsection}}  %
\renewcommand{\thefigure}{\Alph{subsection}.\arabic{figure}}
\renewcommand{\thetable}{\Alph{subsection}.\arabic{table}}
\renewcommand{\theequation}{\Alph{subsection}.\arabic{equation}}
\counterwithin*{figure}{subsection}  %
\counterwithin*{table}{subsection}
\counterwithin*{equation}{subsection}
\setcounter{figure}{0}
\setcounter{table}{0}

\phantomsection  %
\addcontentsline{toc}{section}{Appendix}
\setcounter{subsection}{0}

\FloatBarrier
\subsection{Estimation of the treatment probability}
\label{sec:phat}

For each reform, the treatment probability at actual birth date $x$ is
\begin{align*}
    p^t(x)
    \coloneqq
    \mathbb P(D_i^t=1\mid X_i^t=x)
    =
    \mathbb P(E_i^t\geq x_0\mid X_i^t=x),
\end{align*}
where $x_0$ is the reform-specific cutoff in expected date of birth. We estimate this probability using the Millennium Cohort Study (MCS), which records both actual and expected dates of birth. Let $(X^v,E^v)$ denote a generic validation observation and define $\varepsilon^v=X^v-E^v$. We use $j=1,\ldots,n_v$ to index observations in the usable validation sample.

Applying the validation relationship to the reform cohorts requires two restrictions. We maintain that the distribution of the measurement error is transportable from the validation cohort to each reform cohort and does not vary with actual date of birth. Formally, for the cohorts, birth dates, and values of the measurement error used in the analysis,
\begin{align}
    \mathbb P(\varepsilon_i^t\leq e\mid X_i^t=x)
    =
    \mathbb P(\varepsilon^v\leq e\mid X^v=x)
    =
    \mathbb P(\varepsilon^v\leq e)
    \equiv
    F_\varepsilon(e).
    \label{eq:error_distribution_restriction}
\end{align}
The first equality transports the conditional distribution from the MCS to the reform cohorts. The second allows us to pool validation observations across actual birth dates.

Because $D_i^t=\mathbf 1\{E_i^t\geq x_0\}$ and $E_i^t=X_i^t-\varepsilon_i^t$, restriction~\eqref{eq:error_distribution_restriction} implies
\begin{align*}
    p^t(x)
    =
    \mathbb P(\varepsilon_i^t\leq x-x_0\mid X_i^t=x)
    =
    F_\varepsilon(x-x_0).
\end{align*}
We therefore use the empirical-CDF estimator
\begin{align}
    \widehat p^t(x)
    =
    \widehat F_\varepsilon(x-x_0)
    =
    \frac{1}{n_v}
    \sum_{j=1}^{n_v}
    \mathbf 1\!\left\{X_j^v-E_j^v\leq x-x_0\right\}.
    \label{eq:phat_ecdf}
\end{align}
Evaluating the same empirical CDF at each value of $x-x_0$ yields the complete treatment-probability curve. Panel~B of Figure~\ref{fig:diagnosis} illustrates this curve for the 2007 reform.

\begin{figure}[!htbp]
\begin{center}
    \includegraphics[width=0.8\textwidth]{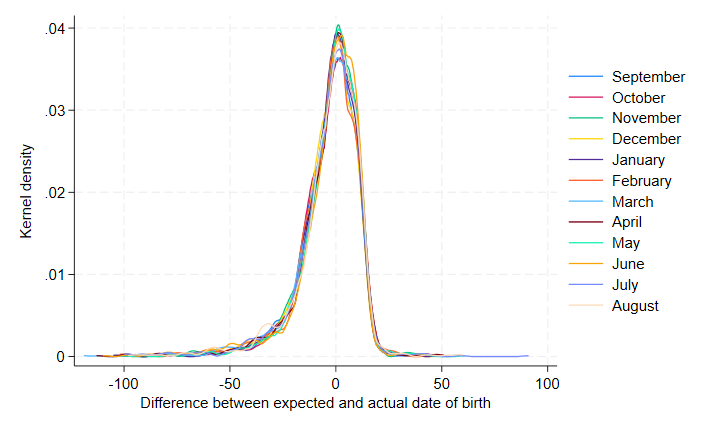}
    \refstepcounter{figure}
    \label{fig:stationarity}

    \vspace{0.35cm}
    \begin{small}
        \textbf{Figure~\thefigure: Distribution of the birth-date measurement error by calendar month}
    \end{small}

    \vspace{0.3cm}
    \begin{minipage}{0.9\textwidth}
        \begin{footnotesize}
            \begin{spacing}{1.15}
                \textit{Notes:} The figure plots kernel-density estimates of $\varepsilon_j^v=X_j^v-E_j^v$ in the MCS, separately by calendar month of actual birth. Negative values indicate births occurring before the expected date of birth. {}
            \end{spacing}
        \end{footnotesize}
    \end{minipage}
\end{center}
\end{figure}

Figure~\ref{fig:stationarity} provides an informal diagnostic for the pooling restriction by comparing the distribution of the measurement error across calendar months of actual birth within the MCS. The close alignment of the curves shows no pronounced monthly seasonality and is therefore consistent with pooling across birth dates. This comparison does not test transportability to the reform cohorts, which remains a maintained restriction.

Because the same empirical CDF is used at every actual birth date, the estimation errors in $\widehat p^t(x)$ are correlated across dates. \ref{sec:standarderrors} derives the corresponding covariance matrix and incorporates this sampling uncertainty into inference for the nonparametric and regression-based estimators.

\FloatBarrier
\subsection{Detailed description of the reforms of interest}
\label{sec:reformsdetails}

The reforms studied in this paper build on entitlements introduced gradually over the preceding decades. Job-protected maternity leave and employer-paid maternity pay were first introduced by the Employment Protection Act 1975 (effective 1976--77), for mothers continuously employed with the same employer for two years at 16 or more hours a week (or five years at 8 to 16 hours); a contribution-based Maternity Allowance had existed since the introduction of the postwar National Insurance system. In April 1987, Statutory Maternity Pay replaced the earlier employer-paid scheme with 18 weeks of pay: six weeks at 90\% of earnings followed by 12 weeks at a flat rate for women with two years' tenure, and 18 weeks at the flat rate for women with 26 weeks' tenure. The 1994 reform, implementing the EU Pregnant Workers Directive, entitled all pregnant employees---regardless of hours worked or length of service---to 14 weeks of maternity leave, of which the two weeks following childbirth were compulsory, and replaced the two-year employer-tenure test for the higher rate of Statutory Maternity Pay (six weeks at 90\% of earnings) with a single test of 26 weeks of continuous employment with the same employer into the 15th week before the expected week of childbirth \parencite{gregg_employment_2007,lourie_family_1999,zabel_eligibility_2009}.

\begin{figure}[!htbp]
  \centering
  \includegraphics[width=0.7\textwidth]{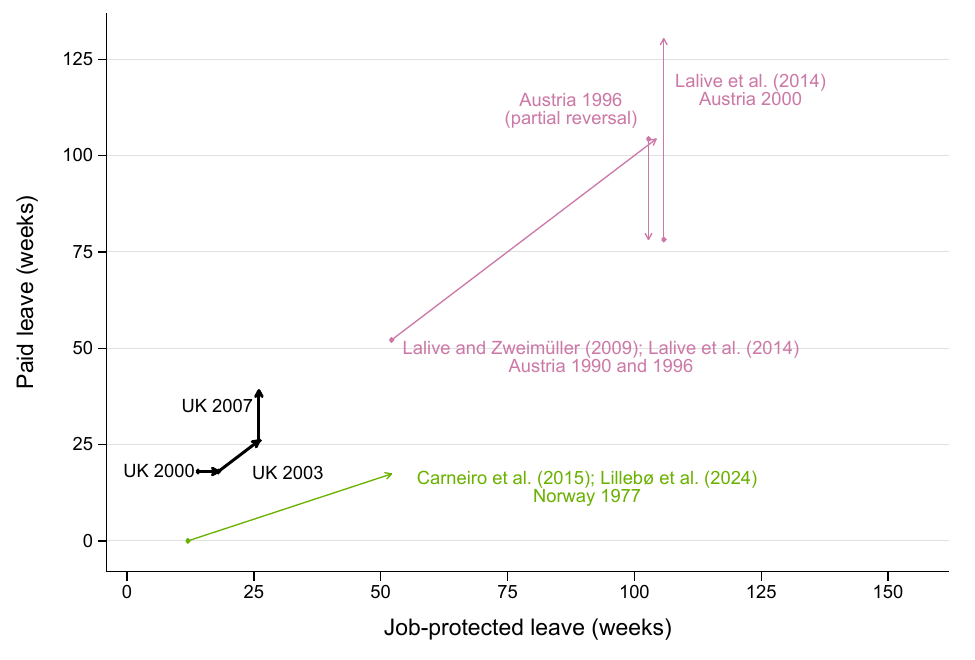}

  \refstepcounter{figure}
  \label{fig:top5}

  \vspace{0.35cm}
  \begin{small}
      \textbf{Figure~\thefigure: Reforms studied in the five leading economics journals}
  \end{small}

  \vspace{0.3cm}
  \begin{minipage}{0.9\textwidth}
      \begin{footnotesize}
          \begin{spacing}{1.15}
          \textit{Notes:} The figure repeats Figure~\ref{fig:literature} for the reforms analysed in studies published in the five leading economics journals; the black arrows are the three UK reforms, and axes and arrows are as in Figure~\ref{fig:literature}.
          \end{spacing}
      \end{footnotesize}
  \end{minipage}
\end{figure}

Two institutional details matter for interpreting the reforms. First, weekly benefit rates are uprated every April and apply by week of payment rather than by the eligibility cutoff: mothers with expected dates of birth shortly before 6 April 2003, for example, also received the new $\pounds 100$ rate for the remaining weeks of their (12-week) flat-rate period. Of the changes to maternity pay, the component assigned by the expected date of birth is therefore the extension of the flat-rate period rather than the rate increases themselves (see Panel B of Table~\ref{tab:reforms_details}); the leave entitlements in Panel A are likewise assigned by the expected date of birth. Second, Maternity Allowance eligibility was reformed \emph{within} the 2000 regime: for expected weeks of childbirth beginning on or after 20 August 2000, contribution conditions were replaced by a test of 26 weeks of employment or self-employment in the 66 weeks before the expected week of childbirth, with average weekly earnings of at least $\pounds 30$ (Welfare Reform and Pensions Act 1999, s.\,53). This change falls at a different expected-birth-date cutoff from the reforms studied here.

The two weeks of paid paternity leave introduced in the 2003 bundle reform were available for children born on or after 6 April 2003 \emph{or} with an expected date of birth on or after that date, conditional on the father having 26 weeks of continuous employment. Because eligibility was triggered by the actual or the expected date of birth, this component of the 2003 bundle was partly assigned by the actual date of birth; its scope---two weeks at the flat rate---is small relative to the maternity components.

Since 2010, successive governments have extended rights to fathers while leaving maternity-specific provision largely unchanged. Regulations made in March 2010 (in force from 6 April 2010) introduced additional paternity leave and additional paternity payments for parents of infants due on or after 3 April 2011. In 2013, unpaid parental leave was extended from 13 to 18 weeks. Shared Parental Leave was introduced for children due on or after 5 April 2015, allowing mothers to share up to 50 weeks of leave and up to 37 weeks of pay with their partner; the benefit rates equal those of the flat-rate period of maternity pay, and eligibility depends on employment status, tenure, and pay (Shared Parental Leave Regulations 2014, SI 2014/3050; Statutory Shared Parental Pay (General) Regulations 2014, SI 2014/3051). Take-up by eligible fathers has been estimated to lie below 5\% \parencite{lee_honey_2021}.

Table~\ref{tab:reforms_details} summarises the eligibility requirements and entitlements in force under each of the four regimes, together with the sources on which our account is based. Figure~\ref{fig:top5} complements Figure~\ref{fig:literature} in the main text by restricting attention to studies published in the five leading economics journals.

\newgeometry{left=1.05cm, right=1.05cm, top=1.2cm, bottom=1.2cm}
\begin{landscape}

\begin{table}
\begin{center}
    \caption{UK maternity-leave entitlements by regime, 1994--2007}
    \label{tab:reforms_details}
    \vspace{0.1cm}
    \begin{adjustbox}{max totalheight=0.9\textheight, max width=\linewidth}
    \begin{minipage}{\linewidth}
\begin{threeparttable}
\footnotesize
\setstretch{1.0}  %
\renewcommand{\arraystretch}{1.05}
\setlength{\tabcolsep}{5pt}
\emph{Panel A: Entitlement and eligibility rules, by regime}\par\vspace{0.15cm}
\begin{tabular}{>{\raggedright\arraybackslash}p{0.20\linewidth} >{\raggedright\arraybackslash}p{0.17\linewidth} >{\raggedright\arraybackslash}p{0.21\linewidth} >{\raggedright\arraybackslash}p{0.17\linewidth} >{\raggedright\arraybackslash}p{0.15\linewidth}}
\toprule
 & 16 October 1994 & 30 April 2000 & 6 April 2003 & 1 April 2007 \\
\midrule
\multicolumn{5}{l}{\emph{Job-protected leave}}\\
\quad Leave for all pregnant employees & 14 weeks & 18 weeks & 26 weeks & 52 weeks \\
\quad Extended leave\tnote{a} & until 29 weeks from the start of the week of childbirth & --- & 52 weeks in total from the start of leave & --- \\
\quad Tenure required for extended leave\tnote{b} & 2 years & 1 year & 26 weeks & none \\
\addlinespace
\multicolumn{5}{l}{\emph{Statutory Maternity Pay (SMP)}}\\
\quad Eligibility & 26 continuous weeks with the same employer into the 15th week before the EWC; earnings at or above the LEL & --- & --- & --- \\
\quad Duration and structure\tnote{c} & 18 weeks: 6 at 90\% of earnings, then 12 flat & --- & 26 weeks: 6 at 90\%, then 20 flat & 39 weeks: 6 at 90\%, then 33 flat \\
\addlinespace
\multicolumn{5}{l}{\emph{Maternity Allowance (MA)}}\\
\quad Eligibility & contribution-based test & 26 weeks of employment or self-employment in the 66 weeks before the EWC, with earnings of at least \pounds 30 per week (from 20 August 2000)\tnote{d} & --- & --- \\
\quad Duration\tnote{c} & 18 weeks & --- & 26 weeks & 39 weeks \\
\addlinespace
\multicolumn{5}{l}{\emph{Paternity leave (fathers)}}\\
\quad Paid paternity leave\tnote{e} & none & --- & 2 weeks at the flat rate & --- \\
\bottomrule
\end{tabular}\par\vspace{0.25cm}
\emph{Panel B: Flat weekly rate of SMP and MA, in force from April of each year}\tnote{c}\par\vspace{0.15cm}
\begin{tabular}{l ccccccccc}
\toprule
 & 1994 & 2000 & 2001 & 2002 & 2003 & 2004 & 2005 & 2006 & 2007 \\
\midrule
Flat weekly rate & \pounds 52.50 & \pounds 60.20 & \pounds 62.20 & \pounds 75 & \pounds 100 & \pounds 102.80 & \pounds 106 & \pounds 108.85 & \pounds 112.75 \\
\bottomrule
\end{tabular}
\end{threeparttable}
\par\vspace{0.18cm}
{\scriptsize
\textit{Notes:} Each column of Panel A describes the rules applying to mothers whose expected week of childbirth (EWC) began on or after the date at the top of the column and before the date of the next column; the first column gives the rules introduced in October 1994. A dash indicates that the rule is unchanged relative to the previous column. In the terminology of the legislation, the leave available to all pregnant employees is ``ordinary maternity leave'' and the extended leave is ``additional maternity leave''; from 2007, both are available to all pregnant employees, and the first row reports their combined length of 52 weeks. LEL: lower earnings limit for National Insurance contributions.\par
\textsuperscript{a}\,Leave could begin up to 11 weeks before the EWC. Until 2003, the extended entitlement was a right to return to the same employer until the end of a 29-week period counted from the beginning of the week of childbirth (Employment Rights Act 1996, s.\,79; SI 1999/3312, reg.\,7); from 2003, ordinary and additional leave each last 26 weeks from their respective start dates, or 52 weeks in total.\par
\textsuperscript{b}\,Measured at the start of the 11th week before the EWC (until 2003) and at the start of the 14th week before the EWC (from 2003). From 2007, extended leave requires no minimum tenure, so that the full 52 weeks are available to all pregnant employees.\par
\textsuperscript{c}\,Rates are uprated every April and apply by week of payment rather than by the EWC cutoff; the April 2003 increase far exceeded routine upratings. Until April 2003 the flat rate was a floor (women whose earnings-related rate fell below it received the flat rate throughout); thereafter it was a cap (flat-rate weeks were paid at the lower of the flat rate and 90\% of average earnings). From October 1994, MA was paid at two rates: a higher rate equal to the SMP flat rate (\pounds 52.50 in 1994) for women employed in the 15th week before the EWC, and a lower rate (\pounds 44.55 in 1994) for the self-employed and for women who had left employment. From August 2000, women with average earnings at or above the LEL received a standard rate equal to the SMP flat rate, and those earning between the MA threshold and the LEL received 90\% of their average earnings up to that rate; from April 2003, all recipients received the lower of the standard rate and 90\% of average earnings.{}\par
\textsuperscript{d}\,Between 30 April and 20 August 2000, the contribution-based test continued to apply; the employment and earnings test applied to expected weeks of childbirth beginning on or after 20 August 2000.\par
\textsuperscript{e}\,Introduced for children born on or after 6 April 2003 or with an expected date of birth on or after that date; requires the father to have 26 weeks of continuous employment ending with the 15th week before the EWC.\par
\textit{Sources:} \textcite{gregg_employment_2007,lourie_family_1999}; CPAG Welfare Benefits and Tax Credits Handbooks (2001/02--2007/08); SI 1999/3312; SI 2002/2788; SI 2002/2789; SI 2006/2014; SI 2006/2379.\par
}
\end{minipage}

    \end{adjustbox}
\end{center}
\end{table}

\end{landscape}
\restoregeometry

\FloatBarrier
\subsection{Donut RD estimates}
\label{sec:donutRD}

\begin{figure}[!htbp]
    \centering
    \includegraphics[width=0.65\textwidth]{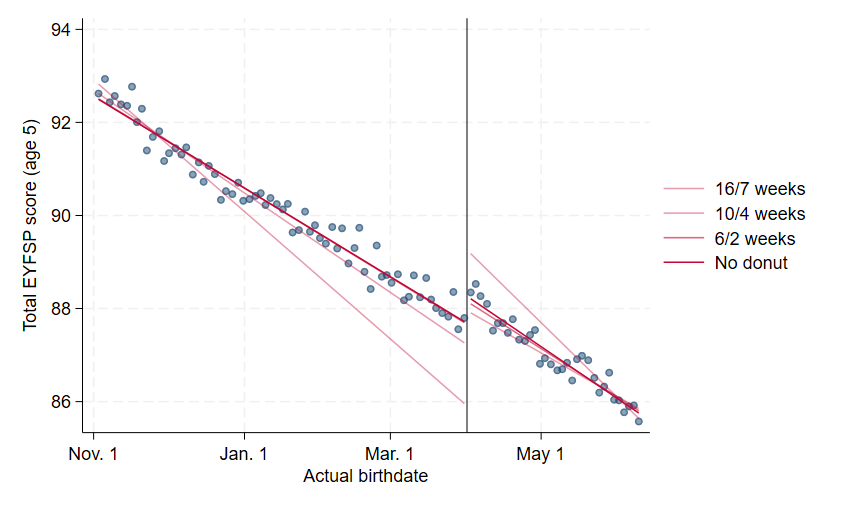}

    \refstepcounter{figure}
    \label{fig:donutRD}

    \vspace{0.35cm}
    \begin{small}
        \textbf{Figure~\thefigure: Donut RD estimates for the 2007 pay reform}
    \end{small}

    \vspace{0.3cm}
    \begin{minipage}{0.9\textwidth}
        \begin{footnotesize}
            \begin{spacing}{1.15}
\textit{Notes:} The figure displays standard RD plots under alternative choices of donut width---that is, the window of discarded observations around the reform introduction date (marked by the vertical line). The y-axis reports the average total EYFSP score at age 5 and the x-axis the actual date of birth; dots are binned averages of test scores by actual date of birth. As indicated in the legend, the different shades of the fitted lines correspond to linear regressions that exclude, respectively, 16/7, 10/4, and 6/2 weeks of births before/after the reform date; the darkest line (``No donut'') is the linear fit with no exclusion.
            \end{spacing}
        \end{footnotesize}
    \end{minipage}
\end{figure}

Figure~\ref{fig:donutRD} reports (donut) linear RD estimates for the 2007 reform. A conservative and well-justified specification would exclude all observations with dates of birth from the point at which $p^t(x)$ becomes positive, around 1 December (see Figure~\ref{fig:diagnosis}), to 7 weeks after the reform date, ensuring correct treatment classification; the widest window in the figure, 16 weeks before and 7 weeks after the cutoff, comes close to this. Specifications with narrower exclusion windows could be justified by the argument that the extreme tails of the measurement-error distribution (e.g., very premature births) are sufficiently rare, and survival among them sufficiently low, that the resulting misclassification would be negligible---for example, excluding 10 weeks before and 4 weeks after the cutoff, or, at the limit of this argument, 6 weeks before and 2 weeks after.

\FloatBarrier
\subsection{Parallel-trends diagnostics}
\label{sec:paralleltrendstest}

\begin{figure}[!htbp]
    \centering
    \begin{adjustbox}{max width=\textwidth}
    \begin{tabular}{c c}
        \multicolumn{2}{c}{\textbf{2000 job-protection reform}} \\
        \includegraphics[width=0.36\textwidth]{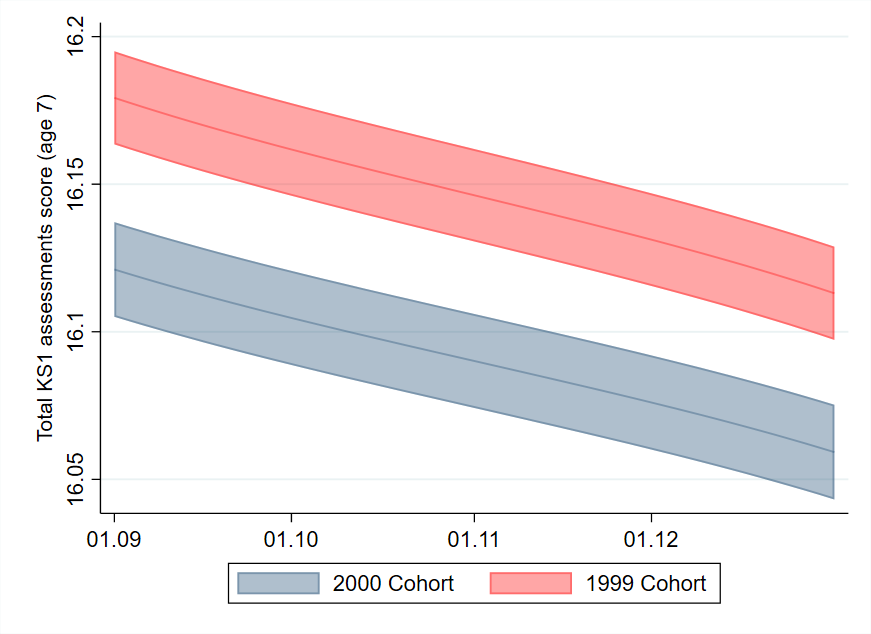} &
        \includegraphics[width=0.36\textwidth]{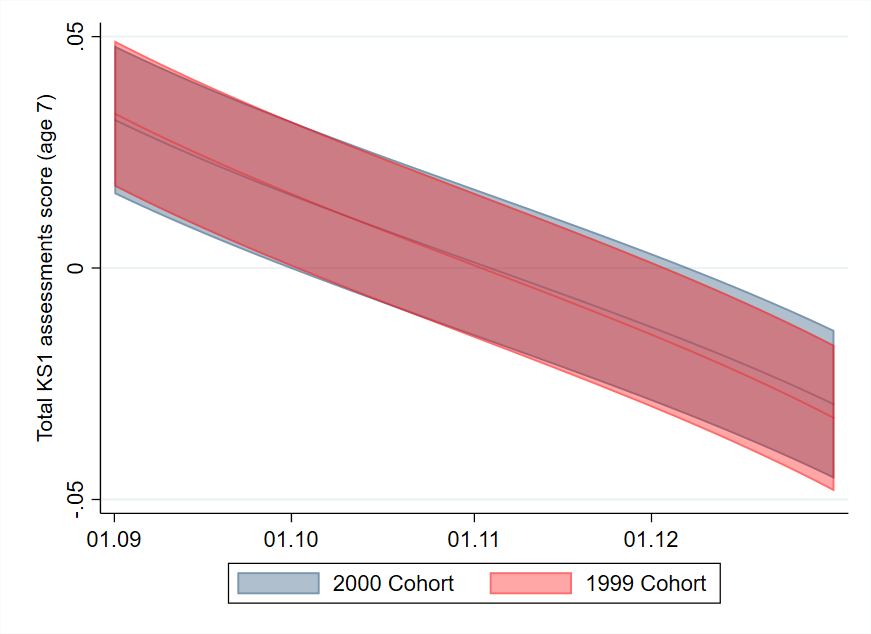} \\
        {\small Panel A. Raw outcomes} & {\small Panel B. Demeaned outcomes} \\[0.3cm]

        \multicolumn{2}{c}{\textbf{2003 bundle reform}} \\
        \includegraphics[width=0.36\textwidth]{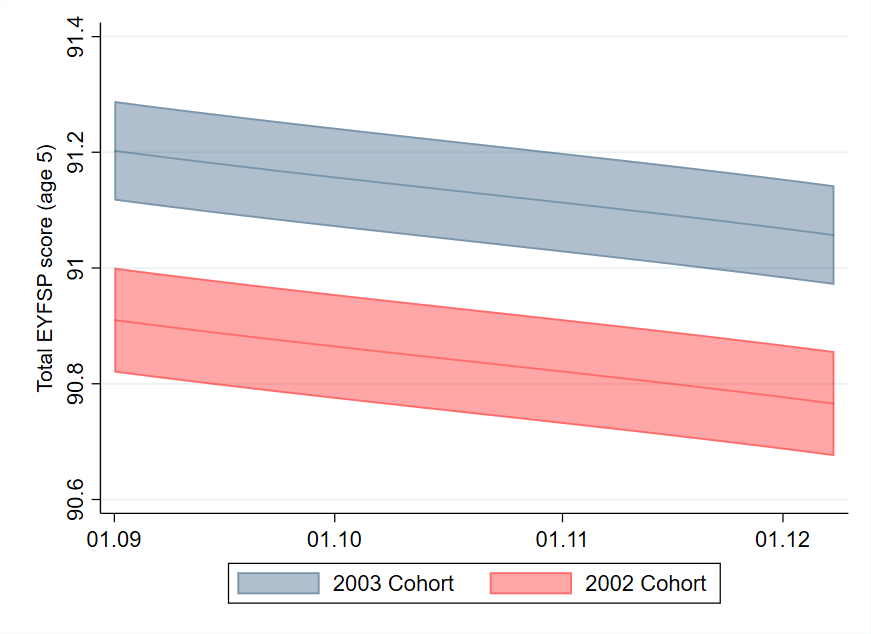} &
        \includegraphics[width=0.36\textwidth]{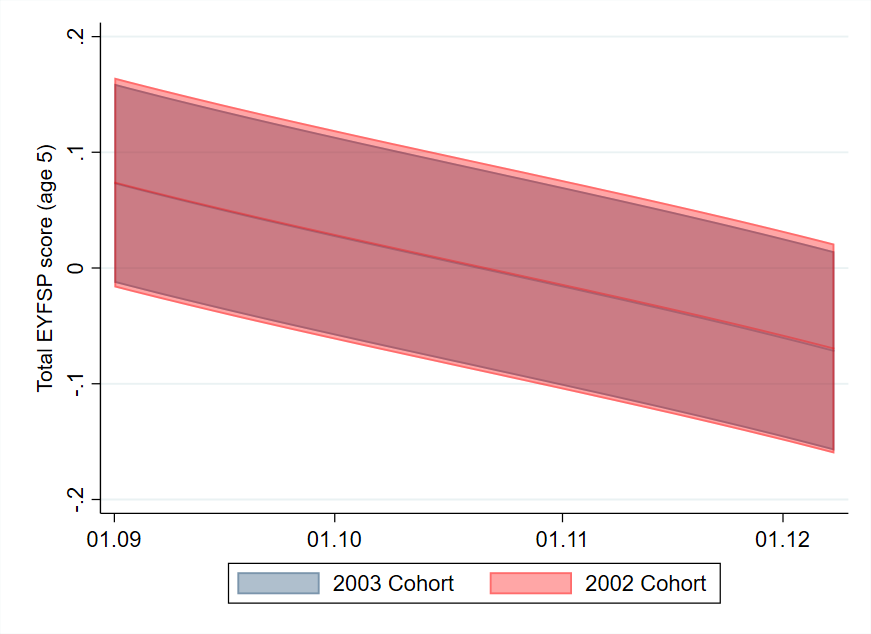} \\
        {\small Panel C. Raw outcomes} & {\small Panel D. Demeaned outcomes} \\[0.3cm]

        \multicolumn{2}{c}{\textbf{2007 pay reform}} \\
        \includegraphics[width=0.36\textwidth]{figures/parallel_trend_test_reform_2007_eyfsp_total_eyfsp_score.png} &
        \includegraphics[width=0.36\textwidth]{figures/parallel_trend_test_reform_2007_eyfsp_total_eyfsp_score_demeaned.png} \\
        {\small Panel E. Raw outcomes} & {\small Panel F. Demeaned outcomes} \\
    \end{tabular}
    \end{adjustbox}

    \refstepcounter{figure}
    \label{fig:paralleltrends_main}

    \vspace{0.35cm}
    \begin{small}
        \textbf{Figure~\thefigure: Parallel-trends diagnostic for the main outcomes}
    \end{small}

    \vspace{0.3cm}
    \begin{minipage}{0.9\textwidth}
        \begin{footnotesize}
            \begin{spacing}{1.15}
                \textit{Notes:} This figure presents the graphical parallel-trends diagnostic of Remark~\ref{rem:over_id} for the main outcomes. In each panel, the later cohort (e.g., the 2007 Cohort) is the treated reform cohort $t$ and the earlier cohort (e.g., the 2006 Cohort) is the untreated control cohort $t-1$. The first column shows raw test scores over the zero-treatment region, while the second column presents each cohort's outcomes demeaned by its own equal-date mean over that region; under the parallel-trends restriction, the two demeaned profiles should coincide up to sampling variation. The earliest available test scores are used: age 7 for the 2000 reform (KS1), and age 5 for the 2003 and 2007 reforms (EYFSP). The series are plotted over the pre-treatment window with $p^t(x)=0$, from 1 September to the last date shown on the horizontal axis; shaded areas are 95\% confidence bands. Horizontal-axis tick labels give the date of birth in day.month format. {}
            \end{spacing}
        \end{footnotesize}
    \end{minipage}
\end{figure}

\begin{figure}[!htbp]
    \centering
    \begin{adjustbox}{max width=\textwidth}
    \begin{tabular}{c c}
        \includegraphics[width=0.45\textwidth]{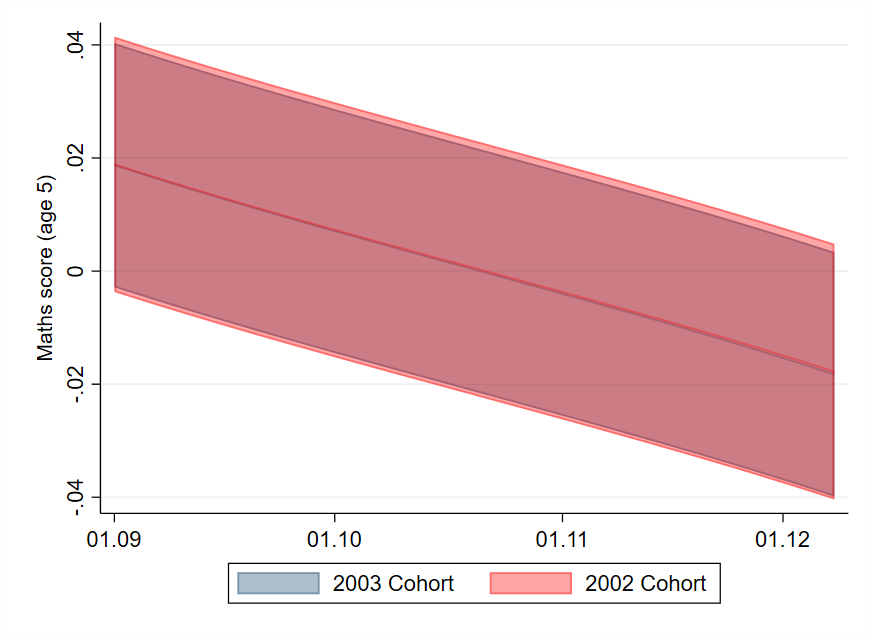} &
        \includegraphics[width=0.45\textwidth]{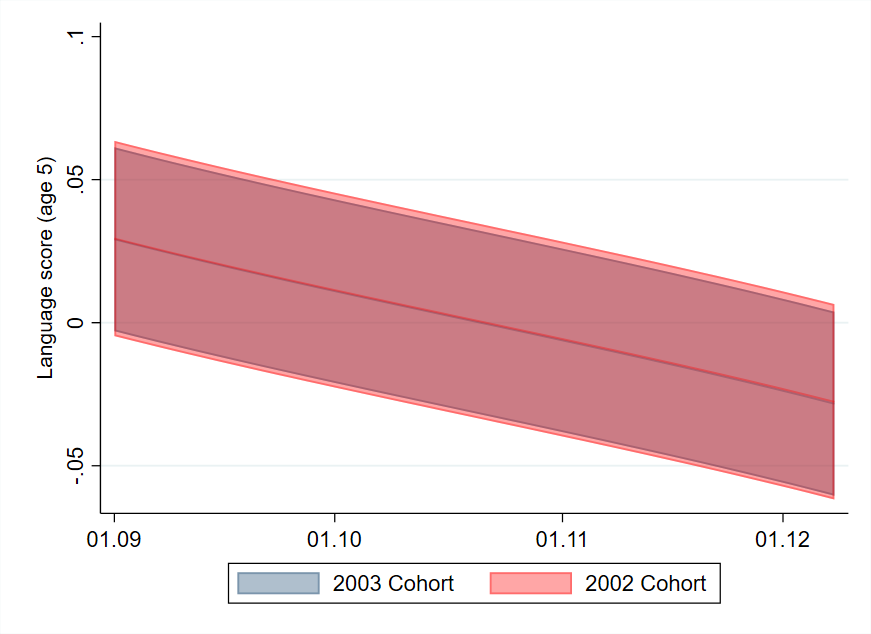} \\
        {\small Panel A. Maths score} & {\small Panel B. Language score} \\
        \includegraphics[width=0.45\textwidth]{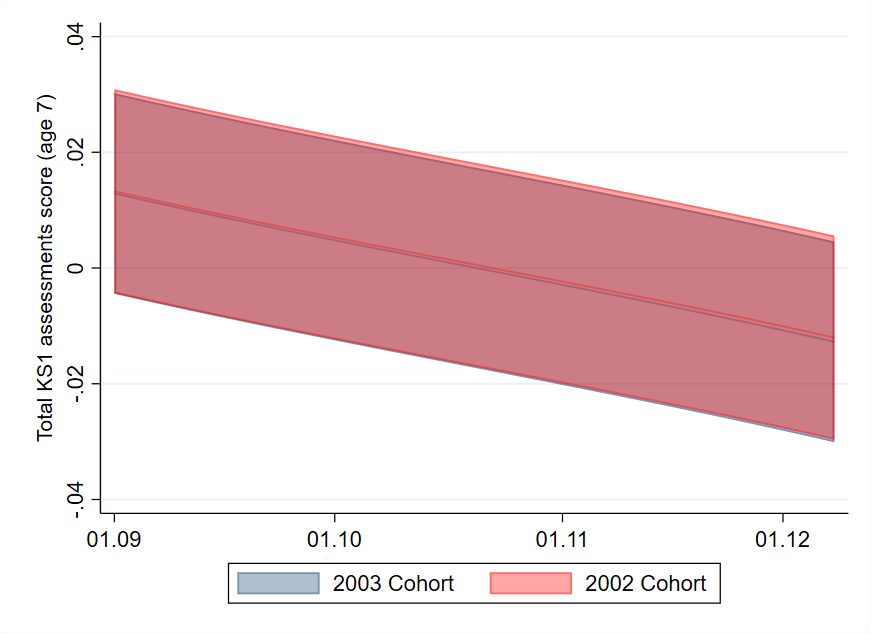} &
        \includegraphics[width=0.45\textwidth]{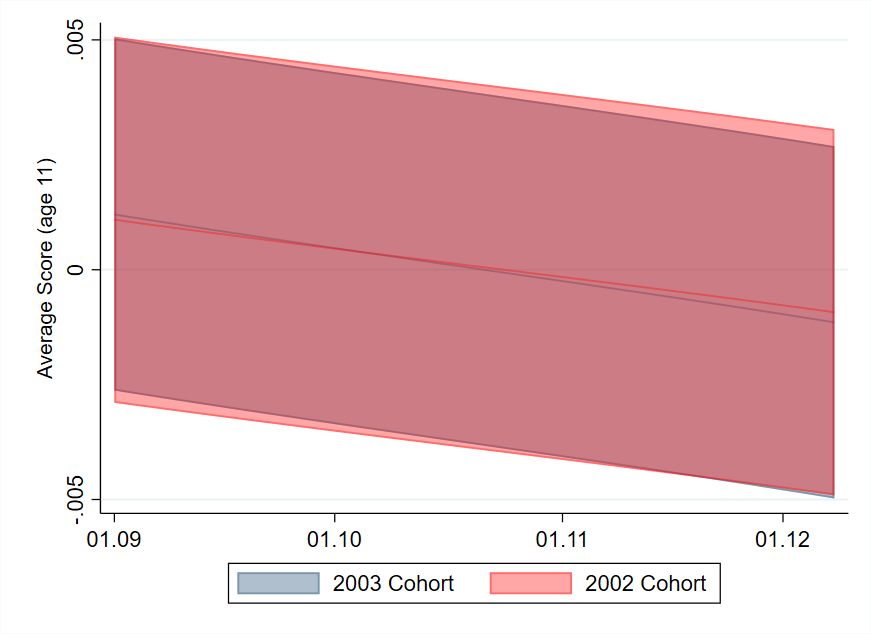} \\
        {\small Panel C. Age 7 score} & {\small Panel D. Average score (age 11)} \\
    \end{tabular}
    \end{adjustbox}

    \refstepcounter{figure}
    \label{fig:paralleltrends_secondary_2003}

    \vspace{0.35cm}
    \begin{small}
        \parbox{0.85\textwidth}{\centering \textbf{Figure~\thefigure: Parallel-trends diagnostic for the secondary outcomes of the 2003 bundle reform}}
    \end{small}

    \vspace{0.3cm}
    \begin{minipage}{0.9\textwidth}
        \begin{footnotesize}
            \begin{spacing}{1.15}
                \textit{Notes:} This figure presents the parallel-trends diagnostic for secondary outcomes of the 2003 reform. The first row presents the maths and language components of the age-5 EYFSP score; the second row presents the total KS1 score at age 7 and the average KS2 score at age 11. All four panels show separately demeaned outcome profiles, as in the second column of Figure~\ref{fig:paralleltrends_main}. Outcomes are in raw assessment points and are not comparable across panels; horizontal-axis tick labels give the date of birth in day.month format.
            \end{spacing}
        \end{footnotesize}
    \end{minipage}
\end{figure}

\begin{figure}[!htbp]
    \centering
    \begin{adjustbox}{max width=\textwidth}
    \begin{tabular}{c c}
        \includegraphics[width=0.45\textwidth]{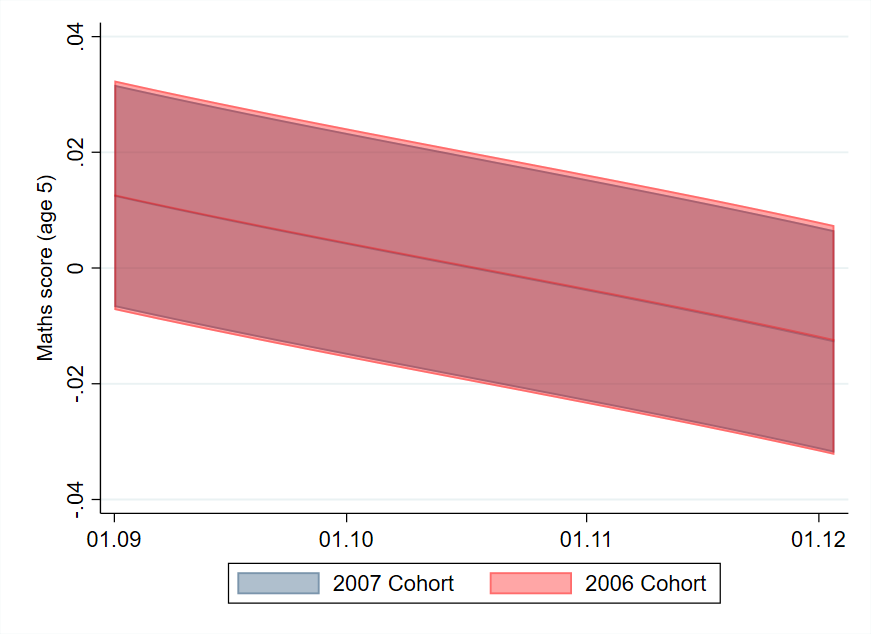} &
        \includegraphics[width=0.45\textwidth]{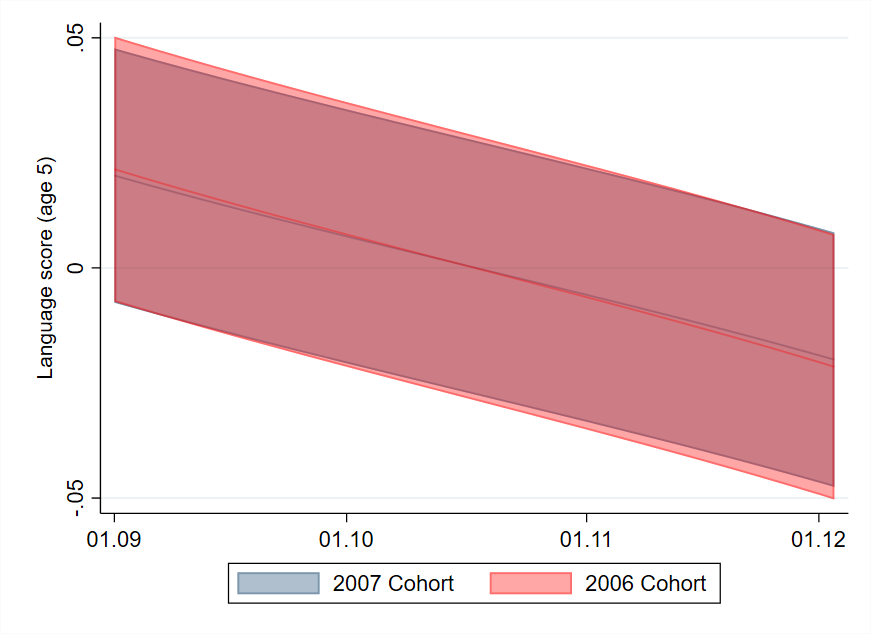} \\
        {\small Panel A. Maths score} & {\small Panel B. Language score} \\
        \includegraphics[width=0.45\textwidth]{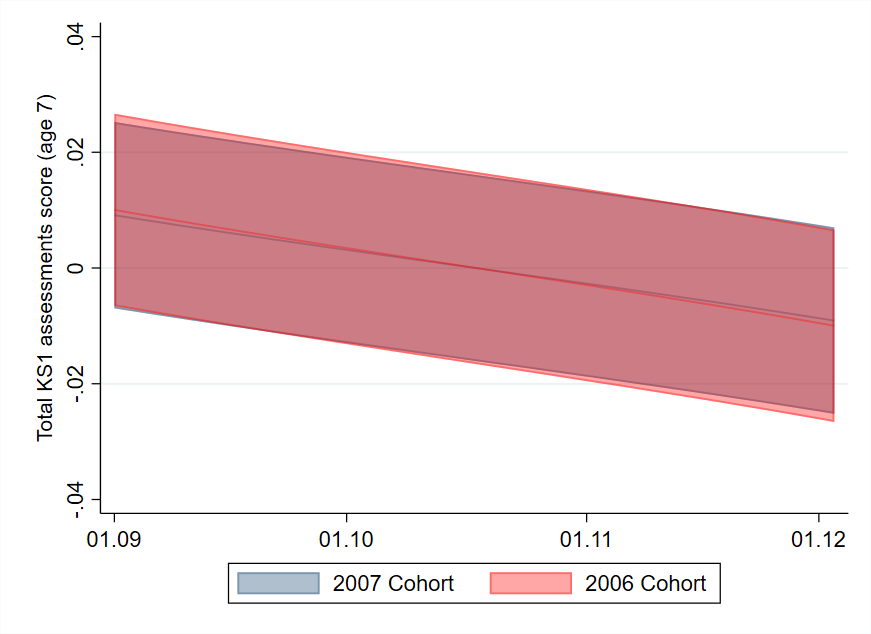} &
        \includegraphics[width=0.45\textwidth]{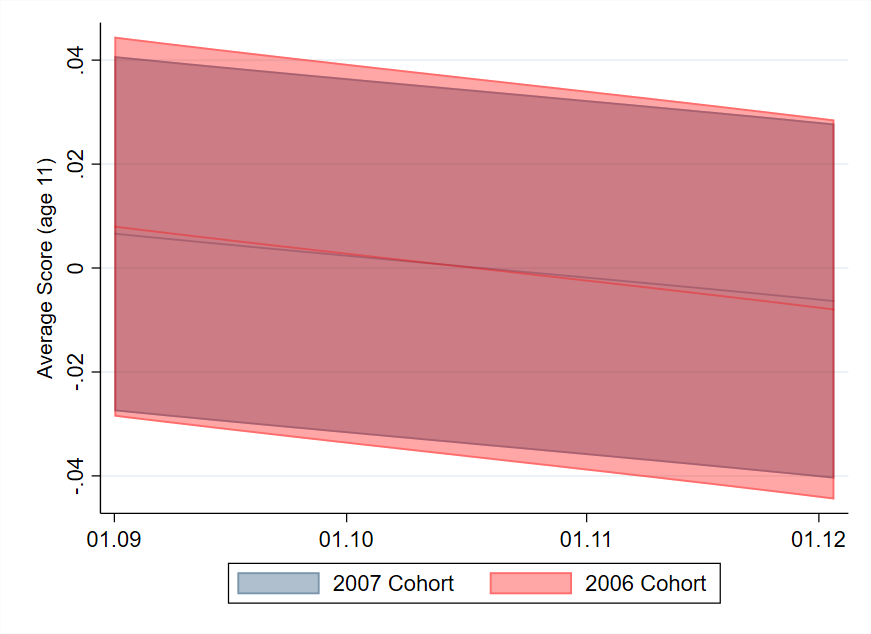} \\
        {\small Panel C. Age 7 score} & {\small Panel D. Average score (age 11)} \\
    \end{tabular}
    \end{adjustbox}

    \refstepcounter{figure}
    \label{fig:paralleltrends_secondary_2007}

    \vspace{0.35cm}
    \begin{small}
        \parbox{0.85\textwidth}{\centering \textbf{Figure~\thefigure: Parallel-trends diagnostic for the secondary outcomes of the 2007 pay reform}}
    \end{small}

    \vspace{0.3cm}
    \begin{minipage}{0.9\textwidth}
        \begin{footnotesize}
            \begin{spacing}{1.15}
                \textit{Notes:} This figure presents the parallel-trends diagnostic for secondary outcomes of the 2007 reform. The first row presents the maths and language components of the age-5 EYFSP score; the second row presents the total KS1 score at age 7 and the average KS2 score at age 11. All four panels show separately demeaned outcome profiles, as in the second column of Figure~\ref{fig:paralleltrends_main}. Outcomes are in raw assessment points and are not comparable across panels; horizontal-axis tick labels give the date of birth in day.month format.
            \end{spacing}
        \end{footnotesize}
    \end{minipage}
\end{figure}

Figures~\ref{fig:paralleltrends_main}--\ref{fig:paralleltrends_secondary_2007} report the graphical parallel-trends diagnostic of Remark~\ref{rem:over_id}. Figure~\ref{fig:paralleltrends_main} covers the main outcome of each reform, and Figures~\ref{fig:paralleltrends_secondary_2003} and \ref{fig:paralleltrends_secondary_2007} the secondary outcomes of the 2003 and 2007 reforms.

\FloatBarrier
\Needspace{6\baselineskip}
\subsection{Standard errors}
\label{sec:standarderrors}

This section derives standard errors for the estimators in Section~\ref{ssec:implementation}. In addition to variation in outcomes, inference must account for the estimated treatment probabilities and, for the nonparametric estimates, the common reform-cohort shift $\widehat\alpha^t$. These common estimates induce dependence across birth dates, which matters when aggregating daily effects.

We consider one reform and outcome at a time and assume that the outcome and validation samples are independent. For children's outcomes, we treat the conversion to days of age at testing as fixed and read $Y_i^\tau$ below as the normalised outcome. No sampling uncertainty is attributed to the conversion factor.

\subsubsection{Estimated treatment probabilities}

Under the restrictions in \ref{sec:phat}, $p^t(x)=F_\varepsilon(x-x_0)$, and $\widehat p^t(x)$ is estimated by the empirical CDF in \eqref{eq:phat_ecdf}. The same validation observations enter this estimate at every birth date. For $n_v$ independent and identically distributed validation observations, the covariance across dates is therefore
\begin{align}
    \Gamma(x,x')
    \coloneqq \operatorname{Cov}\!\left(\widehat p^t(x),\widehat p^t(x')\right)
    = \frac{\min\{p^t(x),p^t(x')\}-p^t(x)p^t(x')}{n_v}.
    \label{eq:validation_covariance}
\end{align}
This follows from the nested indicators in the empirical CDF. We estimate $\Gamma(x,x')$ by replacing $p^t$ with $\widehat p^t$ and denote the result by $\widehat\Gamma(x,x')$. If the validation estimates use survey weights or clustering, the corresponding covariance estimator replaces this expression.

\subsubsection{Daily and aggregated CATT estimates}

As in Section~\ref{ssec:implementation}, $\widehat\mu^\tau(x)$ is the sample mean for children born on date $x$ in cohort $\tau\in\{t-1,t\}$. Write the daily reform--control difference as
\[
    \widehat g(x)=\widehat\mu^t(x)-\widehat\mu^{t-1}(x).
\]
With independent observations within and across cohort--date cells, its estimated variance, conditional on cell counts, is
\[
    \widehat v(x)=\frac{s_t^2(x)}{n_t(x)}+\frac{s_{t-1}^2(x)}{n_{t-1}(x)},
\]
where $n_\tau(x)$ and $s_\tau^2(x)$ are the cell sample size and sample variance.

Let $\mathcal D_0$ denote the fixed set of dates in the zero-treatment region, where we maintain $p^t(x)=0$. The reform-cohort shift is estimated by averaging the daily differences over this region:
\begin{align}
    \widehat\alpha^t
    &=\sum_{x\in\mathcal D_0}\omega_x\widehat g(x),
    &
    \widehat V_\alpha
    &=\sum_{x\in\mathcal D_0}\omega_x^2\widehat v(x).
    \label{eq:alpha_estimator_se}
\end{align}
Here, the fixed, nonnegative weights sum to one, and $\widehat V_\alpha$ estimates the variance of $\widehat\alpha^t$. Our implementation weights each date equally, so $\omega_x=1/|\mathcal D_0|$.

The daily CATT estimator in \eqref{eq:np_estimator} subtracts this common shift from $\widehat g(x)$ and divides by $\widehat p^t(x)$. For dates $x,x'$ with positive treatment probabilities, the delta method gives the covariance estimator
\begin{align}
    \widehat\Sigma(x,x')
    =
    \frac{
        \mathbf 1\{x=x'\}\widehat v(x)
        +\widehat V_\alpha
        +\widehat{\mathrm{CATT}}^t(x)\widehat{\mathrm{CATT}}^t(x')
         \widehat\Gamma(x,x')
    }{\widehat p^t(x)\widehat p^t(x')}.
    \label{eq:daily_catt_covariance_estimator}
\end{align}
The three terms in the numerator capture daily outcome variation, uncertainty in the common cohort shift, and uncertainty in the treatment probabilities. The last two generate covariance even between estimates for different birth dates. The standard error at date $x$ is $\sqrt{\widehat\Sigma(x,x)}$.\footnote{These first-order approximations use fixed date support, with outcome-cell shares bounded away from zero and positive treatment probabilities bounded away from zero at the dates used for CATT estimation, as both sample sizes increase. If outcome sampling induces dependence across cells, replace $\mathbf 1\{x=x'\}\widehat v(x)+\widehat V_\alpha$ in \eqref{eq:daily_catt_covariance_estimator} by an estimate of $\operatorname{Cov}(\widehat g(x)-\widehat\alpha^t,\widehat g(x')-\widehat\alpha^t)$ computed under the relevant sampling design.}

For a weekly, monthly, or full-period average over dates $\mathcal D$, let $w_x$ be fixed, nonnegative weights summing to one. Its standard error is
\begin{align}
    \operatorname{se}\!\left(
        \sum_{x\in\mathcal D}w_x\widehat{\mathrm{CATT}}^t(x)
    \right)
    =
    \left[
        \sum_{x\in\mathcal D}\sum_{x'\in\mathcal D}
        w_xw_{x'}\widehat\Sigma(x,x')
    \right]^{1/2}.
    \label{eq:aggregate_catt_se}
\end{align}
We use equal-date weights, $w_x=1/|\mathcal D|$. Retaining the covariances in this expression accounts for the common estimation errors; averaging daily variances alone would omit them. %

\subsubsection{Placebo estimates}

Within the zero-treatment region, the placebo estimate is
\[
    \widehat\Delta(x)=\widehat g(x)-\widehat\alpha^t,
    \qquad x\in\mathcal D_0.
\]
Its population counterpart is zero under parallel trends. Unlike the CATT estimates, these placebos use observations that also enter $\widehat\alpha^t$, so their covariance with the estimated shift must be retained.

For a placebo window, let $w_x$ sum to one within that window and be zero elsewhere in $\mathcal D_0$. Since
\[
    \sum_{x\in\mathcal D_0}w_x\widehat\Delta(x)
    =\sum_{x\in\mathcal D_0}(w_x-\omega_x)\widehat g(x),
\]
its standard error under independent date cells is
\begin{align}
    \operatorname{se}\!\left(
        \sum_{x\in\mathcal D_0}w_x\widehat\Delta(x)
    \right)
    =
    \left[
        \sum_{x\in\mathcal D_0}(w_x-\omega_x)^2\widehat v(x)
    \right]^{1/2}.
    \label{eq:placebo_aggregate_se}
\end{align}
With dependence across cells, the variance inside the square root becomes
$\sum_{x,x'\in\mathcal D_0}(w_x-\omega_x)(w_{x'}-\omega_{x'})
\widehat{\operatorname{Cov}}(\widehat g(x),\widehat g(x'))$.
For a single date, the independent-cell variance is $\widehat v(x)+\widehat V_\alpha-2\omega_x\widehat v(x)$. The placebo average over the entire zero-treatment region using weights $\omega_x$ is identically zero by construction; the informative comparisons are therefore across dates or smaller windows.

\subsubsection{Regression estimates for mothers' outcomes}

For the regression in \eqref{eq:semiparametric_regression_mothers}, index the $n$ observations in the pooled outcome sample by $i$ and let $\tau_i\in\{t-1,t\}$ denote the cohort. Collect the regressors and coefficients as
\begin{align*}
    Z_i
    &=\left(p^{\tau_i}(X_i^{\tau_i}),\,1,\,
        \mathbf 1\{\tau_i=t\},\,X_i^{\tau_i},\,W_i^{\tau_i\prime}\right)',\\
    \vartheta
    &=\left(\beta,\,\gamma_0,\,\alpha^t,\,\lambda,\,\delta'\right)'.
\end{align*}
Writing $Y_i=Y_i^{\tau_i}$, the population regression is $Y_i=Z_i'\vartheta+u_i$, with $\mathbb E[u_i\mid\tau_i,X_i^{\tau_i},W_i^{\tau_i}]=0$. The feasible regressor vector $\widehat Z_i$ replaces $p^{\tau_i}$ with $\widehat p^{\tau_i}$; both probabilities are zero for the control cohort.

Using the estimated probability changes the regression error, evaluated at the true coefficients, to
\[
    u_i-\beta\left[\widehat p^{\tau_i}(X_i^{\tau_i})-p^{\tau_i}(X_i^{\tau_i})\right].
\]
The additional error is shared by reform-cohort observations born on the same date, and is correlated across dates through $\Gamma(x,x')$. Its contribution to the coefficient variance therefore depends on the sum of the regressors at each date.

Let $\widehat u_i=Y_i-\widehat Z_i'\widehat\vartheta$ and define
\begin{align*}
    \widehat Q
    &=\frac{1}{n}\sum_{i=1}^n\widehat Z_i\widehat Z_i',
    &
    \widehat\Omega_u
    &=\frac{1}{n}\sum_{i=1}^n
        \widehat Z_i\widehat Z_i'\widehat u_i^2,\\
    \widehat S_x
    &=\sum_{i:\,\tau_i=t,\,X_i^t=x}\widehat Z_i.
\end{align*}
A first-order covariance estimator is
\begin{align}
    \widehat{\operatorname{Var}}(\widehat\vartheta)
    =
    \frac{1}{n}\widehat Q^{-1}
    \left[
        \widehat\Omega_u
        +\frac{\widehat\beta^2}{n}
         \sum_x\sum_{x'}\widehat S_x\widehat S_{x'}'
         \widehat\Gamma(x,x')
    \right]\widehat Q^{-1}.
    \label{eq:regression_variance}
\end{align}
Here the sums run over dates in the reform cohort. This approximation uses fixed date support, the usual moment and rank conditions, and $n,n_v\to\infty$ with $n/n_v$ converging to a finite limit. The conditional-mean restriction and independence of the outcome and validation samples make the two variance contributions additive.

The first term is the usual heteroskedasticity-robust covariance contribution; the second accounts for estimation of the treatment probabilities. Coefficient standard errors are the square roots of the diagonal entries, with the first entry corresponding to $\beta$. Because the regression jointly estimates all coefficients, the usual term already accounts for estimation of $\alpha^t$: no separate cohort-shift correction is needed.

If the regression uses estimation weights, they must enter $\widehat Q$, $\widehat\Omega_u$, and $\widehat S_x$ consistently. Clustering or repeated observations on the same mother should be reflected in $\widehat\Omega_u$. The validation-sample correction remains additive.

\FloatBarrier
\subsection{Robustness}
\label{sec:robustness}

\begin{figure}[!htbp]
    \centering
    \begin{tabular}{cc}
        \includegraphics[width=0.45\textwidth]{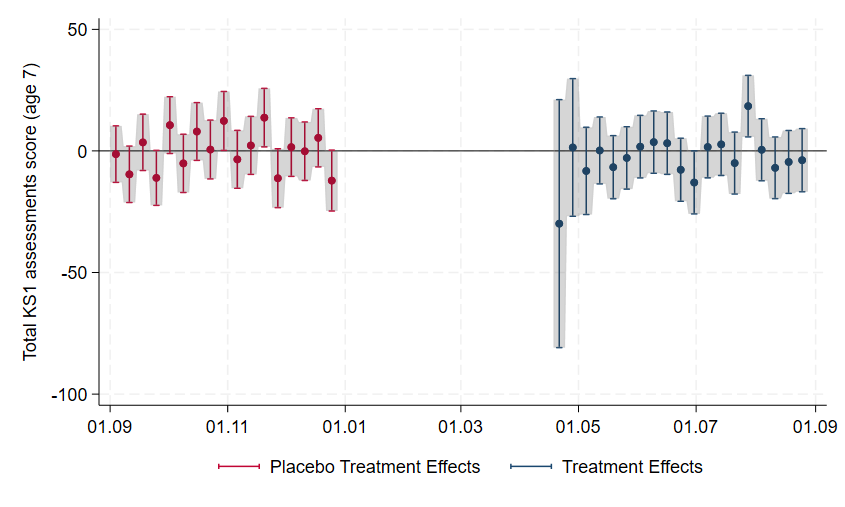} &
        \includegraphics[width=0.45\textwidth]{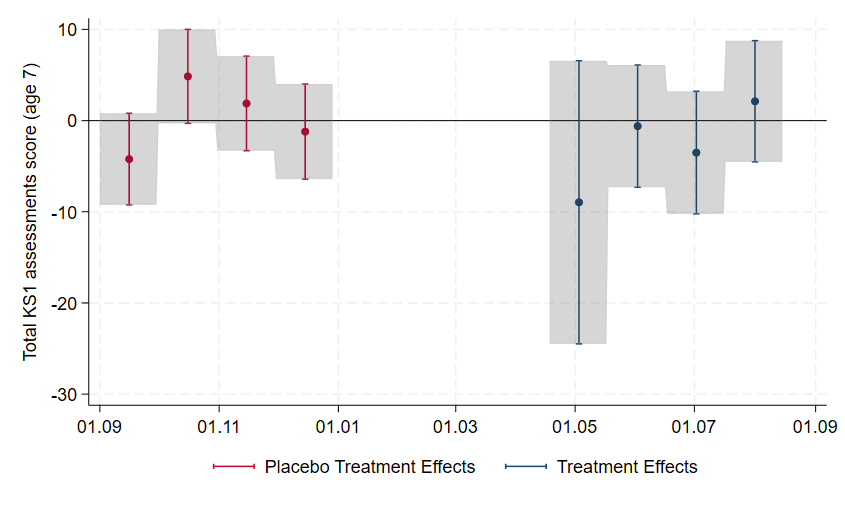} \\
        {\small Panel A. Week aggregation} & {\small Panel B. Month aggregation} \\
    \end{tabular}

    \refstepcounter{figure}  %
    \label{fig:cohortrobustness2000}  %

    \vspace{0.35cm}
    \begin{small}
        \parbox{0.85\textwidth}{\centering \textbf{Figure~\thefigure: Estimated effects of the 2000 job-protection reform \\ (alternative control cohort)}}
    \end{small}

    \vspace{0.12cm}
    \begin{minipage}{0.9\textwidth}
        \begin{footnotesize}
            \begin{spacing}{1.15}
                \textit{Notes:} The figure presents the effects of the 2000 job-protection reform, repeating the analysis of Section~\ref{sec:findings} but using the second preceding cohort, rather than the immediately preceding one, as the control cohort. The left panel aggregates at the weekly level, while the right panel aggregates at the monthly level. The outcome variable is the total score at age 7 (KS1). Placebo estimates for pre-reform birth dates with $p^t(x) = 0$, constructed as described in the notes to Figure~\ref{fig:fullsample2000}, are shown in red. Estimates are expressed in days of age (see the notes to Figure~\ref{fig:coefplot}); vertical lines and shaded areas indicate pointwise 95\% confidence intervals. Horizontal-axis tick labels give the date of birth in day.month format.
            \end{spacing}
        \end{footnotesize}
    \end{minipage}
\end{figure}

\begin{figure}[!htbp]
    \centering
    \begin{tabular}{cc}
        \includegraphics[width=0.45\textwidth]{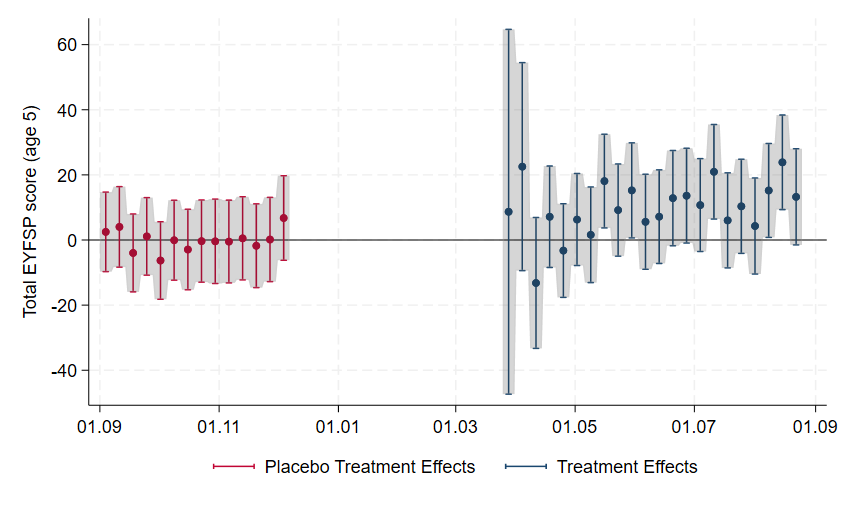} &
        \includegraphics[width=0.45\textwidth]{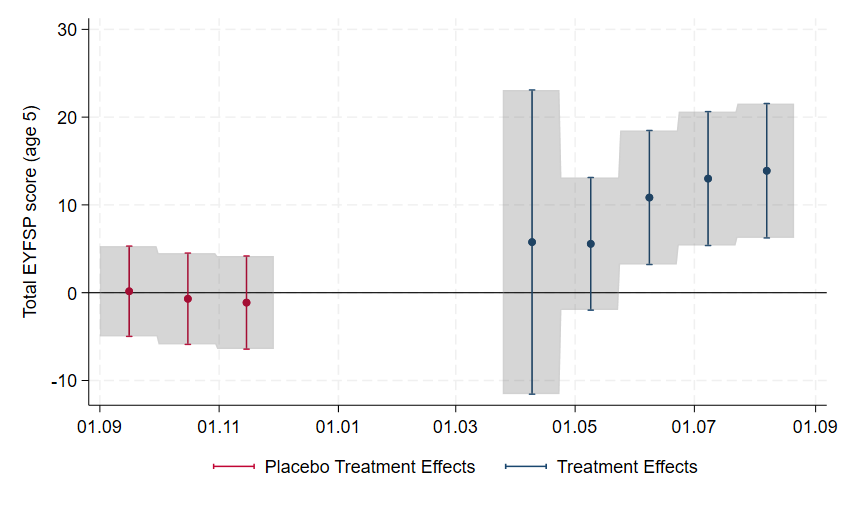} \\
        {\small Panel A. Week aggregation} & {\small Panel B. Month aggregation} \\
    \end{tabular}

    \refstepcounter{figure}  %
    \label{fig:cohortrobustness2003}  %

    \vspace{0.35cm}
    \begin{small}
        \parbox{0.85\textwidth}{\centering \textbf{Figure~\thefigure: Estimated effects of the 2003 bundle reform \\ (alternative control cohort)}}
    \end{small}

    \vspace{0.12cm}
    \begin{minipage}{0.9\textwidth}
        \begin{footnotesize}
            \begin{spacing}{1.15}
                \textit{Notes:} As Figure~\ref{fig:cohortrobustness2000}, for the 2003 bundle reform. The left panel aggregates at the weekly level and the right panel at the monthly level. The outcome variable is the total score at age 5 (EYFSP). Placebo estimates for pre-reform birth dates with $p^t(x) = 0$, constructed as described in the notes to Figure~\ref{fig:fullsample2000}, are shown in red. Estimates are expressed in days of age (see the notes to Figure~\ref{fig:coefplot}); vertical lines and shaded areas indicate pointwise 95\% confidence intervals. Horizontal-axis tick labels give the date of birth in day.month format.
            \end{spacing}
        \end{footnotesize}
    \end{minipage}
\end{figure}

\begin{figure}[!htbp]
    \centering
    \begin{tabular}{cc}
        \includegraphics[width=0.45\textwidth]{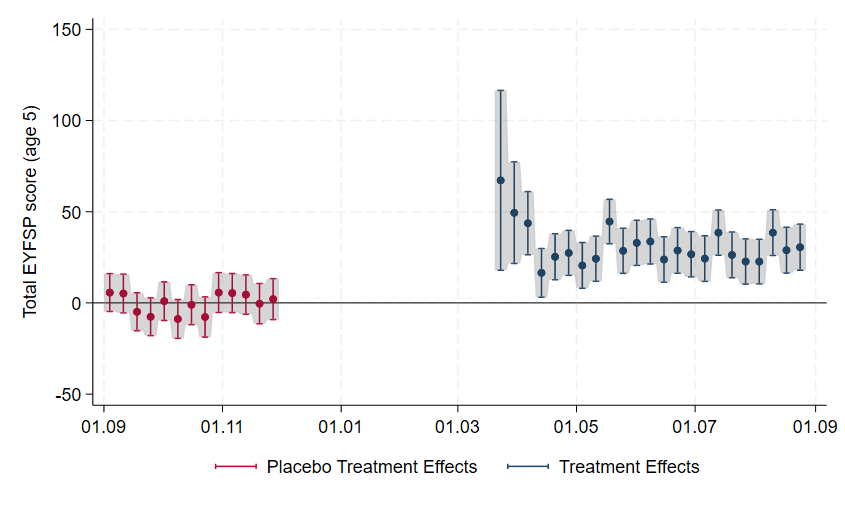}  &
        \includegraphics[width=0.45\textwidth]{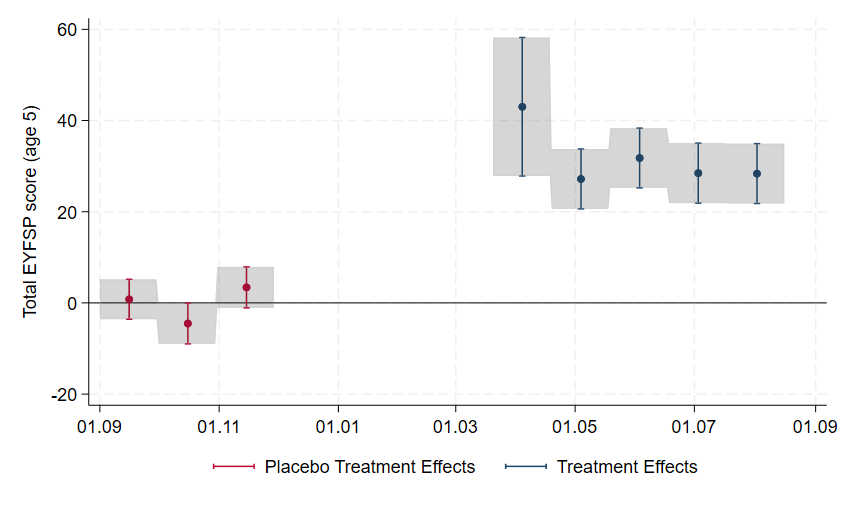} \\
        {\small Panel A. Week aggregation} & {\small Panel B. Month aggregation} \\
    \end{tabular}

    \refstepcounter{figure}
    \label{fig:cohortrobustness2007}

    \vspace{0.35cm}
    \begin{small}
        \parbox{0.85\textwidth}{\centering \textbf{Figure~\thefigure: Estimated effects of the 2007 pay reform \\ (alternative control cohort)}}
    \end{small}

    \vspace{0.12cm}
    \begin{minipage}{0.9\textwidth}
        \begin{footnotesize}
            \begin{spacing}{1.15}
                \textit{Notes:} As Figure~\ref{fig:cohortrobustness2000}, for the 2007 pay reform. The left panel aggregates at the weekly level and the right panel at the monthly level. The outcome variable is the total score at age 5 (EYFSP). Placebo estimates for pre-reform birth dates with $p^t(x) = 0$, constructed as described in the notes to Figure~\ref{fig:fullsample2000}, are shown in red. Estimates are expressed in days of age (see the notes to Figure~\ref{fig:coefplot}); vertical lines and shaded areas indicate pointwise 95\% confidence intervals. Horizontal-axis tick labels give the date of birth in day.month format.
            \end{spacing}
        \end{footnotesize}
    \end{minipage}
\end{figure}

\begin{figure}[!htbp]
    \centering
    \begin{tabular}{cc}
        \includegraphics[width=0.45\textwidth]{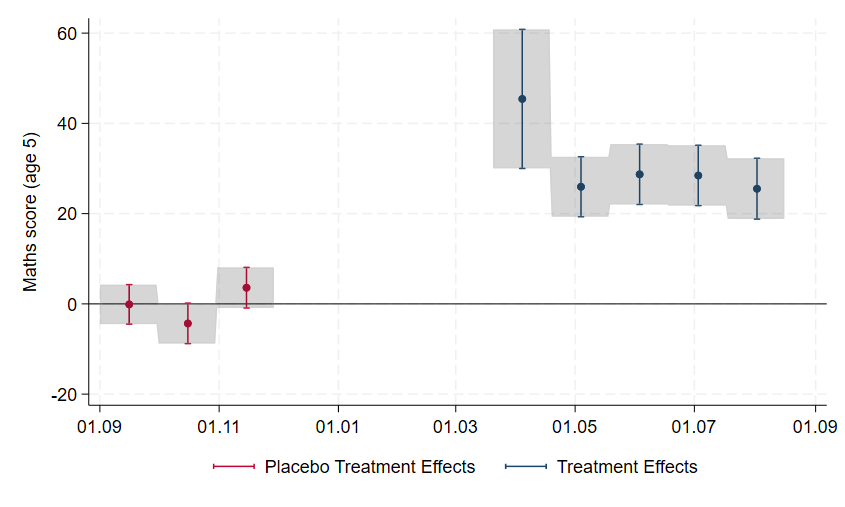}  &
        \includegraphics[width=0.45\textwidth]{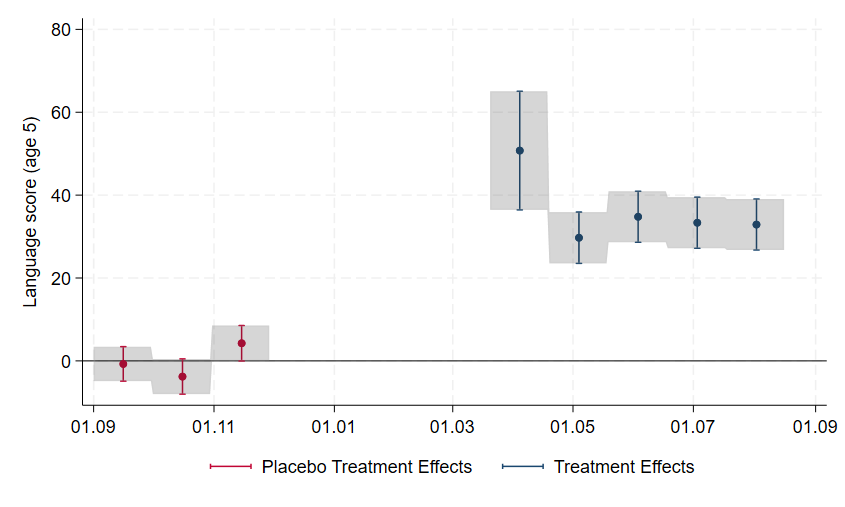} \\
        {\small Panel A. Maths score (age 5)} & {\small Panel B. Language score (age 5)} \\
        \includegraphics[width=0.45\textwidth]{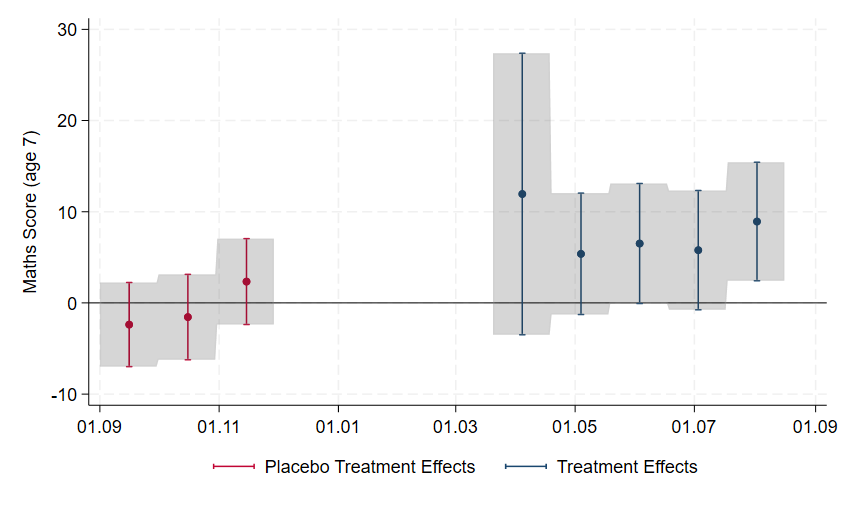}  &
        \includegraphics[width=0.45\textwidth]{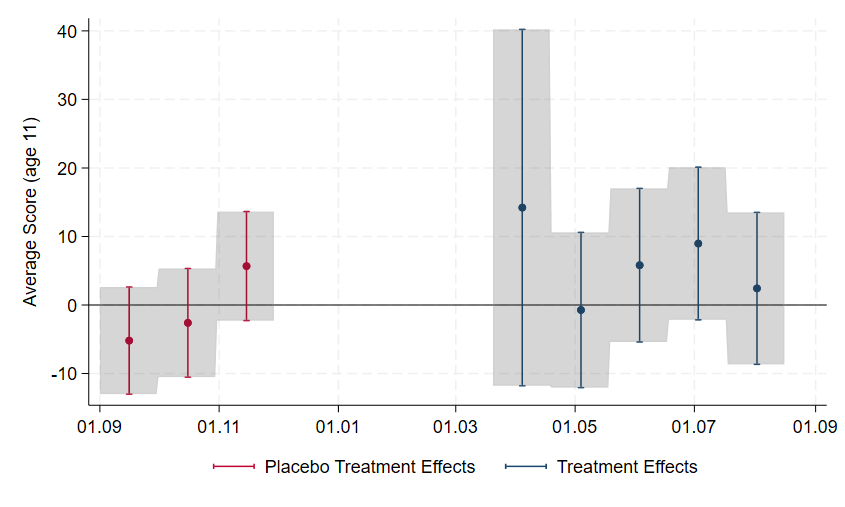} \\
        {\small Panel C. Maths score (age 7)} & {\small Panel D. Average score (age 11)} \\
    \end{tabular}

    \refstepcounter{figure}
    \label{fig:robustness2007robustness}

    \vspace{0.35cm}
    \begin{small}
        \parbox{0.85\textwidth}{\centering \textbf{Figure~\thefigure: Outcome and age robustness for the 2007 pay reform \\ (alternative control cohort)}}
    \end{small}

    \vspace{0.3cm}
    \begin{minipage}{0.9\textwidth}
        \begin{footnotesize}
            \begin{spacing}{1.15}
                \textit{Notes:} As Figure~\ref{fig:cohortrobustness2000}, for the 2007 pay reform. The first row reports the treatment effects separately for maths and language outcomes. The second row shows how the effects evolve at later ages (KS1 maths score at age 7 and KS2 average score at age 11). Estimates are aggregated at the monthly level and expressed in days of age (see the notes to Figure~\ref{fig:coefplot}); placebo estimates for pre-reform birth dates with $p^t(x) = 0$, constructed as described in the notes to Figure~\ref{fig:fullsample2000}, are shown in red. Vertical lines and shaded areas indicate pointwise 95\% confidence intervals. Horizontal-axis tick labels give the date of birth in day.month format.
            \end{spacing}
        \end{footnotesize}
    \end{minipage}
\end{figure}

Figures~\ref{fig:cohortrobustness2000}--\ref{fig:robustness2007robustness} repeat the analyses of Section~\ref{sec:findings} using the second preceding cohort, instead of the immediately preceding one, as the control cohort.

\FloatBarrier
\subsection{Additional results}
\label{sec:additional}

\begin{figure}[!htbp]
    \centering
    \begin{tabular}{cc}
        \includegraphics[width=0.45\textwidth]{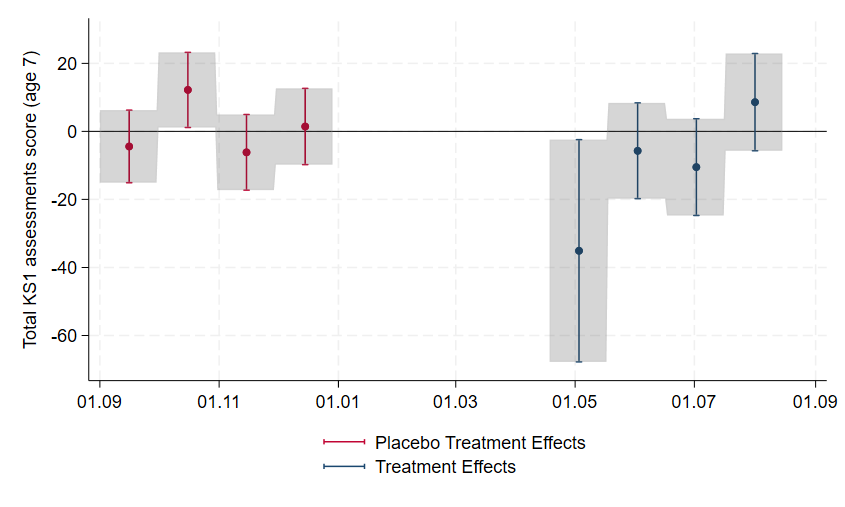} &
        \includegraphics[width=0.45\textwidth]{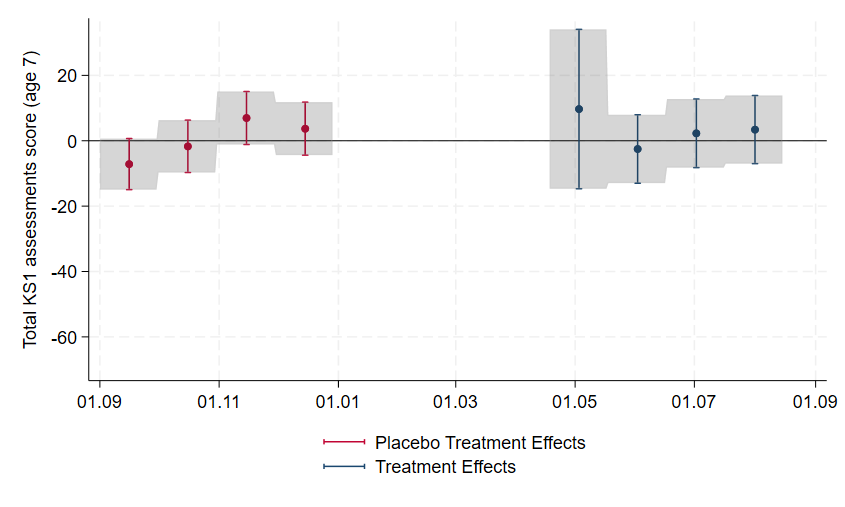}   \\
        {\small Panel A. Rural regions} & {\small Panel B. Urban regions} \\
        \includegraphics[width=0.45\textwidth]{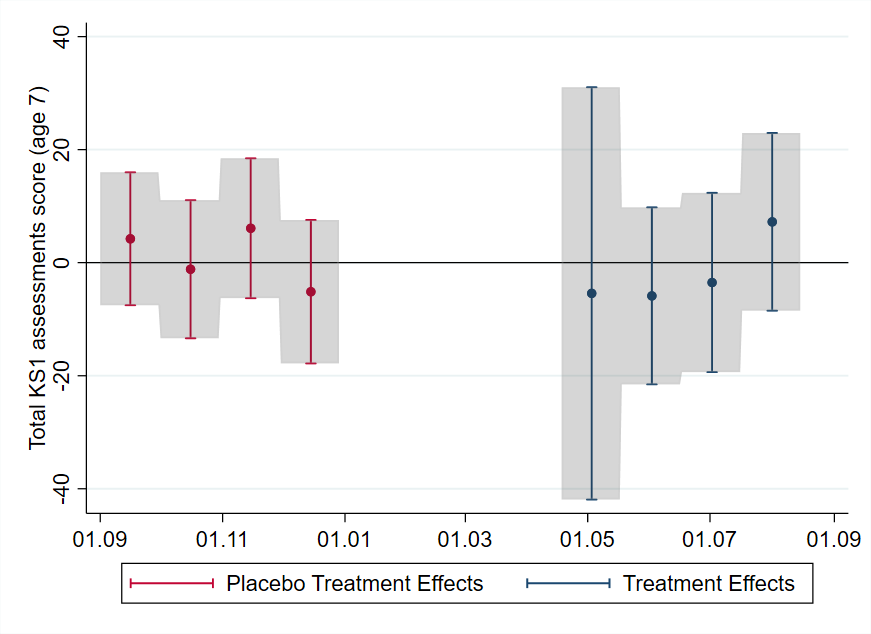}  &
        \includegraphics[width=0.45\textwidth]{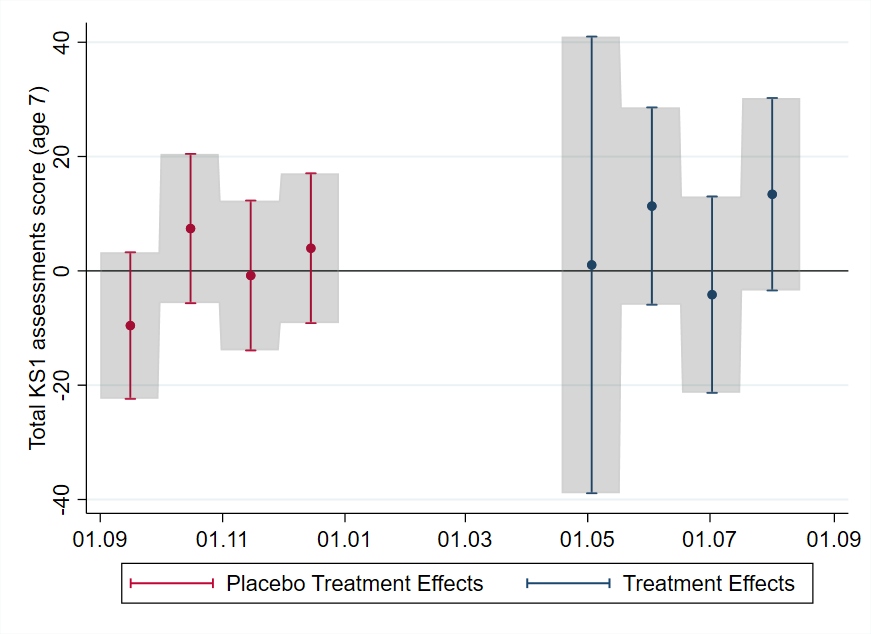} \\
        {\small Panel C. Bottom $20\%$ child poverty} & {\small Panel D. Top $20\%$ child poverty} \\
    \end{tabular}

    \refstepcounter{figure}
    \label{fig:geo2000}

    \vspace{0.35cm}
    \begin{small}
        \parbox{0.85\textwidth}{\centering \textbf{Figure~\thefigure: Geographic heterogeneity in the effects \\ of the 2000 job-protection reform}}
    \end{small}

    \vspace{0.3cm}
    \begin{minipage}{0.9\textwidth}
        \begin{footnotesize}
            \begin{spacing}{1.15}
                \textit{Notes:} This figure presents geographic heterogeneity in the effects of the 2000 job-protection reform. The top row compares rural and urban areas, while the bottom row contrasts areas with low and high child poverty, defined as the bottom and top quintiles of the IDACI distribution. The outcome variable is the total score at age 7 (KS1); estimates are conditional average treatment effects aggregated at the monthly level. Placebo estimates for pre-reform birth dates with $p^t(x) = 0$, constructed as described in the notes to Figure~\ref{fig:fullsample2000}, are shown in red. Estimates are expressed in days of age (see the notes to Figure~\ref{fig:coefplot}); vertical lines and shaded areas indicate pointwise 95\% confidence intervals. The vertical scales differ across panels. Horizontal-axis tick labels give the date of birth in day.month format.
            \end{spacing}
        \end{footnotesize}
    \end{minipage}
\end{figure}

\begin{figure}[!htbp]
    \centering
    \begin{tabular}{cc}
        \includegraphics[width=0.45\textwidth]{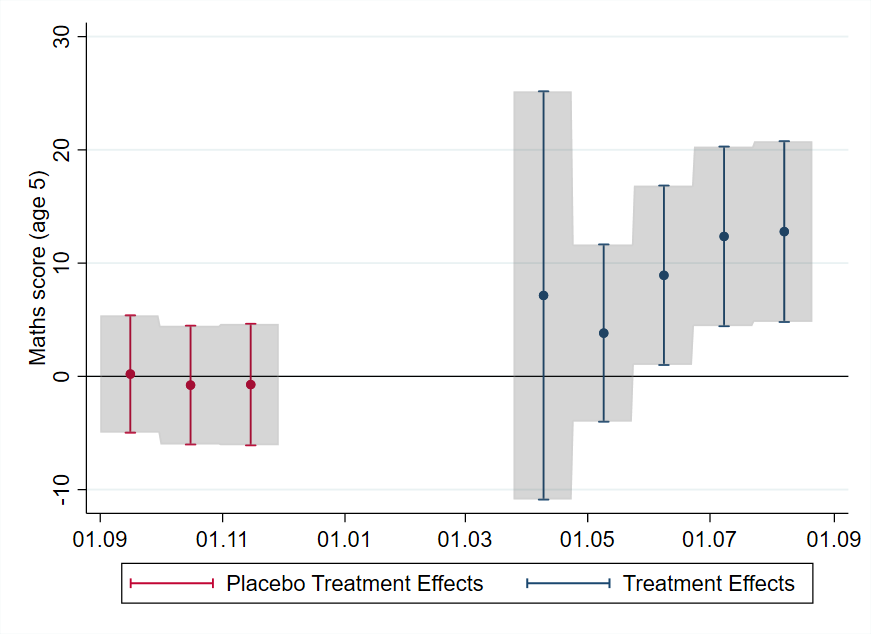}  &
        \includegraphics[width=0.45\textwidth]{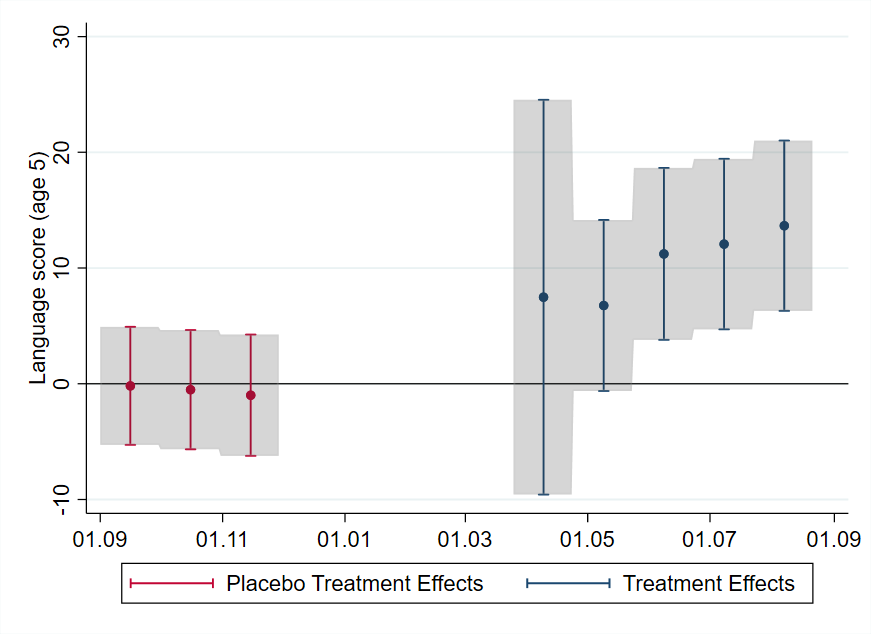} \\
        {\small Panel A. Maths score (age 5)} & {\small Panel B. Language score (age 5)} \\
        \includegraphics[width=0.45\textwidth]{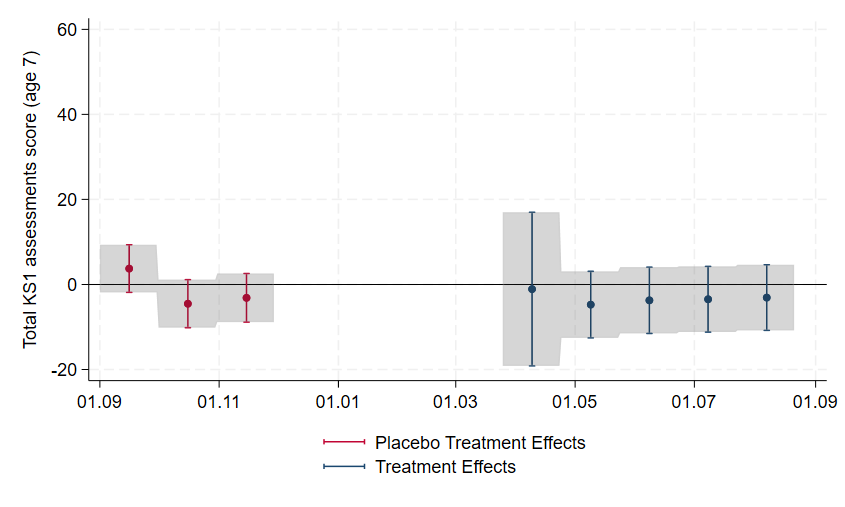} &
        \includegraphics[width=0.45\textwidth]{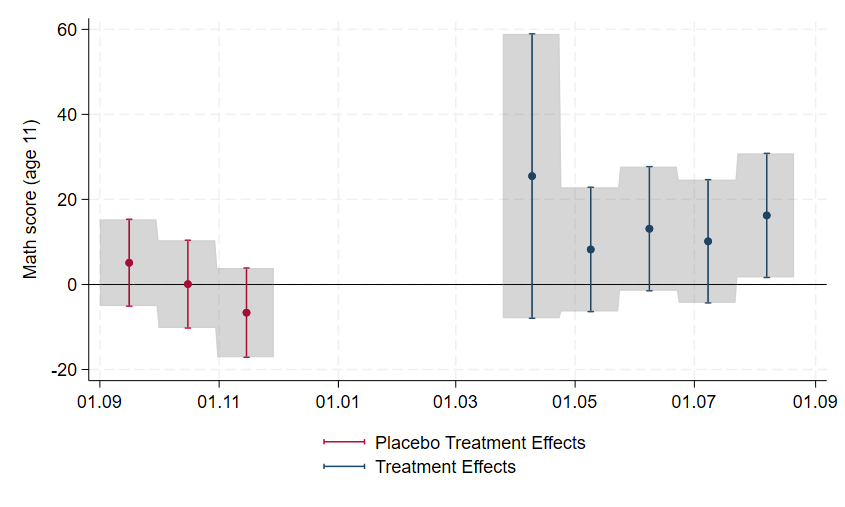} \\
        {\small Panel C. Total score (age 7)} & {\small Panel D. Maths score (age 11)} \\
    \end{tabular}

    \refstepcounter{figure}
    \label{fig:robustness2003}

    \vspace{0.35cm}
    \begin{small}
        \textbf{Figure~\thefigure: Outcome and age robustness for the 2003 bundle reform}
    \end{small}

    \vspace{0.3cm}
    \begin{minipage}{0.9\textwidth}
        \begin{footnotesize}
            \begin{spacing}{1.15}
                \textit{Notes:} The first row presents the effects of the 2003 bundle reform on maths and language outcomes at age 5. The second row (total score at age 7 and maths score at age 11) examines the persistence of the effects at later ages. Estimates are aggregated at the monthly level and expressed in days of age (see the notes to Figure~\ref{fig:coefplot}); placebo estimates for pre-reform birth dates with $p^t(x) = 0$, constructed as described in the notes to Figure~\ref{fig:fullsample2000}, are shown in red. Vertical lines and shaded areas indicate pointwise 95\% confidence intervals. The vertical scales differ across rows. Horizontal-axis tick labels give the date of birth in day.month format.
            \end{spacing}
        \end{footnotesize}
    \end{minipage}
\end{figure}

\begin{figure}[!htbp]
    \centering
    \begin{tabular}{cc}
        \includegraphics[width=0.45\textwidth]{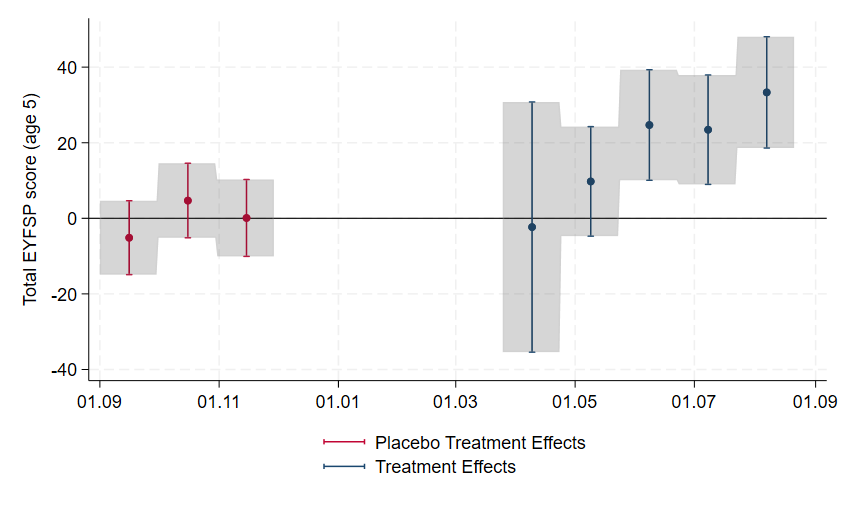} &
        \includegraphics[width=0.45\textwidth]{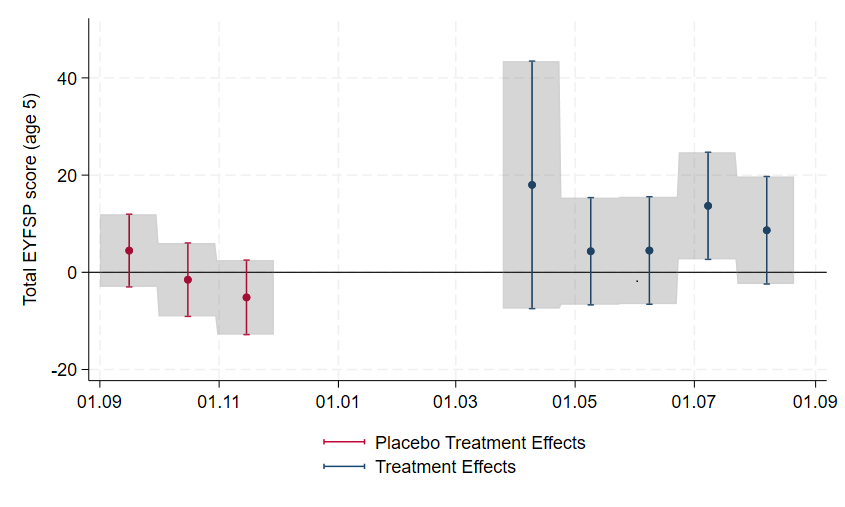}   \\
        {\small Panel A. Rural regions} & {\small Panel B. Urban regions} \\
        \includegraphics[width=0.45\textwidth]{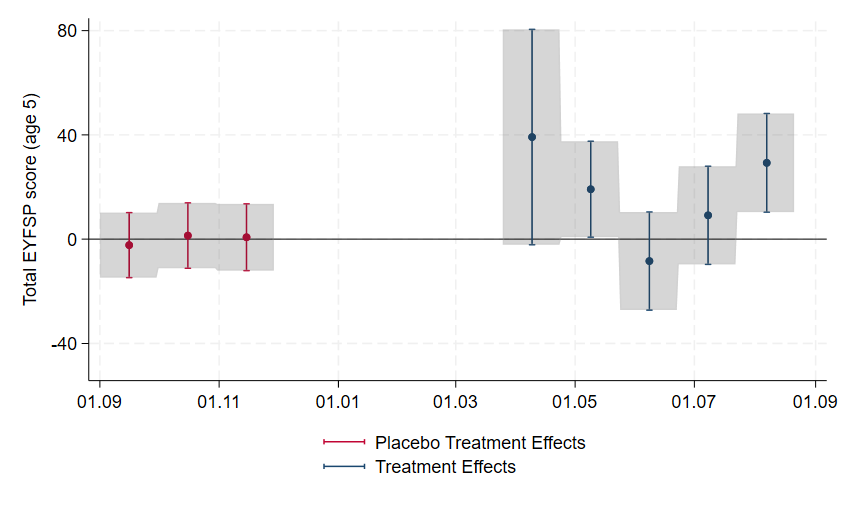} &
        \includegraphics[width=0.45\textwidth]{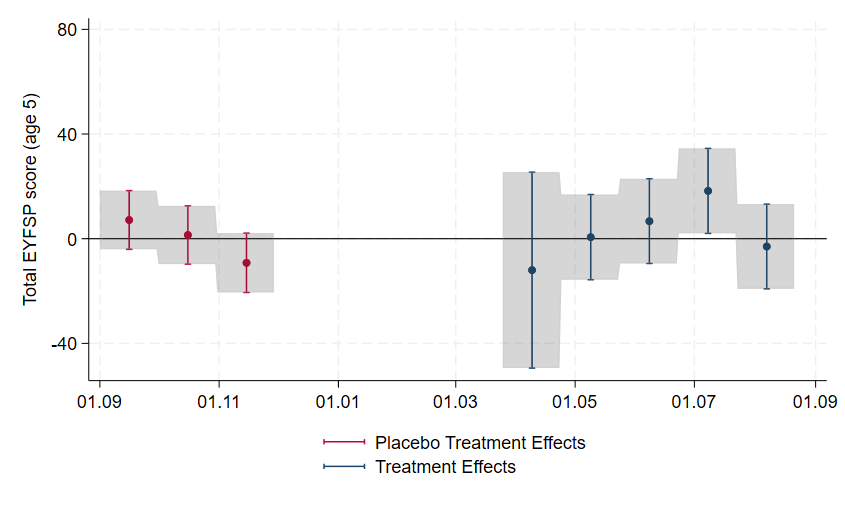}   \\
        {\small Panel C. Bottom $20\%$ child poverty} & {\small Panel D. Top $20\%$ child poverty} \\
    \end{tabular}

    \refstepcounter{figure}
    \label{fig:geo2003}

    \vspace{0.35cm}
    \begin{small}
        \textbf{Figure~\thefigure: Geographic heterogeneity in the effects of the 2003 bundle reform}
    \end{small}

    \vspace{0.3cm}
    \begin{minipage}{0.9\textwidth}
        \begin{footnotesize}
            \begin{spacing}{1.15}
                \textit{Notes:} Similar to Figure~\ref{fig:geo2000}, this figure presents geographic heterogeneity in the effects of the 2003 bundle reform. The top row compares rural and urban areas, while the bottom row contrasts areas with low and high child poverty. The outcome variable is the total score at age 5 (EYFSP); estimates are aggregated at the monthly level. Placebo estimates for pre-reform birth dates with $p^t(x) = 0$, constructed as described in the notes to Figure~\ref{fig:fullsample2000}, are shown in red. Estimates are expressed in days of age (see the notes to Figure~\ref{fig:coefplot}); vertical lines and shaded areas indicate pointwise 95\% confidence intervals. The vertical scales differ across panels. Horizontal-axis tick labels give the date of birth in day.month format.
            \end{spacing}
        \end{footnotesize}
    \end{minipage}
\end{figure}

\begin{figure}[!htbp]
    \centering
    \begin{tabular}{cc}
        \includegraphics[width=0.45\textwidth]{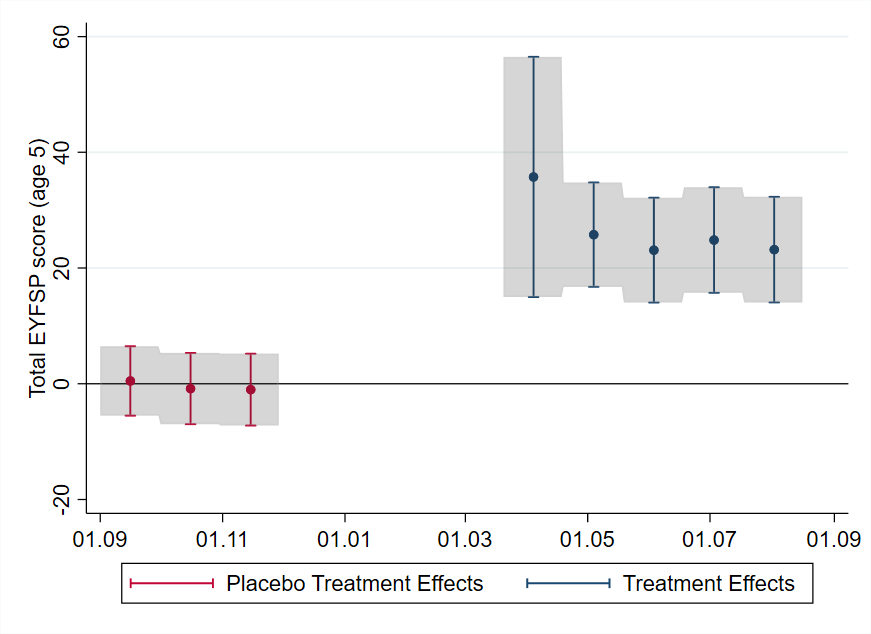}  &
        \includegraphics[width=0.45\textwidth]{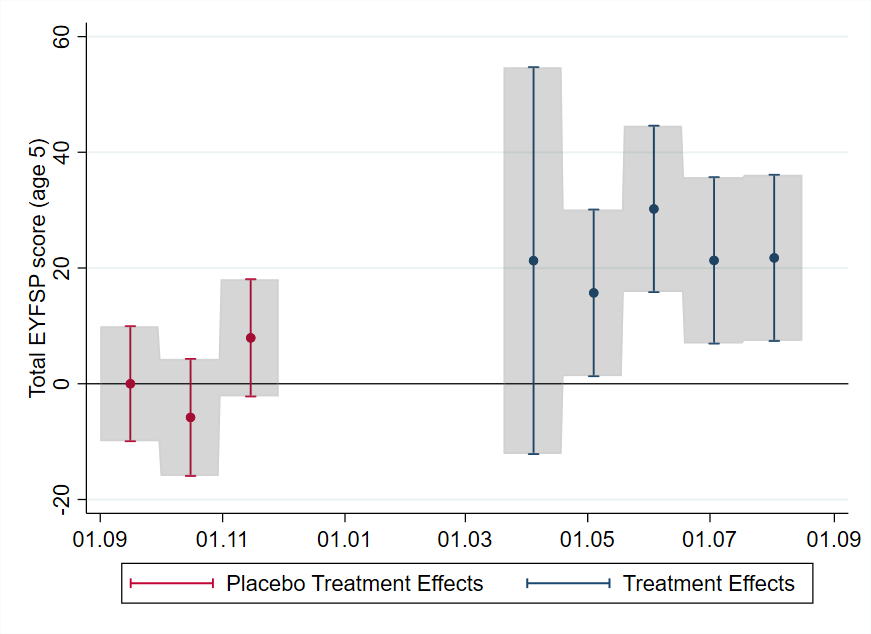} \\
        {\small Panel A. Ethnicity: white} & {\small Panel B. Ethnicity: non-white} \\
        \includegraphics[width=0.45\textwidth]{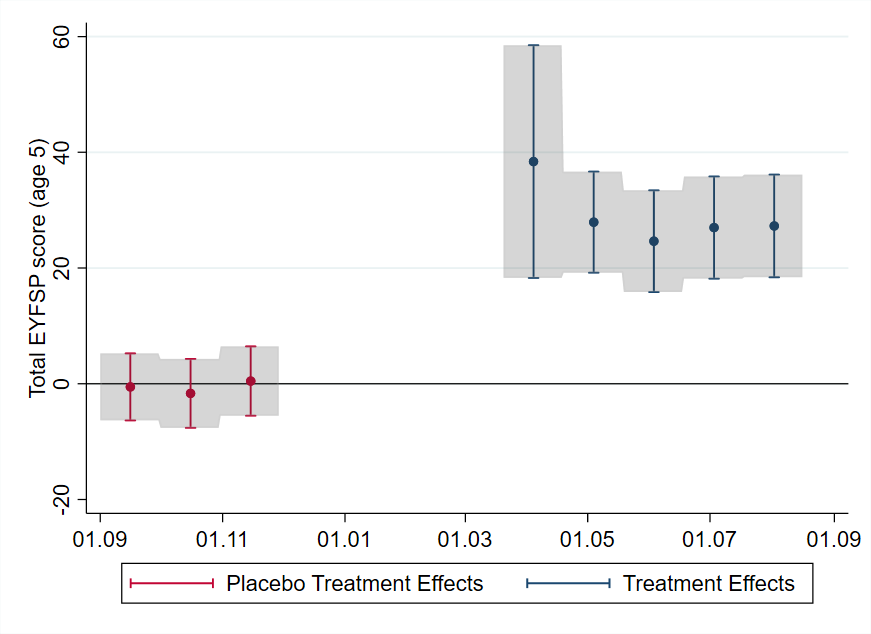}  &
        \includegraphics[width=0.45\textwidth]{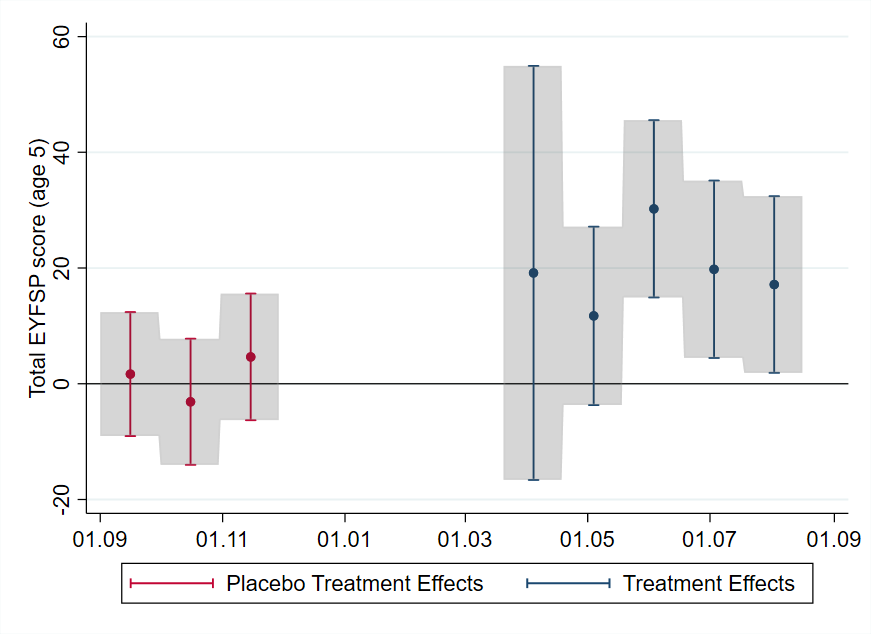} \\
        {\small Panel C. English first language} & {\small Panel D. Foreign first language} \\
        \includegraphics[width=0.45\textwidth]{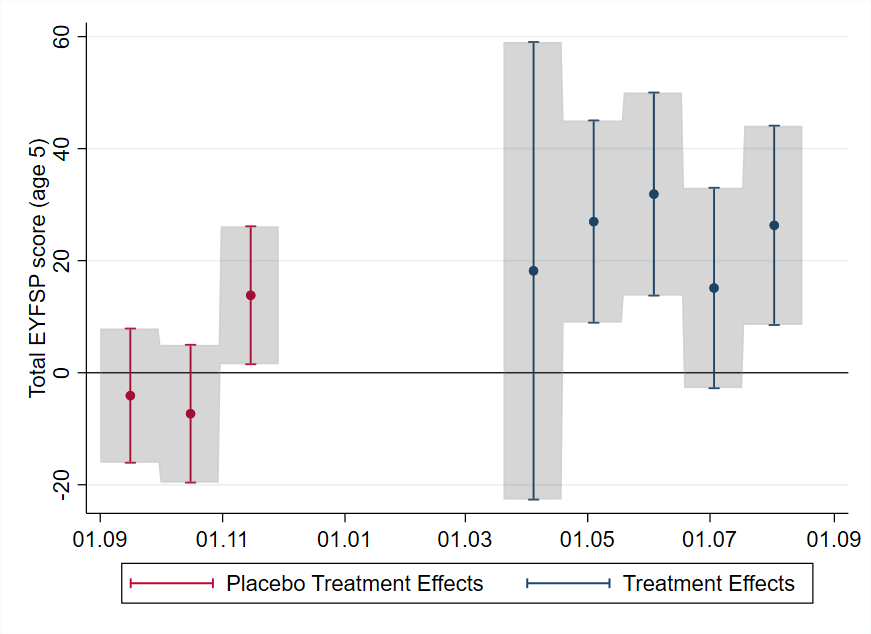}  &
        \includegraphics[width=0.45\textwidth]{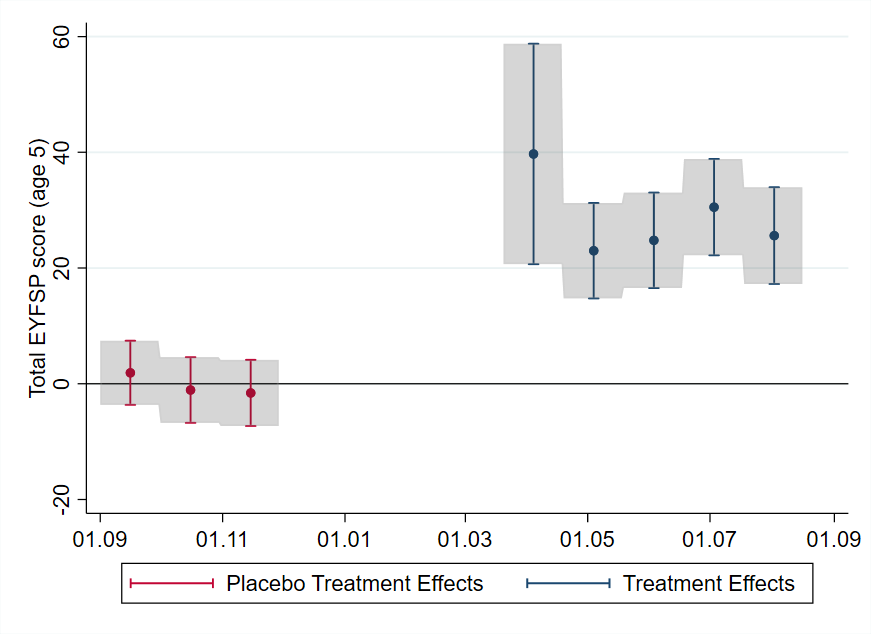} \\
        {\small Panel E. FSM eligible} & {\small Panel F. FSM ineligible} \\
    \end{tabular}

    \refstepcounter{figure}
    \label{fig:heterogeneity2007}

    \vspace{0.35cm}
    \begin{small}
        \textbf{Figure~\thefigure: Individual-level heterogeneity in the effects of the 2007 pay reform}
    \end{small}

    \vspace{0.3cm}
    \begin{minipage}{0.9\textwidth}
        \begin{footnotesize}
            \begin{spacing}{1.15}
                \textit{Notes:} This figure analyses heterogeneity in the effects of the 2007 pay reform. The top row splits the sample by ethnicity, the middle row by first language, and the bottom row by free school meal (FSM) eligibility. The outcome variable is the total score at age 5 (EYFSP); estimates are aggregated at the monthly level. Placebo estimates for pre-reform birth dates with $p^t(x) = 0$, constructed as described in the notes to Figure~\ref{fig:fullsample2000}, are shown in red. Estimates are expressed in days of age (see the notes to Figure~\ref{fig:coefplot}); vertical lines and shaded areas indicate pointwise 95\% confidence intervals. Horizontal-axis tick labels give the date of birth in day.month format.
            \end{spacing}
        \end{footnotesize}
    \end{minipage}
\end{figure}

Figure~\ref{fig:geo2000} reports geographic heterogeneity in the effects of the 2000 reform, Figure~\ref{fig:robustness2003} examines the robustness of the 2003 estimates across outcomes and test ages, Figure~\ref{fig:geo2003} reports geographic heterogeneity in the effects of the 2003 reform, and Figure~\ref{fig:heterogeneity2007} reports individual-level heterogeneity in the effects of the 2007 reform, complementing the analyses of Section~\ref{sec:findings}.

\end{document}